\documentclass[a4paper,fleqn]{cas-dc}
\usepackage[numbers]{natbib}
\usepackage{threeparttablex, booktabs, url, amsmath, xr, subfiles}

\def\tsc#1{\csdef{#1}{\textsc{\lowercase{#1}}\xspace}}
\tsc{WGM}
\tsc{QE}

\begin{document}

\shorttitle{Zhang et al.}
\title[mode = title]{Probing Globular Cluster with MeerKAT and FAST: A Pulsar Polarization Census}

\author[1,2]{Lei Zhang}
\cormark[1]
\ead{leizhang996@nao.cas.cn}

\author[3,4]{Federico Abbate}
\cormark[1]
\ead{federico.abbate@inaf.it}

\author[5,1]{Di Li}
\cormark[1]
\ead{dili@tsinghua.edu.cn}

\author[3]{Andrea Possenti}
\author[2,6]{Matthew Bailes}
\author[7]{Alessandro Ridolfi}
\author[4]{ Paulo C. C. Freire}
\author[8]{Scott M. Ransom}
\author[1]{Yong-Kun Zhang}
\author[9,10]{Meng Guo}
\author[10]{Meng-Meng Ni}
\author[10]{Jia-Le Hu}
\author[11,12]{Yi Feng}
\author[1]{Pei Wang}
\author[13]{Jie Zhang}
\author[14]{Qi-Jun Zhi}

\affiliation[1]{organization={National Astronomical Observatories},
            addressline={Chinese Academy of Sciences},
            city={Beijing},
            postcode={100101},
            country={China}}

\affiliation[2]{organization={Centre for Astrophysics and Supercomputing},
            addressline={Swinburne University of Technology},
            city={VIC},
            postcode={3122},
            country={Australia}}

\affiliation[3]{organization={INAF -- Osservatorio Astronomico di Cagliari},
            addressline={Via della Scienza 5},
            city={Selargius},
            postcode={I-09047},
            country={Italy}}

\affiliation[4]{organization={Max-Planck-Institut f\"{u}r Radioastronomie},
            addressline={Auf dem H\"{u}gel 69},
            city={Bonn},
            postcode={D-53121},
            country={Germany}}

\affiliation[5]{organization={Department of Astronomy},
            addressline={Tsinghua University},
            city={Beijing},
            postcode={100190},
            country={China}}

\affiliation[6]{organization={ARC Center of Excellence for Gravitational Wave Discovery (OzGrav)},
            addressline={Swinburne University of Technology},
            city={VIC},
            postcode={3122},
            country={Australia}}

\affiliation[7]{organization={Faculty of Physics},
            addressline={University of Bielefeld},
            city={Bielefeld},
            postcode={33501},
            country={Germany}}

\affiliation[8]{organization={National Radio Astronomy Observatory},
            addressline={520 Edgemont Road},
            city={Charlottesville},
            postcode={22903},
            country={USA}}

\affiliation[9]{organization={National Supercomputing Center in Jinan},
            addressline={Qilu University of Technology},
            city={Jinan},
            postcode={250103},
            country={China}}

\affiliation[10]{organization={Jinan Institute of Supercomputing Technology},
            addressline={28666 East Jingshi Road},
            city={Jinan},
            postcode={250103},
            country={China}}

\affiliation[11]{organization={Research Center for Intelligent Computing Platforms},
            addressline={Zhejiang Laboratory},
            city={Hangzhou},
            postcode={311100},
            country={China}}

\affiliation[12]{organization={Institute for Astronomy, School of Physics},
            addressline={Zhejiang University},
            city={Hangzhou},
            postcode={310027},
            country={China}}

\affiliation[13]{organization={College of Physics and Electronic Engineering},
            addressline={Qilu Normal University},
            city={Jinan},
            postcode={250200},
            country={China}}

\affiliation[14]{organization={Guizhou Provincial Key Laboratory of Radio Astronomy and Data Processing},
            addressline={Guizhou Normal University},
            city={Guiyang},
            postcode={550001},
            country={China}}
            
\cortext[1]{Corresponding author}

\maketitle

Magnetic fields are pervasive throughout the Universe. They are integral to a wide array of astrophysical processes that span various physical scales and field strengths. The Galactic magnetic field, in particular, holds significant importance in shaping the evolution of our Galaxy. However, our understanding of its behavior on small scales remains poor, especially when considering its penetration into the Galactic halo~\citep{Widrow02}.

Pulsars in Globular Clusters (GCs) are ideal probes to study the small-scale structure of the Galactic magnetic field as GCs typically host large numbers of pulsars with small angular separation, from arcseconds to arcminutes, and their distances are known with great accuracy. To date, only one GC, 47 Tucanane (hereafter 47 Tuc), has been found to contain intracluster medium, with an electron density 100 times higher than that of the interstellar medium (ISM) in its vicinity~\citep{Freire01, Abbate18}. The characteristics of this intracluster medium are closely related to GC evolution and the compact objects within. However, significant knowledge gaps remain regarding the ionized gas content of GCs, particularly in Galactic halo clusters.

Recent advances in radio astronomy, particularly the advent of sensitive facilities such as the Five-hundred-meter Aperture Spherical radio Telescope (FAST)~\citep{Nan11,Li18} in the northern hemisphere and the MeerKAT radio telescope~\citep{Jonas16} in the southern hemisphere, have revitalized the field. These telescopes enable high-quality polarization studies of GC pulsars, providing unprecedented opportunities to probe their emission properties, the Galactic magnetic field, and the elusive intracluster gas.

\begin{figure*}
\centering
\includegraphics[width=0.9\linewidth]{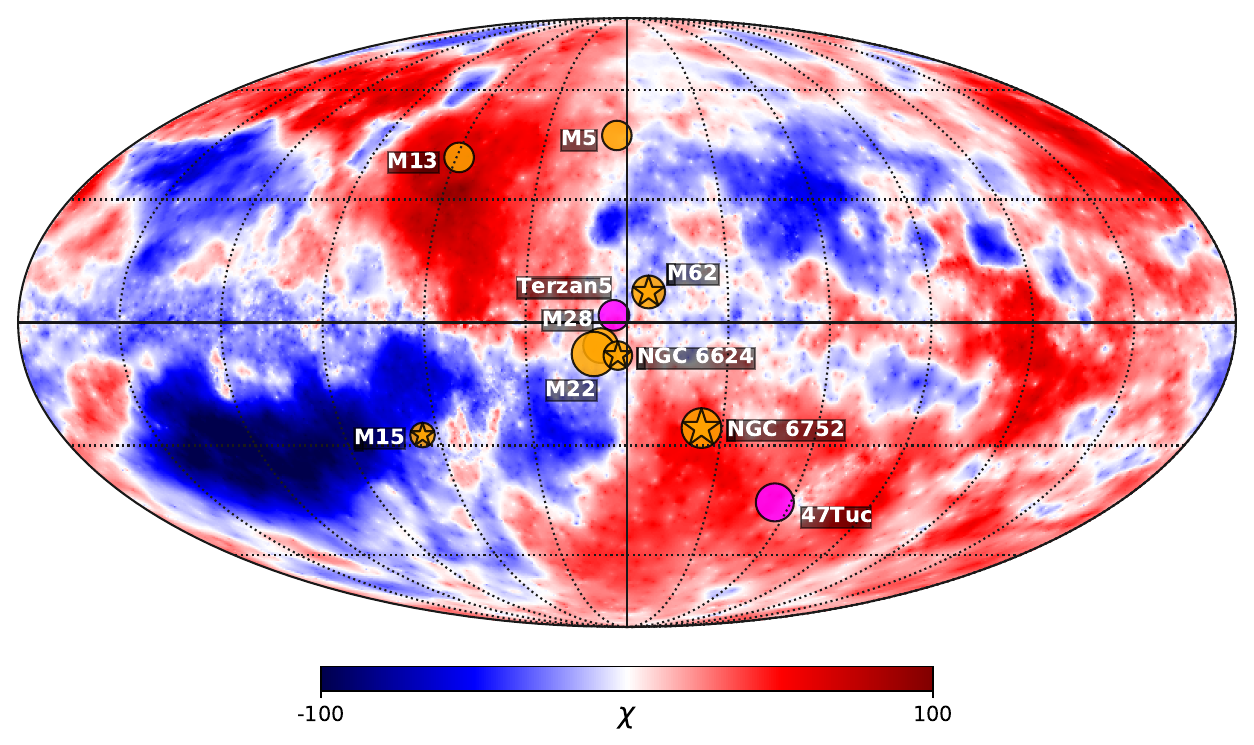}
\caption{Distribution of 10 GCs in the Milky Way. Core-collapsed GCs are marked with circles containing a star, while non-core-collapsed GCs are represented by plain circles. The size of each circle is inversely proportional to the cluster's distance, ranging from M15 (10.7\,kpc) to M22 (3.3\,kpc). Orange points mark the eight GCs analyzed for polarization profiles and RMs using MeerKAT and FAST in this study, while magenta points denote two well-studied GCs from previous work~\citep{Abbate2020,Abbate23,Martsen22}. The background image is adapted from the Galactic Faraday rotation sky 2020. The sign field $\chi$ is dimensionless and represents the sign or orientation of the magnetic field along the line of sight~\citep{Hutschenreuter2022}}.
\label{fig.MagGCs}
\end{figure*}

In this study, we carried out a polarization census of GC pulsars using MeerKAT and FAST. Table~S1 (online) lists each GC's main characteristics and the distribution of these GCs in the Milk way shows in Fig.~\ref{fig.MagGCs}. The details of the observations and data reduction can be found in Supplementary materials A. This first combined effort of observations from these two major radio telescopes resulted in high signal-to-noise ratio, full polarization pulse profiles for 43 pulsars in 8 GCs, doubling the number of rotation measures (RMs) known in these clusters (Supplementary materials Section B and Table S4 online). The accuracy of dispersion measures (DMs) was improved by a factor of 8 compared to previous publications.

\begin{figure*}
\centering
\includegraphics[height=12cm,width=12cm]{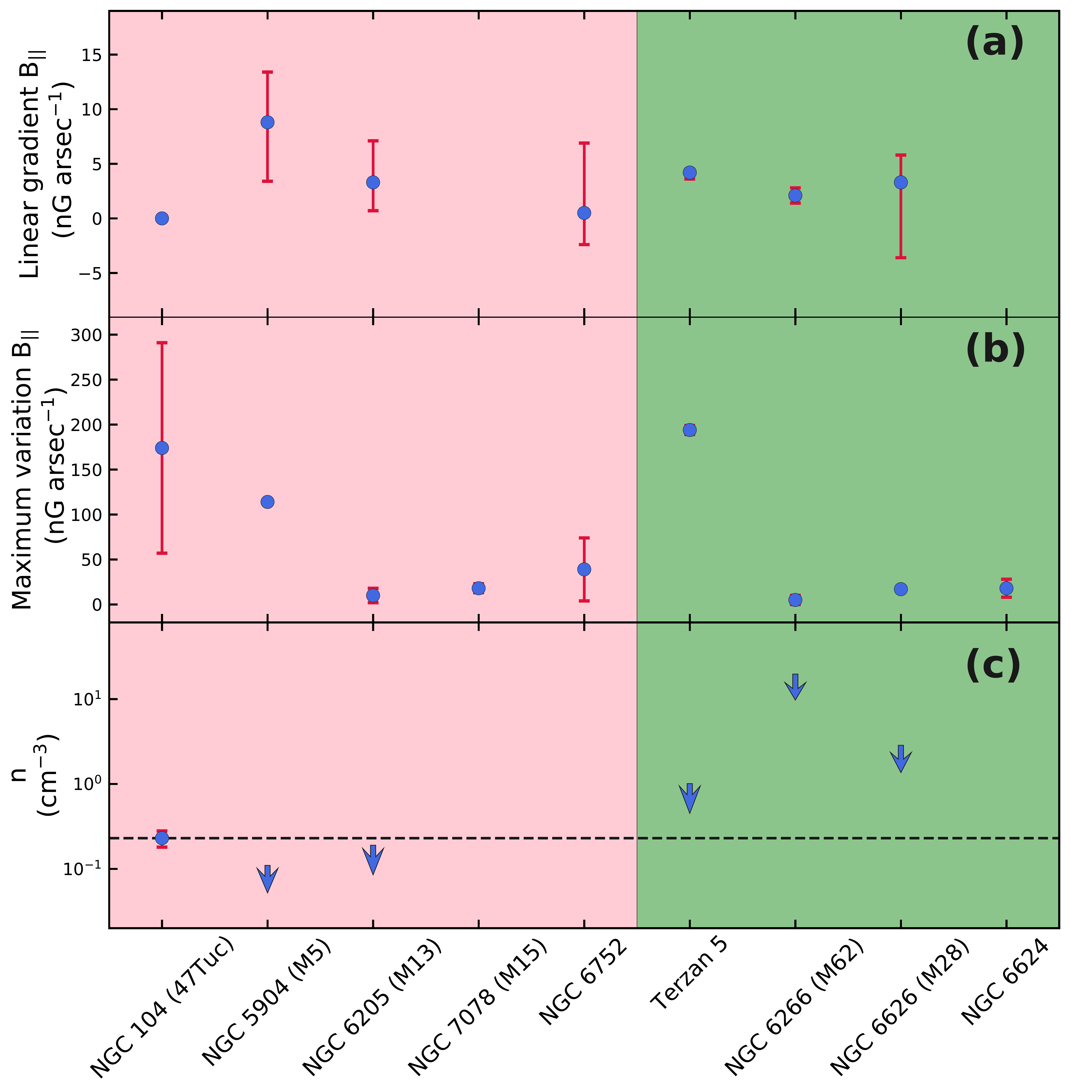}
\caption{The linear gradient and maximum variation of the magnetic in front of the cluster (a,b) and the upper limit of ionized gas in the cluster (c). The value of 47\,Tuc shows in plot (c) is the ionized gas density detection from \protect\cite{Abbate18} and marks as the dark horizontal dashed line. The shaded regions indicate: halo clusters (pink) and bulge clusters (green).}
\label{fig.B-gas}
\end{figure*}

The RMs, combined with DMs derived from the summed total intensity profiles, have enabled us to map the projected and averaged parallel magnetic field strength across the spatial extent of the clusters (Supplementary materials C). We look for the presence of a linear gradient of $\left<B_{||}\right>$ across the GCs by performing a fit over the direction in the plane of the sky, intensity and value of $\left<B_{||}\right>$ in the center of the GC. The results of the fits are reported in Table~S6 (online) and Fig.~\ref{fig.B-gas}a.
The clusters that appear to show a linear gradient are Terzan 5 and M62. In these cases the linear gradients are $4.2_{-0.6}^{+0.3}$ nG arcsec$^{-1}$ and $2.1_{-0.7}^{+0.7}$ nG arcsec$^{-1}$ respectively.
The presence of these gradients implies that that the injection scale of turbulence is larger than the area covered by the GC pulsars. This allows us to put lower limits on the energy injection scale in the direction of Terzan 5 and M62 of 1 pc and 0.5 pc respectively.
These results are consistent with the values reported in Refs. \citep{Minter96,Haverkorn08}.

The variations of $\left<B_{||}\right>$ between single pulsars can be informative on the strengths contained in the fluctuations at small scales. Table~S6 (online) and Fig.~\ref{fig.B-gas}b shows the largest difference of $\left<B_{||}\right>$ measured between pairs of pulsars in the GC. The GCs M5 and Terzan 5 stand out as the only ones where this quantity is statistically larger than the linear gradient. This quantity does not seem to be affected by the location of the GCs within the Galaxy. However, these results are affected by the low number of pulsars in many of the selected clusters. Figs.~S12 and  S13 (online) present the measured RM and DM against position perpendicular offset from the center of GC, illustrating the variations observed for pulsars within the host cluster.

Additionally, we investigated the presence of ionized gas within GCs. None of the GCs in the current sample, with the exception of 47 Tuc, show any trend in the plot of DM versus line-of-sight position indicating the presence of ionized gas. By performing a fit that includes the ionized gas and the variations due to the ISM, assumed to be gaussian, we can derive upper limits on the density of ionized gas. These are listed in Table~S7 (online) and Fig.~\ref{fig.B-gas}c. The trend of smaller upper limits for ionized gas in halo GCs likely result from lower ambient electron densities in the halo compared to the denser ISM environment of bulge GCs. These upper limits can be compared to the simulations run by Ref. \cite{Chantereau20} to determine the dominant gas clearing processes.

Simliar to Ref.~\cite{Freire01}, our study reveals that electron density of GCs are much below the expected intra-cluster medium (ICM) accumulation based on stellar evolution~\cite{McDonald15}. The surprising barrenness of GCs suggests effective gas removal mechanisms, such as strong winds from ionizing radiation from post-AGB stars and young white dwarfs~\cite{Chantereau20, McDonald15}. Such remarkably low electron densities provide valuable benchmarks for theoretical models of intracluster gas dynamics and Galactic ISM interaction.

Searches and monitoring of pulsars in GCs with MeerKAT and FAST are ongoing. GCs located at high Galactic latitudes and hosting a large number of pulsars are the most promising targets for investigating internal gas and magnetic fields, as well as for probing turbulence in electron density and magnetic fields along the line of sight.

\section*{Conflict of interest}
The authors declare that they have no conflict of interest.

\section*{Acknowledgments}
This work was supported by the National Natural Science Foundation of China (NSFC) (11988101, 12103069, 11725313, 12273008, 12373109, and 12203045), the National Key R\&D Program of China (2017YFA0402600 and 2023YFB4503300), the National SKA Program of China (2022SKA0130100 and 2022SKA0130104), the Shandong Provincial Key R\&D Program (2022CXGC020106), and the Zhejiang Provincial grants (2023R01008 and 2024SSYS0012). Additional support came from the Pilot Project for Integrated Innovation of Science, Education and Industry of Qilu University of Technology (2022JBZ01-01). Di Li is a New Cornerstone investigator. Federico Abbate acknowledges funding from NextGenerationEU under the Italian PNRR (Project IR0000034 – STILES). This work utilized data from MeerKAT and FAST, with data processing conducted primarily on the OzSTAR national facility at Swinburne University of Technology. We thank Dong-Zi Li and Marcus Lower for valuable discussions.

\section*{Data availability}
Our processed data collection is publicly available via the Science Data Bank\footnote{\url{https://doi.org/10.57760/sciencedb.Fastro.00019}}, including polarization-calibrated, full-polarization, DM- and RM-corrected MeerKAT and FAST data in PSRFITS format. Full polarization profiles from this work will also be shared through the European Pulsar Network (EPN) database. FAST raw data will be accessible via the FAST Data Access Portal\footnote{\url{https://fast.bao.ac.cn/}}  after a 12-month embargo. MeerKAT data can be obtained through the MEERTIME Data Access portal\footnote{\url{https://pulsars.org.au}}.

\section*{Author contributions}
Lei Zhang and Federico Abbate led the data analysis, interpretation, and manuscript preparation. Di Li launched the FAST observations and contributed to writing. Andrew Possenti, Matthew Bailes, Alessandro Ridolfi, and Paulo C. C. Freire conducted the MeerKAT observations, while Scott M. Ransom provided Green Bank Telescope polarization profiles for Terzan 5 pulsars. Yong-Kun Zhang, Meng Guo, Meng-Meng Ni, Jia-Le Hu, Yi Feng, Pei Wang, Jie Zhang, and Qi-Jun Zhi contributed to FAST data analysis. All authors contributed to data interpretation and manuscript finalization.
\printcredits

\bibliographystyle{model3-num-names}
\bibliography{cas-refs}

\clearpage

\shorttitle{Zhang et al.}
\centerline{\bf\huge Supplementary Text}

\setcounter{section}{0}
\renewcommand{\thesection}{S\arabic{section}}
\setcounter{equation}{0}
\renewcommand{\theequation}{S\arabic{equation}}
\setcounter{figure}{0}
\renewcommand{\thefigure}{S\arabic{figure}}
\setcounter{table}{0}
\renewcommand{\thetable}{S\arabic{table}}

\appendix
\section{Observations and data reduction} \label{sec:obs}
\subsection{Source selection}
Polarization studies of the pulsars in GCs can be challenging, because most of them are at large distances, which makes their flux density typically weak and signals strongly distorted by propagation through the interstellar medium (ISM)~\citep{Zhang19}. Moreover, the pulsars in GCs are often members of tight binary systems, causing large changes in their observed spin period and sometimes periodic eclipsing of the radio signal~\citep{Li23}. Among all the GCs observed with the MeerKAT and FAST, we selected a sample of five from MeerKAT and three from FAST. The clusters were chosen to study in this work based on two main criteria: 
\begin{enumerate}[1.]
\item  Each of these clusters hosts more than one known pulsar. This requirement allows us to get the polarization properties and RM values of the cluster pulsars to study the magnetic field in the clusters and the small-scale magnetic field in the Galactic disk. 

\item  To make much higher signal-to-noise polarization-calibrated pulse profiles of many of the GC pulsars, each of these clusters needs enough full polarization information data and the best known timing ephemeris for a given pulsar, allowed us to integrate each observation in time.
\end{enumerate}

\subsection{Observations} 
\subsubsection{MeerKAT}
The five selected GCs were observed from 2019 July to 2024 March with the MeerKAT radio telescope in South Africa using the UHF (544–1088\,MHz) receivers and the L-band (856–1712\,MHz) receivers. The campaign made use of the PTUSE (Pulsar Timing User Supplied Equipment) machines~\citep{Bailes20} and their ability to synthesize four different tied-array beams to observe different parts of the cluster simultaneously. The full list of MeerKAT observations used for this work is given in Table~\ref{tab:MeerKATobs}, which lists the number of antennas, number of channels, observation length, value of DM used for each beam coherent de-dispersion, sampling time, central frequency and bandwidth.

The number of antennas used for the timing beams changed during the campaign based on their availability. For the search beams, the number of antennas was reduced in order to increase the field of view and thus cover a larger area of the cluster. Using simulations of the size and orientation of the beams as they evolved during the observation, we derived an optimal number of antennas to use in order to observe the largest number of pulsars. All of the observations were recorded in full-Stokes so as to recover the polarimetric information. The polarization calibration of pulsar observations at MeerKAT is described in~\cite{Serylak21}.

\subsubsection{FAST}
The three selected GCs were observed from 2020 September to 2023 October with the FAST in China using the central beam (the beam width is 3\,arcmin at 1250\,MHz) of the FAST 19-beam receiver. A summary of all the FAST observations used in this paper is given in Table~\ref{tab:FASTobs}, which lists the point of the telescope, sampling time, number of channels, observation length and project ID.

All the FAST observations were recorded with 8-bit sampling in pulsar search mode along with full-Stokes information. Each pulsar observation started with a one-minute calibration noise diode for the polarization calibration, while the flux calibrator observations were not performed. In all the observations, the observing band from 1000\,MHz to 1500\,MHz and due to bandpass roll-off the effective band is from 1050\,MHz to 1450\,MHz.

\subsection{Data Reduction} 
First, we folded each search-mode observation with the best known timing ephemeris for a given pulsar using \textsc{dspsr}\footnote{\url{http://dspsr.sourceforge.net}}~\citep{vanS11}. We removed data affected by radio frequency interference (RFI) both in the frequency and time domains for each observations using the \texttt{paz} routine of \textsc{psrchive}\footnote{\url{http://psrchive.sourceforge.net}\label{psrchive}}~\citep{Hotan04} package. After that, we combined each set of observations using the same backend and receiver using the \texttt{psradd} of \textsc{psrchive}. We then utilized the \texttt{pac} routine from \textsc{psrchive} to correct for variations in the parallactic angle. The Stokes parameters were calibrated following the astronomical conventions outlined by~\cite{vanS10}. For pulsars exhibiting eclipses during the observations, we excluded the orbital phases near the eclipse to minimize their potential impact on the linear polarization percentage as well as the DM and RM values~\cite{Li23}.

To determine the DM for each pulsar, we generated times of arrival (ToAs) from all combined observations using the noise-free template. The DM was measured with \textsc{tempo2}~\citep{Hobbs06} using ToAs from multifrequency subbands. The DM measurements of each pulsars are reported in Table~\ref{tab:RMs_GCs}. We installed these updated DMs into the original data files and performed the summation one final time. This last iteration improved the S/N of the final sums for many of the faintest pulsars, particularly for those that previously had less accurate DMs due to low S/N.

To measure the RMs, we again used programs from the {\sc psrchive} package. Once the correct DM is established we then use the same data and the routine \textsc{rmfit} to calculate the RM. This
routine performs trial RMs and finds the value at which the linear polarization is maximized across the profile as a whole. The RM measurements of each pulsars are also listed in Table~\ref{tab:RMs_GCs} with the errors on the last digit of RM given in brackets. We did not correct the measured RMs for ionospheric contributions for two main reasons: (1) Pulsar RMs are known to vary over time due to inaccuracies in models of the ionospheric Faraday rotation (e.g., \citealt{Yan11}), and not all RMs reported in the ATNF pulsar catalogue\footnote{\url{https://www.atnf.csiro.au/research/pulsar/psrcat/}} are corrected for the ionosphere; (2) the Ionospheric Faraday rotation likely introduces a systematic bias for each of the measurements. However, these biases should be consistent for each pulsar within the same cluster, meaning that relative comparisons of the RMs remain unaffected. 

To obtain the polarization pulse profiles, we used the final time- and frequency-averaged data, which were corrected for dispersion and Faraday rotation using the DM and RM measurements reported in Table~\ref{tab:RMs_GCs}. The measured RM was then used to refer all measured position angles (PAs) to the overall band center, before summing in frequency to form the average polarization pulse profiles. For each pulsar we also measure the percentage of linear polarization (L/I), which listed in the first column of Table~\ref{tab:RMs_GCs}, from the average polarization pulse profile. The measured linear polarization is overestimated in the presence of noise. To remove the bias, we determined the off-pulse standard deviation of Stokes I using $\sigma_{I}=\text{rms(I)}\sqrt{n_{\text{pulse}}}$, where rms(I) is the root mean square of the total intensity profile in the off-pulse region, and $n_{\text{pulse}}$ is the number of bins within the pulse region~\cite{Wardle74}. We then use the frequency-averaged, de-biased total linear polarization~\citep{Everett01}:
\begin{equation}
L_{de-bias}=
\begin{cases}
 &\sqrt{(\frac{L_{meas}}{\sigma_{I}})^{2}-\sigma_{I}} \text{~~~~if~~} \frac{L_{meas}}{\sigma_{I}}>1.57\\
 &  0  \text{~~~~~~~~~~~~~~~~~~~~~~~~~~~~~~otherwise},
\end{cases}
\end{equation}
Where $L_{meas}$ is the measured frequency-averaged linear polarization, it is a positive definite quantity. The summed DM- and RM-corrected polarization pulse profiles are shown in Figure~\ref{fig.M5_polprof} to \ref{fig.M15_polprof}.

\section{DM, RM Measurements and Polarization Profiles}
\subsection{DM and RM Measurements}\label{sec:dm-rm}
We obtained polarization pulse profiles for 46 pulsars in eight different GCs using MeerKAT and FAST data with full-Stokes information. The DM- and RM-corrected polarization pulse profiles for the pulsars we analyzed are shown in Figure~\ref{fig.M5_polprof} to Figure~\ref{fig.M15_polprof}, and further discussed in Section~\ref{sec:Polprof}.
These high-quality data, enabling us to measure accurate ISM parameters, explore each pulsar’s emission properties and probe the small-scale structure of the Galactic magnetic field. Table~\ref{tab:RMs_GCs} provides a summary of the DM and RM measurements, $\left< B_{||}\right>$ estimates along with the percentage of linear polarization (L/I), circular polarisation (V/I) and absolute circular polarisation ($|V|$/I) towards each pulsar.

The DMs we measured for GC pulsars in this work range from 29.5\,pc cm$^{-3}$ (towards pulsars in M5) to 119.7\,pc cm$^{-3}$ (towards pulsars in M28). The mean of the uncertainties on the DM measurements is 0.005\,pc cm$^{-3}$. Comparison the mean uncertainty on the DMs from the ATNF Pulsar Catalogue (v2.4) is 0.04\,pc cm$^{-3}$. This factor of 8 improvement is largely due to due to the high signal-to-noise profiles, which result from long-duration, combined data. 

\begin{figure*}
\centering 
\includegraphics[height=10cm,width=18cm]{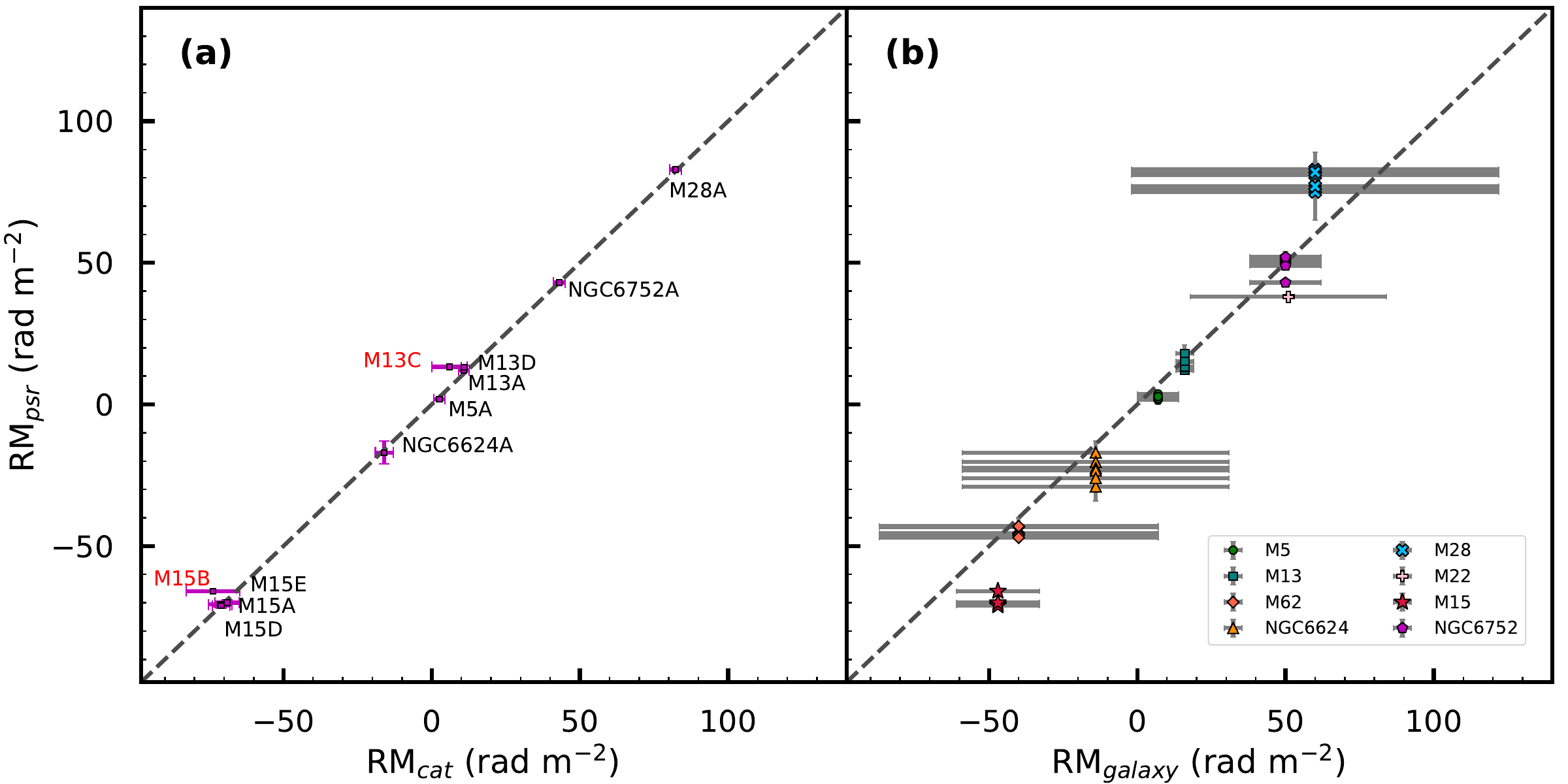}
\caption{(a) A comparison between the RMs measured in this work (RM$_{psr}$) and the catalogue value (RM$_{cat}$) for ten pulsars in GCs. The black line corresponds to the RM$_{psr}$ = RM$_{cat}$ trend, which we expect to see. M13C and M15B are highlighted as they significantly deviate from the trend due to the large uncertainties in the previously published RMs for these two pulsars. (b) A comparison between the RMs measured in this work (RM$_{psr}$) and the Galactic contribution to RM (RM$_{galaxy}$) estimated from the background sources in the area surrounding the GC  obtained from the Galactic Faraday rotation sky 2020~\citep{Hutschenreuter2023}, shown by the data points with error bars representing the uncertainties. Pulsars in M15 show large deviations from the dashed line RM$_{psr}$ = RM$_{galaxy}$.}
\label{fig.RMpsr-RMcat_galaxy}
\end{figure*}

In total, we measure 43 non-zero RMs, increasing the number of pulsars in GCs with known RMs by 33. To verify the reliability of our measurements, we compare the RMs in the ATNF Pulsar Catalogue (v2.4) for the ten GC pulsars with previously published values. As can be seen from Figure~\ref{fig.RMpsr-RMcat_galaxy}(a), most of our measured RM (RM$_{psr}$) agree with the catalogue value (RM$_{cat}$), which is reassuring. Furthermore, we measure RMs with substantially lower uncertainties than those in the catalogue for these ten pulsars due to the high signal-to-noise profiles, which result from long-duration, combined data. 

The measurements we obtain towards the GC pulsars in this work probe a variety of LOS within the Galaxy over the Galactic disc to the Galactic halo. For comparison, we calculated the RM expected due to the entire LOS through the Galaxy towards each GC using the full sky Faraday Sky maps and associated uncertainties, reconstructed using RM measurements of polarized extraGalactic sources from~\cite{Hutschenreuter2023}\footnote{\url{https://wwwmpa.mpa-garching.mpg.de/\%7Eensslin/research/data/faraday2020.html}}. Figure~\ref{fig.RMpsr-RMcat_galaxy}(b) shows this comparison.

\subsection{Polarization Profiles}\label{sec:Polprof}
\begin{figure*}
\centering 
\includegraphics[width=1.0\linewidth]{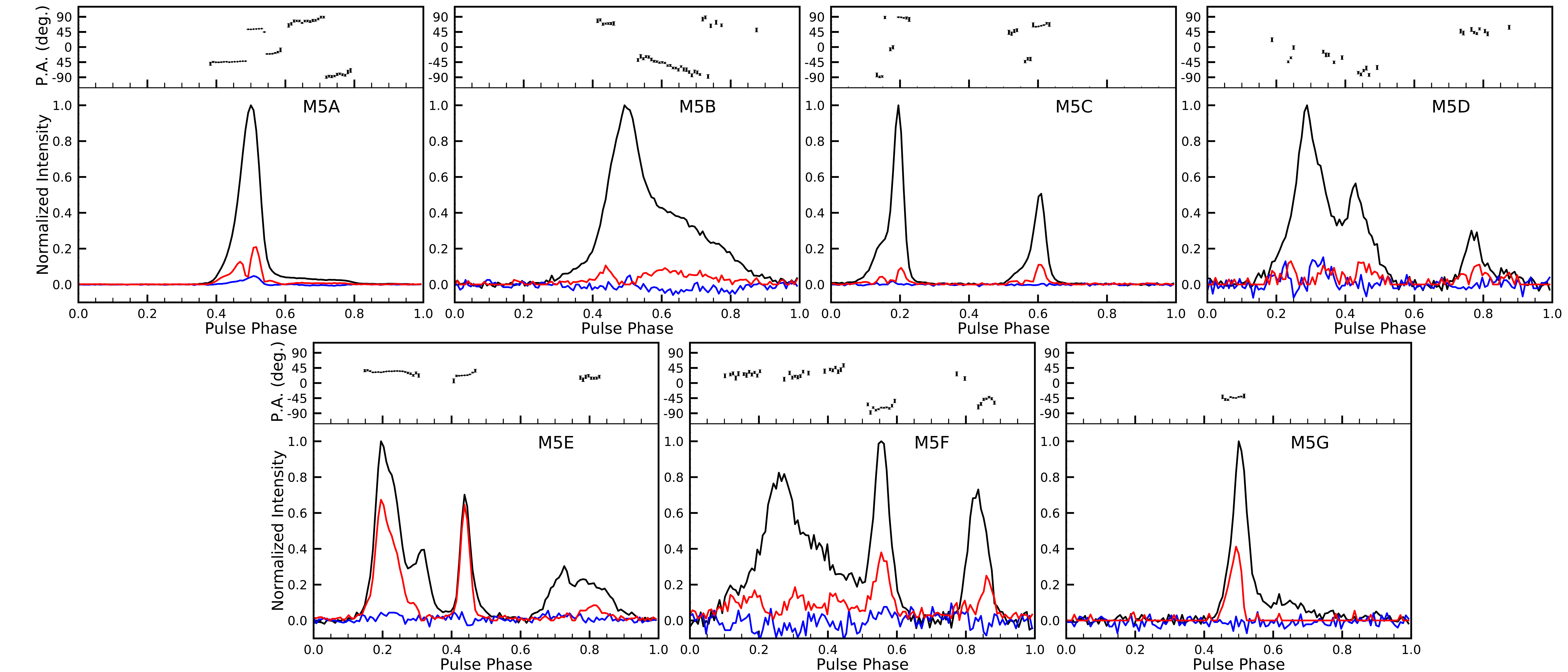}
\caption{Integrated polarization pulse profiles of the 7 pulsars in NGC~5904 (M5) from FAST data at 1250\,MHz (L Band). The red line is the linear polarization profile, the blue line is the circular polarization profile, and the black line is the normalized intensity profile. Black dots in the top panel give the linear position angle (PA) referred to the overall band center of the integrated profiles.}
\label{fig.M5_polprof}
\end{figure*}

\begin{figure*}
\centering 
\includegraphics[width=1.0\linewidth]{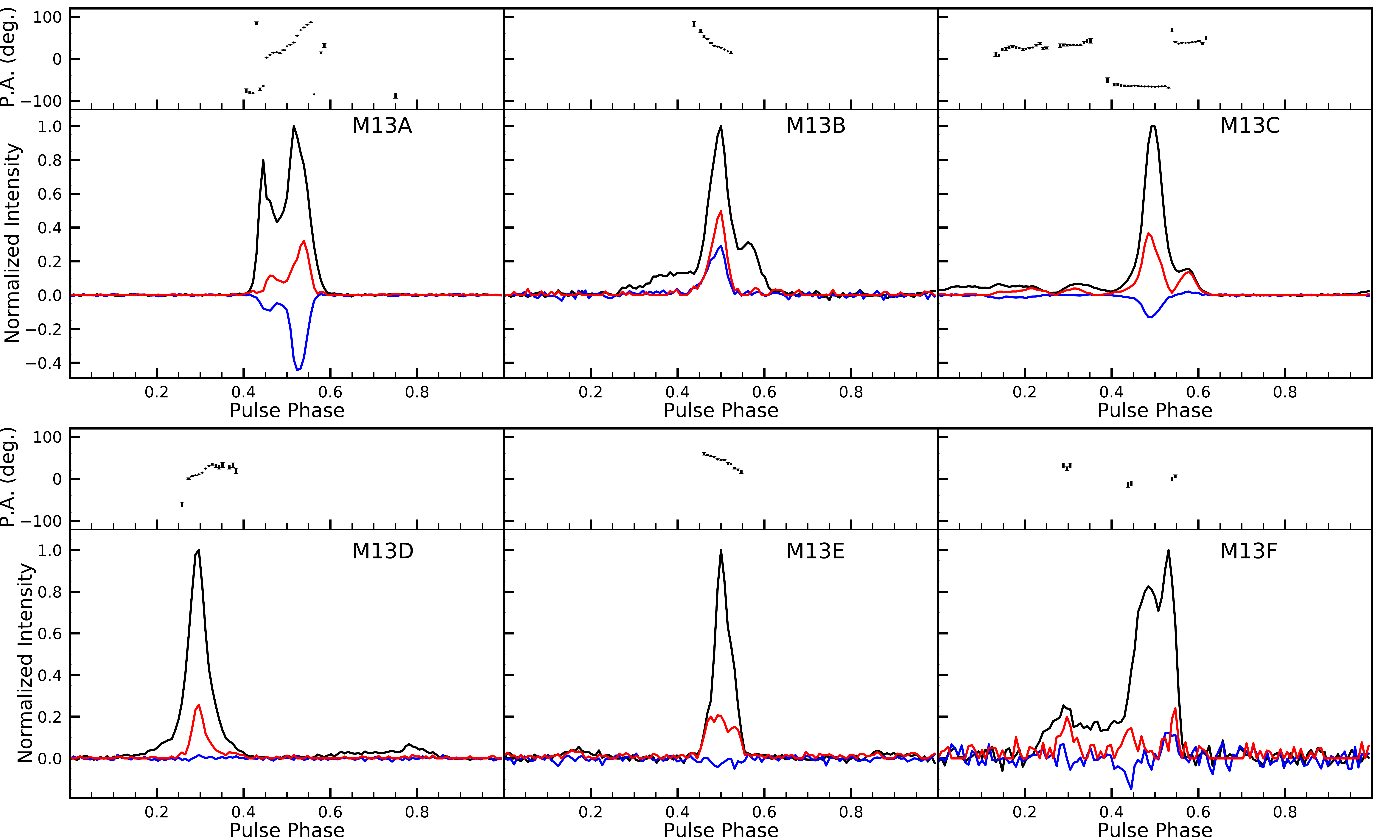}
\caption{Integrated polarization pulse profiles of the 6 pulsars in NGC~6205 (M13) from FAST data at 1250\,MHz (L Band). The red line is the linear polarization profile, the blue line is the circular polarization profile, and the black line is the normalized intensity profile. Black dots in the top panel give the linear position angle (PA) referred to the overall band center of the integrated profiles.}
\label{fig.M13_polprof}
\end{figure*}

\begin{figure*}
\centering 
\includegraphics[width=1.0\linewidth]{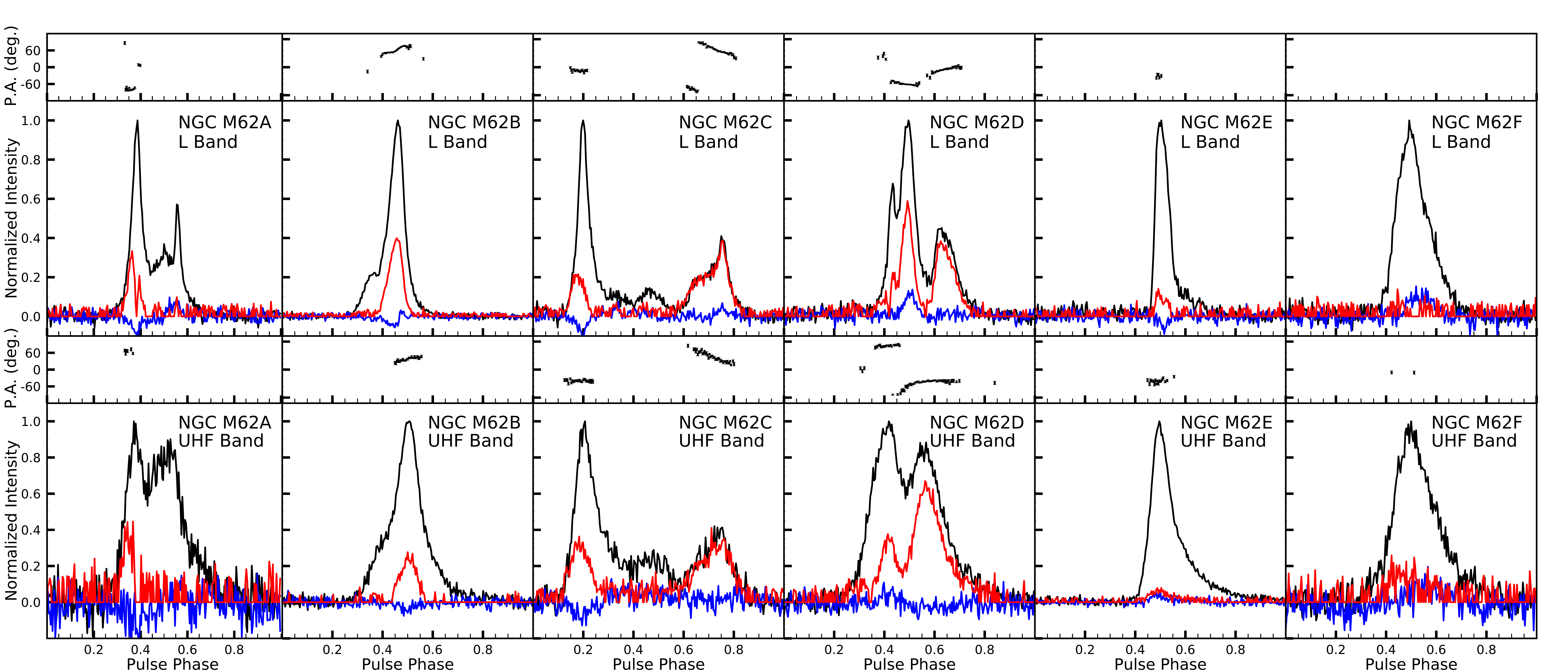}
\caption{Integrated polarization pulse profiles of the 6 pulsars in NGC~6266 (M62) from MeerKAT data at 544\,MHz (UHF Band) and 1284\,MHz (L Band). The red line is the linear polarization profile, the blue line is the circular polarization profile, and the black line is the normalized intensity profile. Black dots in the top panel give the linear position angle (PA) referred to as the overall band center of the integrated profiles.}
\label{fig.M62_polprof}
\end{figure*}

\begin{figure*}
\centering 
\includegraphics[width=1.0\linewidth]{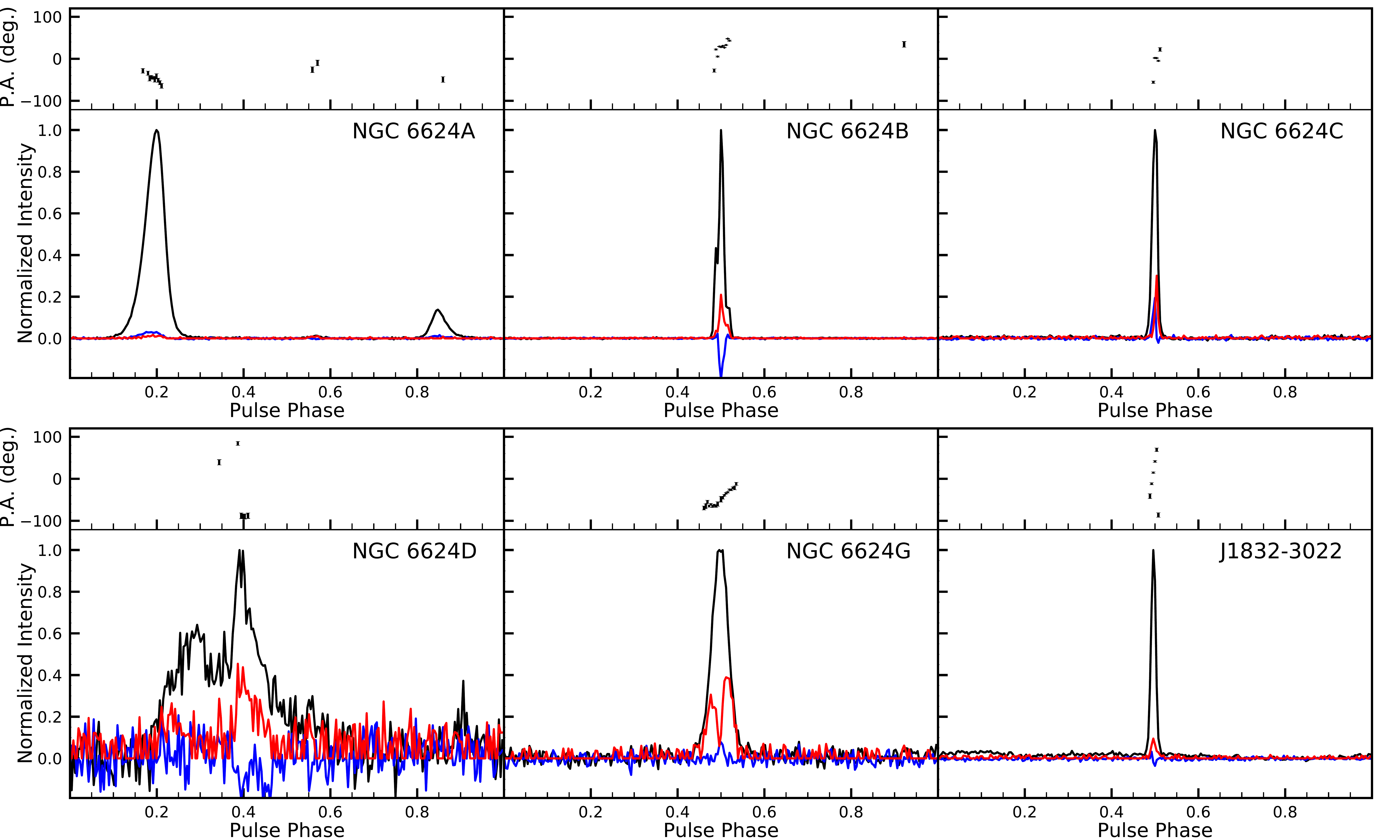}
\caption{Integrated polarization pulse profiles of the 6 pulsars in NGC~6624 from MeerKAT data at 1284\,MHz (L Band). The red line is the linear polarization profile, the blue line is the circular polarization profile, and the black line is the normalized intensity profile. Black dots in the top panel give the linear position angle (PA) referred to the overall band center of the integrated profiles.}
\label{fig.NGC6624_polprof}
\end{figure*}

\begin{figure*}
\centering 
\includegraphics[width=1.0\linewidth]{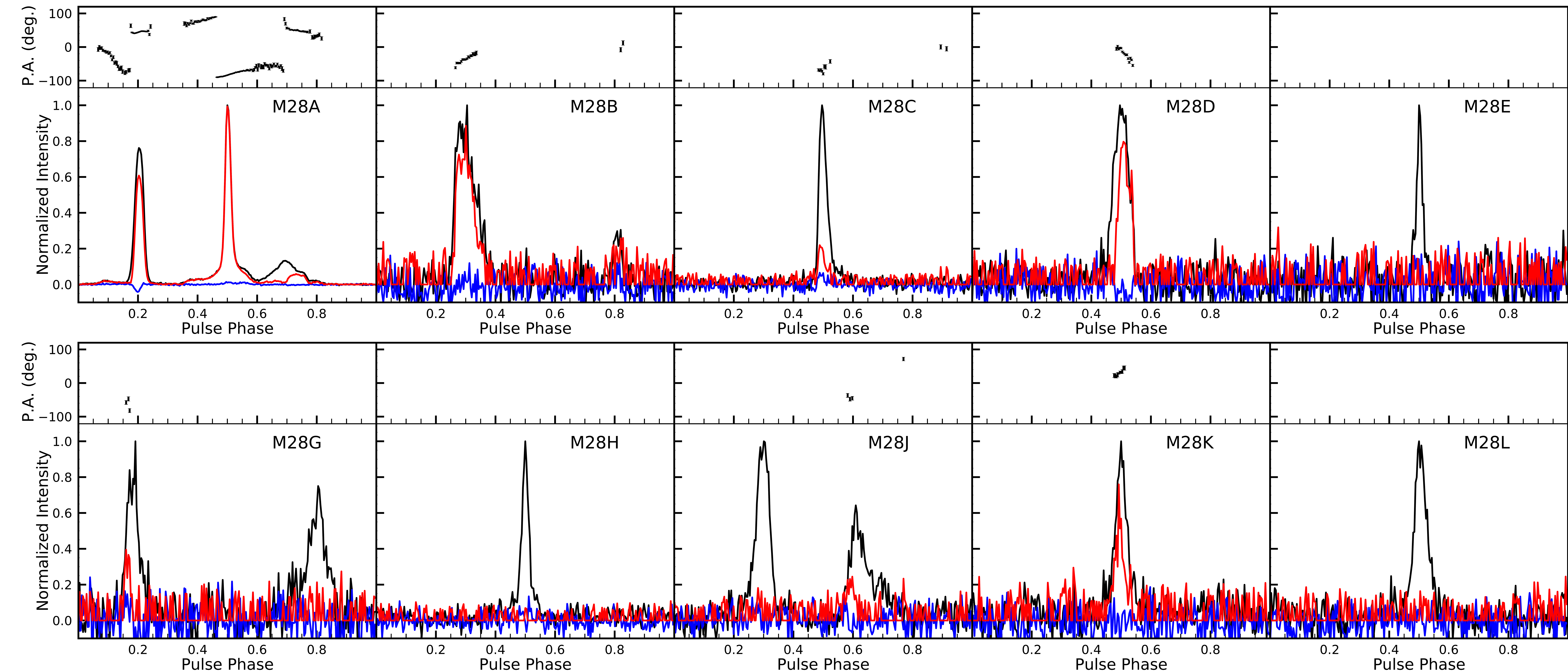}
\caption{Integrated polarization pulse profiles of the 10 pulsars in NGC~6626 (M28) from MeerKAT data at 1284\,MHz (L Band). The red line is the linear polarization profile, the blue line is the circular polarization profile, and the black line is the normalized intensity profile. Black dots in the top panel give the linear position angle (PA) referred to the overall band center of the integrated profiles.}
\label{fig.M28_polprof}
\end{figure*}

\begin{figure*}
\centering 
\includegraphics[width=0.5\linewidth]{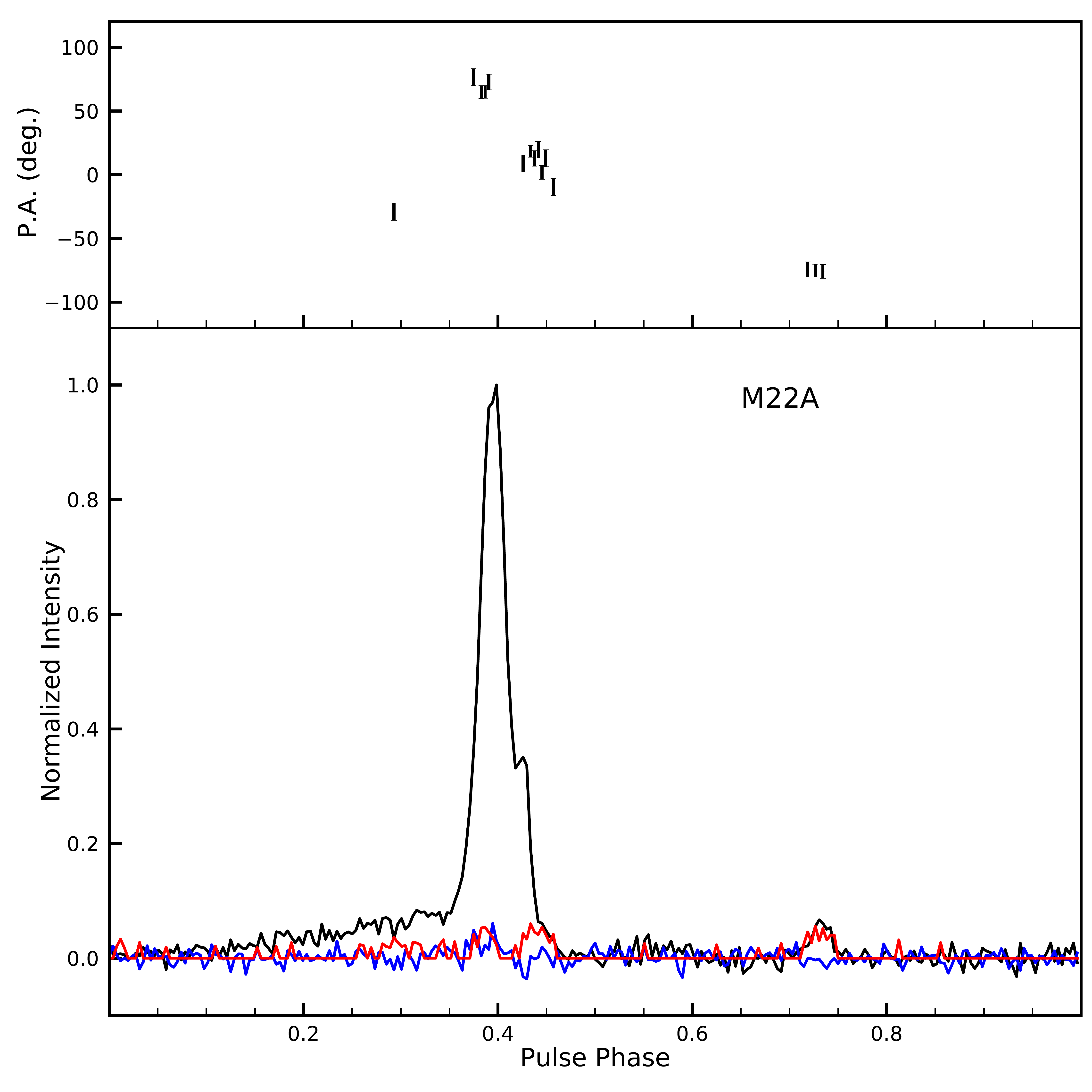}
\caption{Integrated polarization profile of the one pulsar in NGC~6656 (M22) from MeerKAT data at 1284\,MHz (L Band). The red line is the linear polarization profile, the blue line is the circular polarization profile, and the black line is the normalized intensity profile. Black dots in the top panel give the linear position angle (PA) referred to the overall band center of the integrated profiles.}
\label{fig.M22_polprof}
\end{figure*}

\begin{figure*}
\centering 
\includegraphics[width=1.0\linewidth]{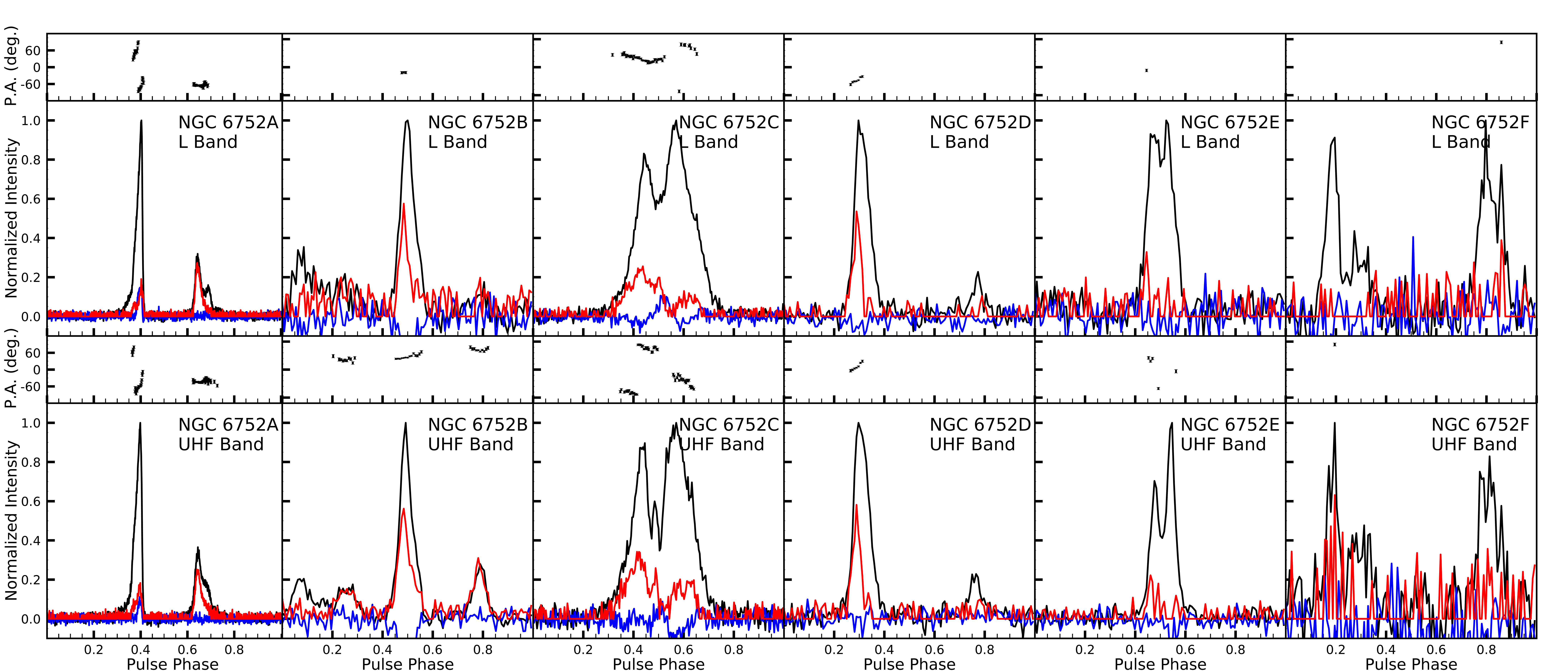}
\caption{Integrated polarization pulse profiles of the 6 pulsars in NGC~6752 from MeerKAT data at 544\,MHz (UHF Band) and 1284\,MHz (L Band). The red line is the linear polarization profile, the blue line is the circular polarization profile, and the black line is the normalized intensity profile. Black dots in the top panel give the linear position angle (PA) referred to as the overall band center of the integrated profiles.}
\label{fig.NGC6752_polprof}
\end{figure*}

\begin{figure*}
\centering 
\includegraphics[width=1.0\linewidth]{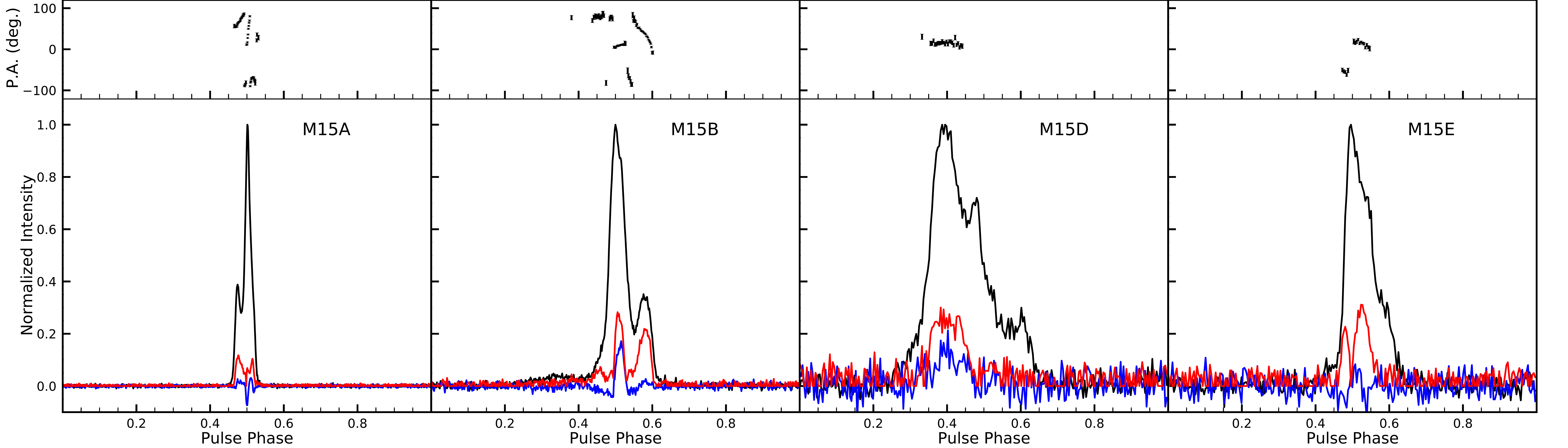}
\caption{Integrated polarization pulse profiles of the 4 pulsars in NGC~7078 (M15) from FAST data at 1250\,MHz (L Band). The red line is the linear polarization profile, the blue line is the circular polarization profile, and the black line is the normalized intensity profile. Black dots in the top panel give the linear position angle (PA) referred to the overall band center of the integrated profiles.}
\label{fig.M15_polprof}
\end{figure*}

As can be seen from Figure~\ref{fig.M5_polprof} to Figure~\ref{fig.M15_polprof}, there is a wide variety of pulse shapes and polarization fractions. Since all pulsars were observed in the same data and calibrated with the same method, the differences are attributable to the pulsars themselves, not to observational effects. 

\begin{figure*}
\centering 
\includegraphics[height=12cm,width=16cm]{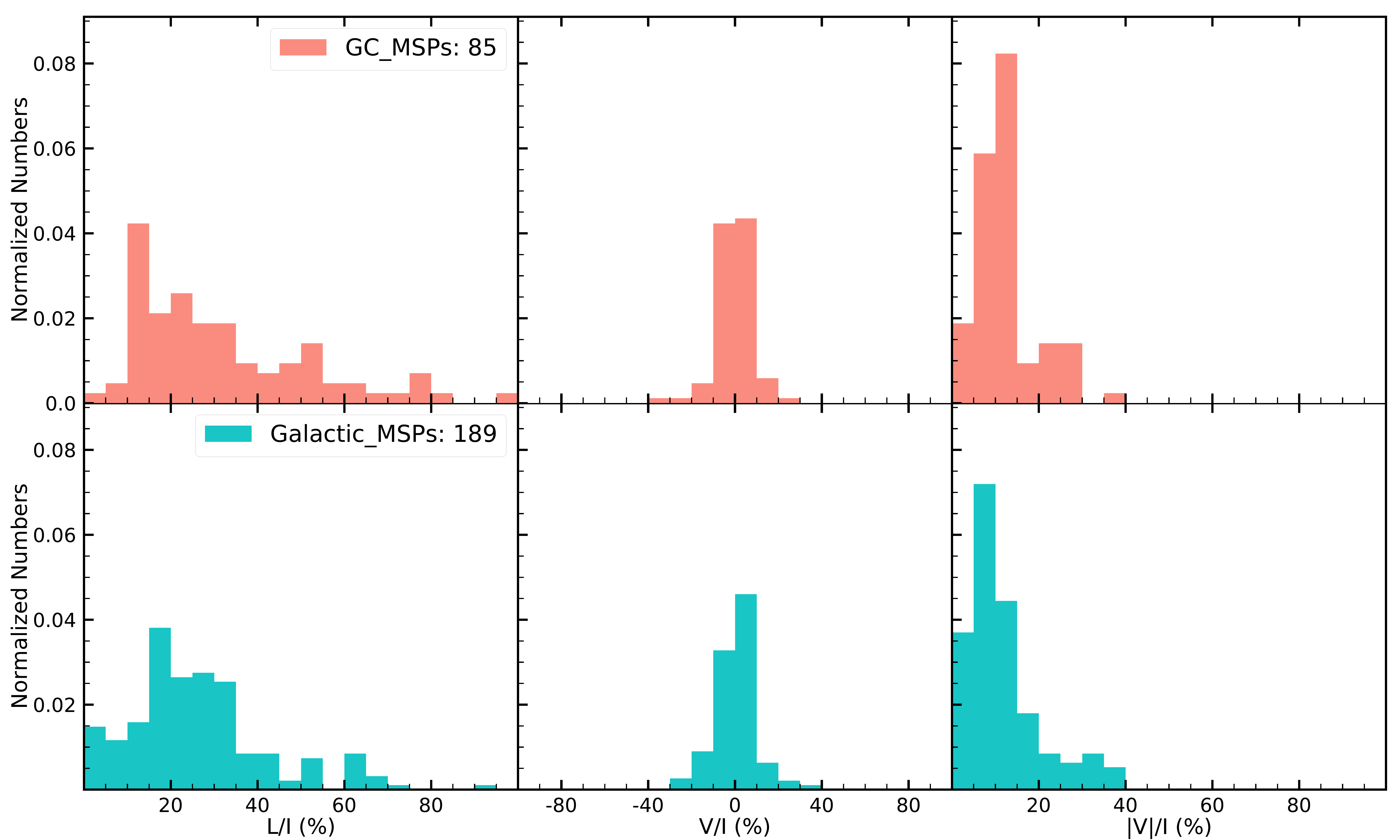}
\caption{Histograms of the percent linear (L/I), circular (V/I) and absolute circular ($|V|$/I) polarization for 85 MSPs in GCs (upper panel, salmon), and 189 Galactic MSPs (bottom panle, cyan).}
\label{fig.GC-Gal_MSPs}
\end{figure*}

Linear polarization is commonly seen in MSPs (e.g., \citealt{Dai2015,Serylak21}) although the observed PA variations seldom follow a rotating-vector-model pattern~\citep{Radhakrishnan69}. In Figure~\ref{fig.GC-Gal_MSPs}, the distribution of percent linear, circular and absolute circular polarization for 85 GC Millisecond pulsars (MSPs\footnote{We use the same definition with~\cite{Serylak21} that an MSP is characterised by short spin periods (P$<$50\,ms) and low magnetic field strengths (B$_{\text{surf}}<10^{10}$\,G) to make the comparison between pulsars in GCs and Galaxy more reliable.}), listed in Table~\label{tab:RMs_GCs} and Table~\ref{tab:RMs_47Tuc-Ter5} are shown. 
We performed a Kolmogorov-Smirnov test to determine whether the distributions of percent linear, circular, and absolute circular polarization of MSPs in globular clusters and the Galactic field can be described by a common relation. For the percent linear and circular polarization distributions, the p-values are 0.11 and 0.56, respectively greater than 0.05, indicating insufficient evidence to conclude that the samples originate from different distributions. However, for the absolute circular polarization distributions, the p-value is 0.009, which is less than 0.05, revealing statistically significant differences. Compared to Galactic MSPs~\citep{Serylak21}, the pulsars in globular clusters exhibit a relatively higher degree of absolute circular polarization.

Pulsars in M62 show visibly significant interstellar scattering effects, evidenced by the long exponential scattering tails in their pulse profiles at the MeerKAT UHF band (550--1050\,MHz) observed and also show generally decreasing fractional linear polarization with decreasing frequency (Figure~\ref{fig.M62_polprof}). Depolarization can arise in highly scattered pulsars due to propagation through turbulent plasma components with irregular magnetization (e.g. PSRs J2113+4644, J0742$-$2822, J1721$-$3532; \citealt{Noutsos15,Xue19,Sobey2019} ). Therefore, we conclude that the observed low degree of linear polarization for pulsars in M62 at low frequency attributed to stochastic Faraday rotation across the scattering disk.

\section{Magnetic Field and Internal Ionized Gas}\label{sec:MagField}
\subsection{Parallel Magnetic Field}\label{sec:MagField}
Pulsars are among the most polarized astronomical sources. The radio waves from pulsars probe the line-of-sight ISM and its magnetic field between the Earth and the pulsar via two frequency-dependent effects: rotation measure (RM, the line-of-sight election column density and weighted by the line-of-sight parallel component of the magnetic field strength) and dispersion measure (DM, the line-of-sight election column density). The ratio of the RM to the DM provides an estimate of the averaged magnetic field strength along the parallel LOS by ~\citep{Lorimer04}:
\begin{equation}
\left< B_{||}\right> = 1.23\mu \text{G}\left(\frac{\text{RM}}{\text{rad}~\text{m}^{-2}}\right) \left(\frac{\text{DM}}{\text{pc}~\text{cm}^{-3}}\right)^{-1}. 
\end{equation}

The Galactic magnetic field was first measured over 75 years ago~\citep{Hall49,Hiltner49}, and using  values from Galactic pulsars to measure the $\left<B_{||}\right>$ and model the large-scale structure of the Galactic magnetic field in scales of degrees or tens of arcminutes  has been attempted several times (e.g., \citealt{Han99,Han06,Han18,Hutschenreuter2022,Hutschenreuter2023}), but its small-scale behaviour is still poorly known.

Compared to Galactic pulsars, using globular cluster pulsars allows us to trace changes in $\left<B_{||}\right>$ on arcsecond scales, as well as make direct comparisons of RMs and $\left<B_{||}\right>$ between different sight lines. This is because the pulsars within a cluster are closely spaced, and their distances are effectively the same. To date, polarization studies of GC pulsars to map the $\left<B_{||}\right>$ toward the clusters only have been presented for two pulsar-rich clusters, 47\,Tuc~\citep{Abbate23} and Terzan 5~\citep{Martsen22}. 

\begin{figure*}
    \centering
    \includegraphics[height=5.7cm,width=5.7cm]{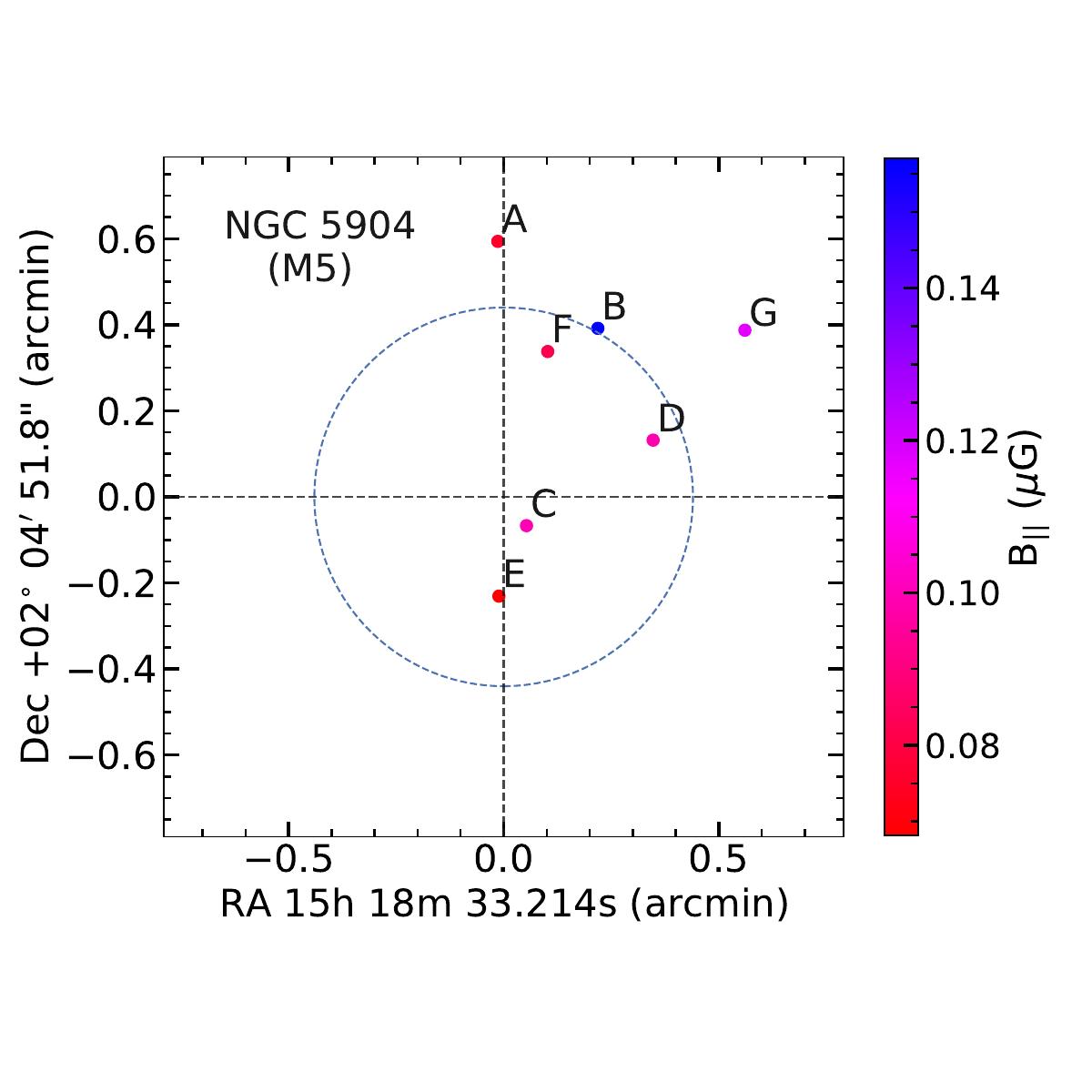}
    \includegraphics[height=5.7cm,width=5.7cm]{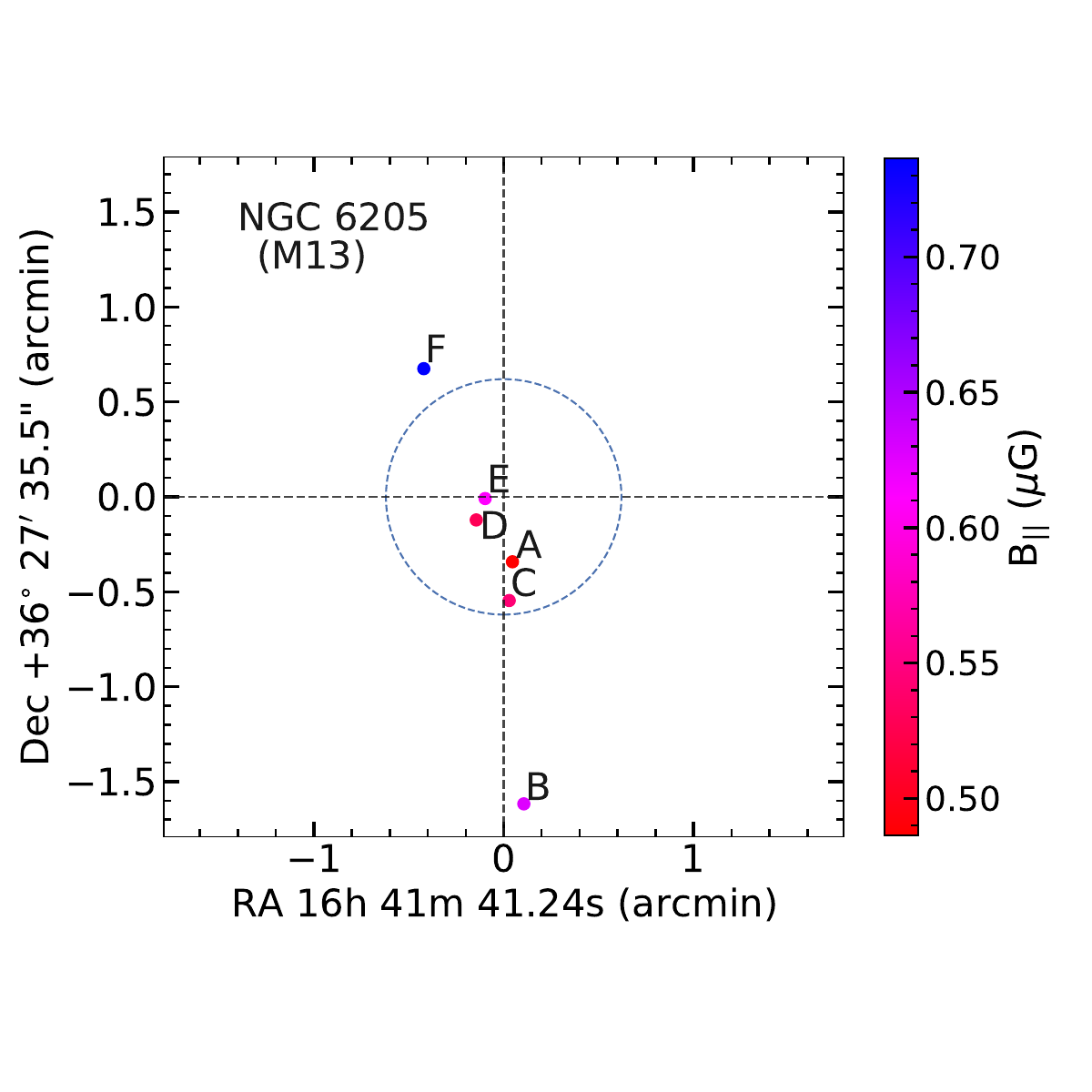}
    \includegraphics[height=5.7cm,width=5.7cm]{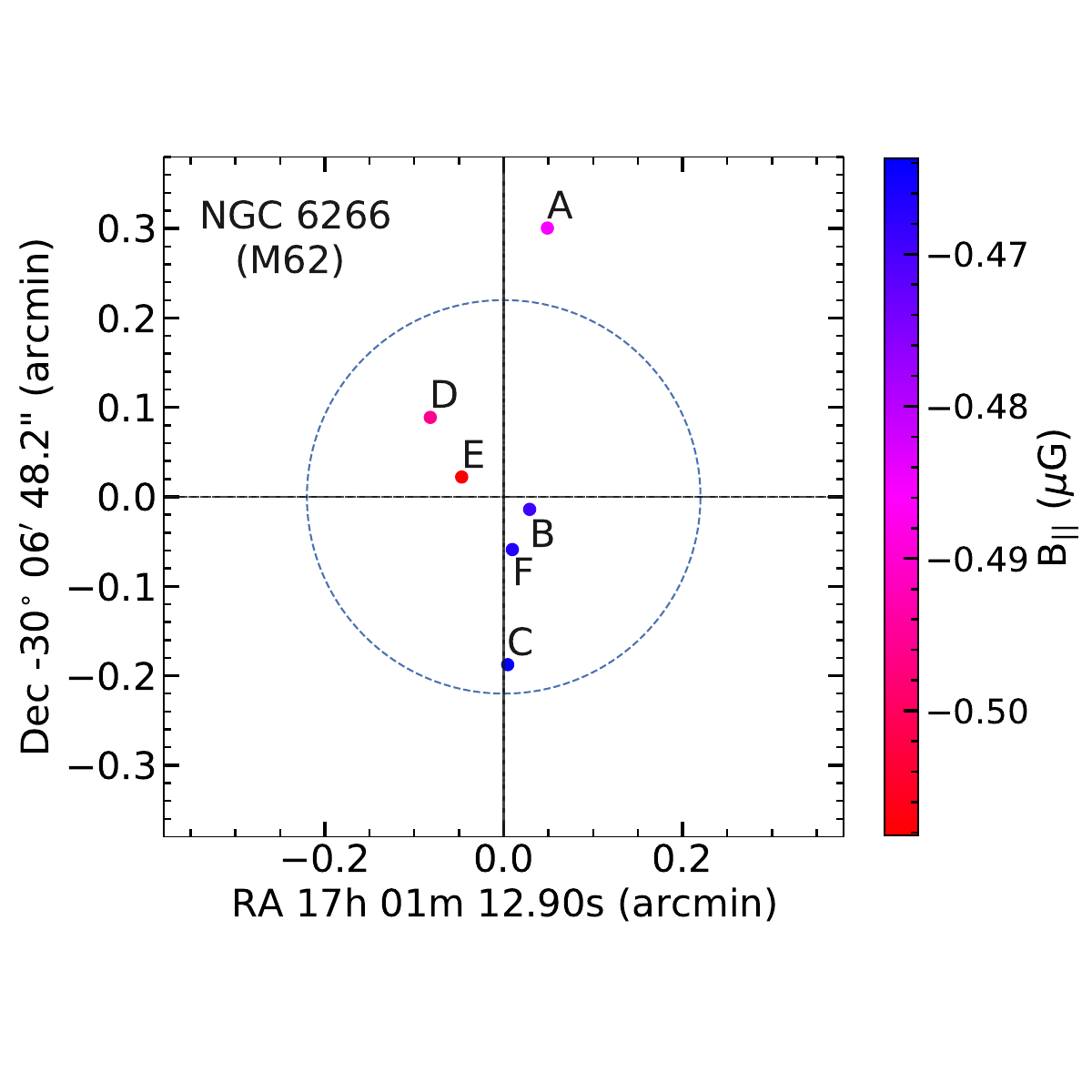}\\
    \includegraphics[height=5.7cm,width=5.7cm]{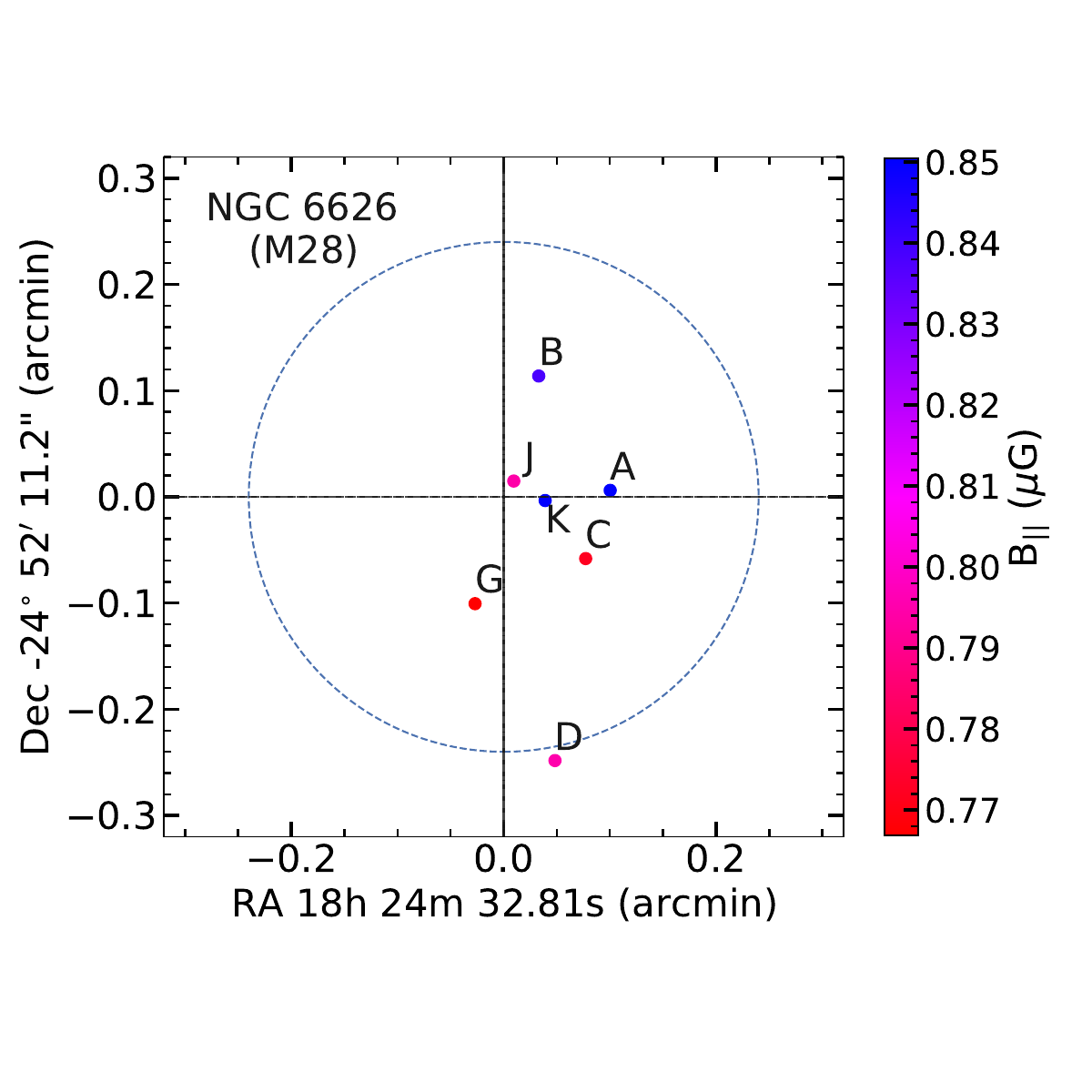}
    \includegraphics[height=5.7cm,width=5.7cm]{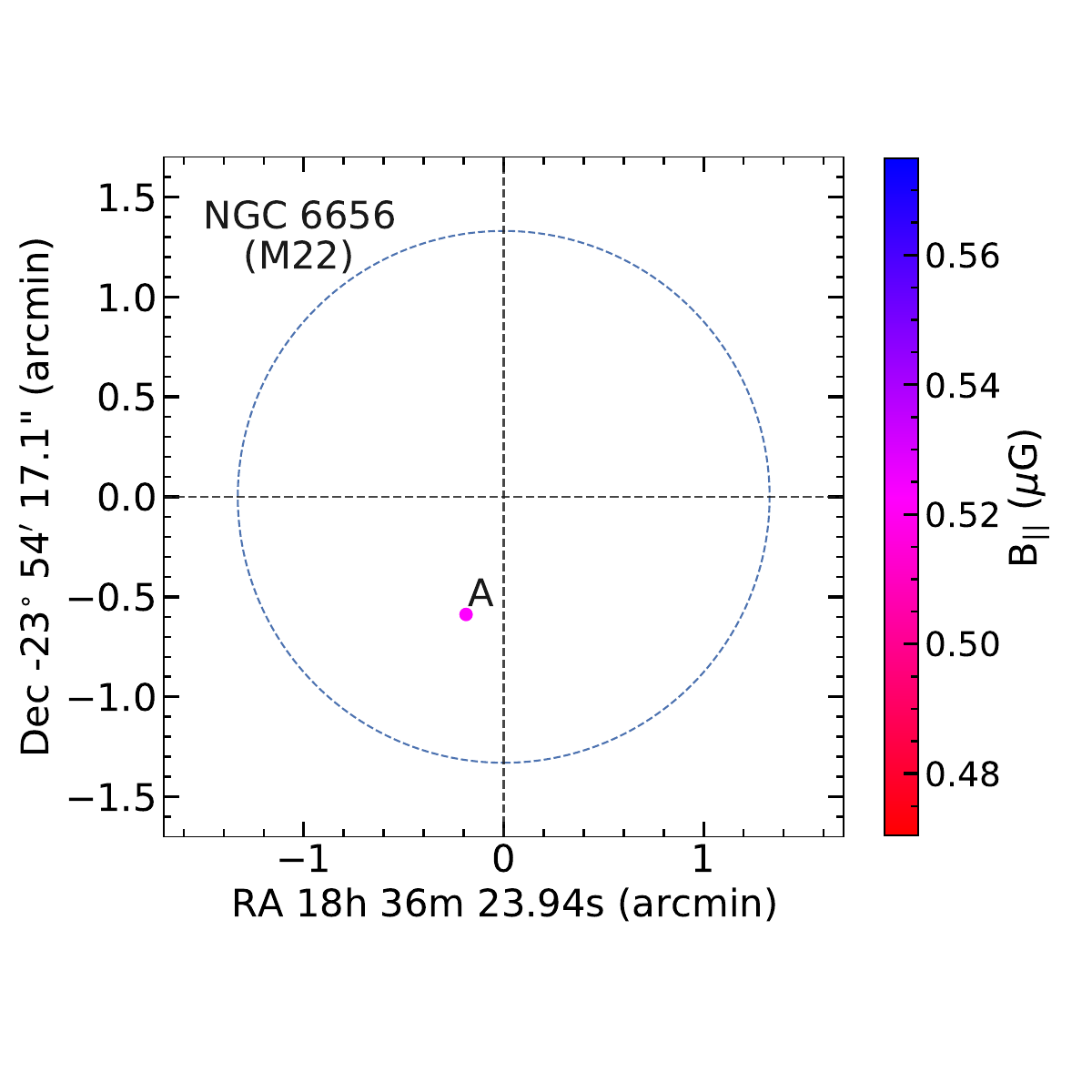}
    \includegraphics[height=5.7cm,width=5.7cm]{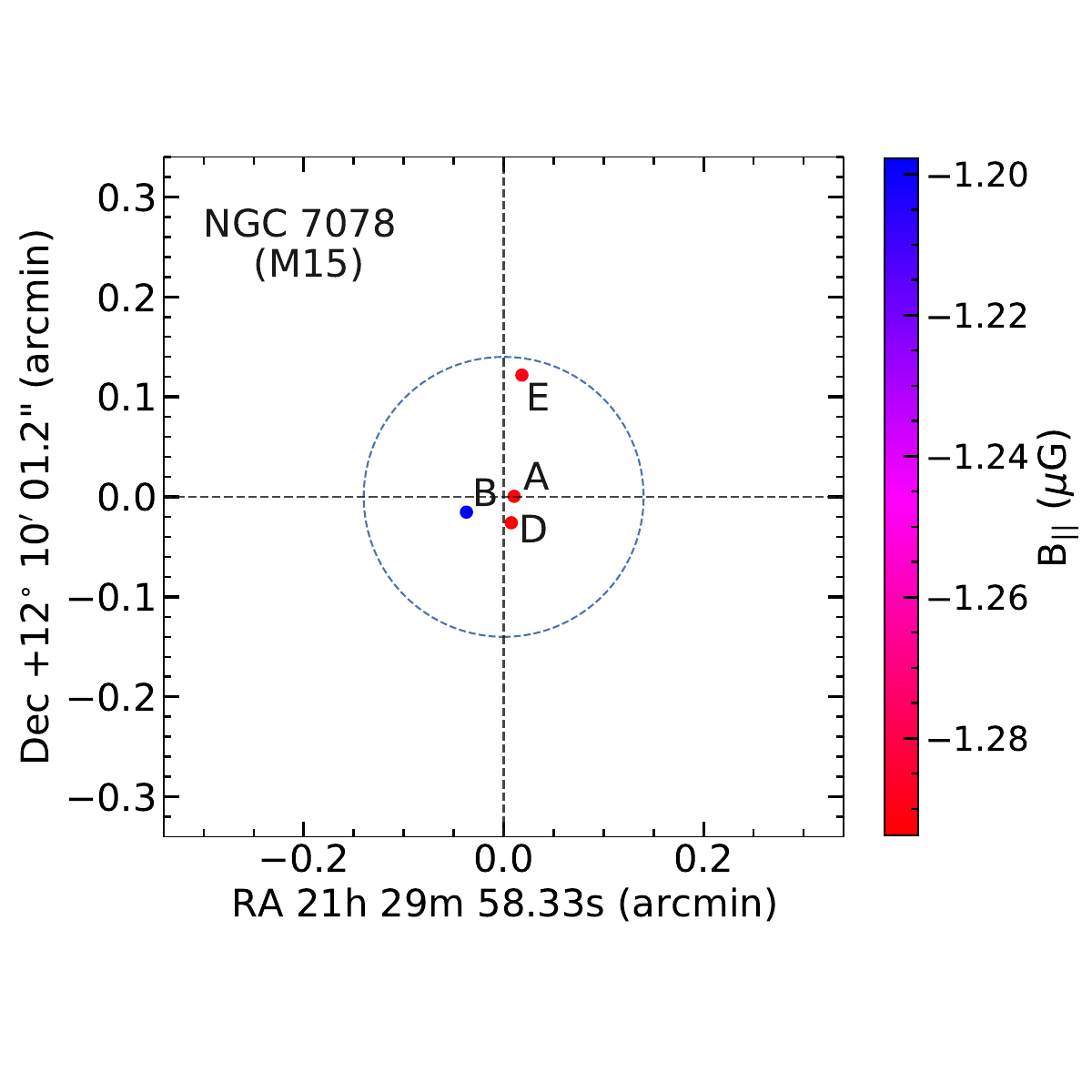}\\
    \includegraphics[height=6.4cm,width=6.4cm]{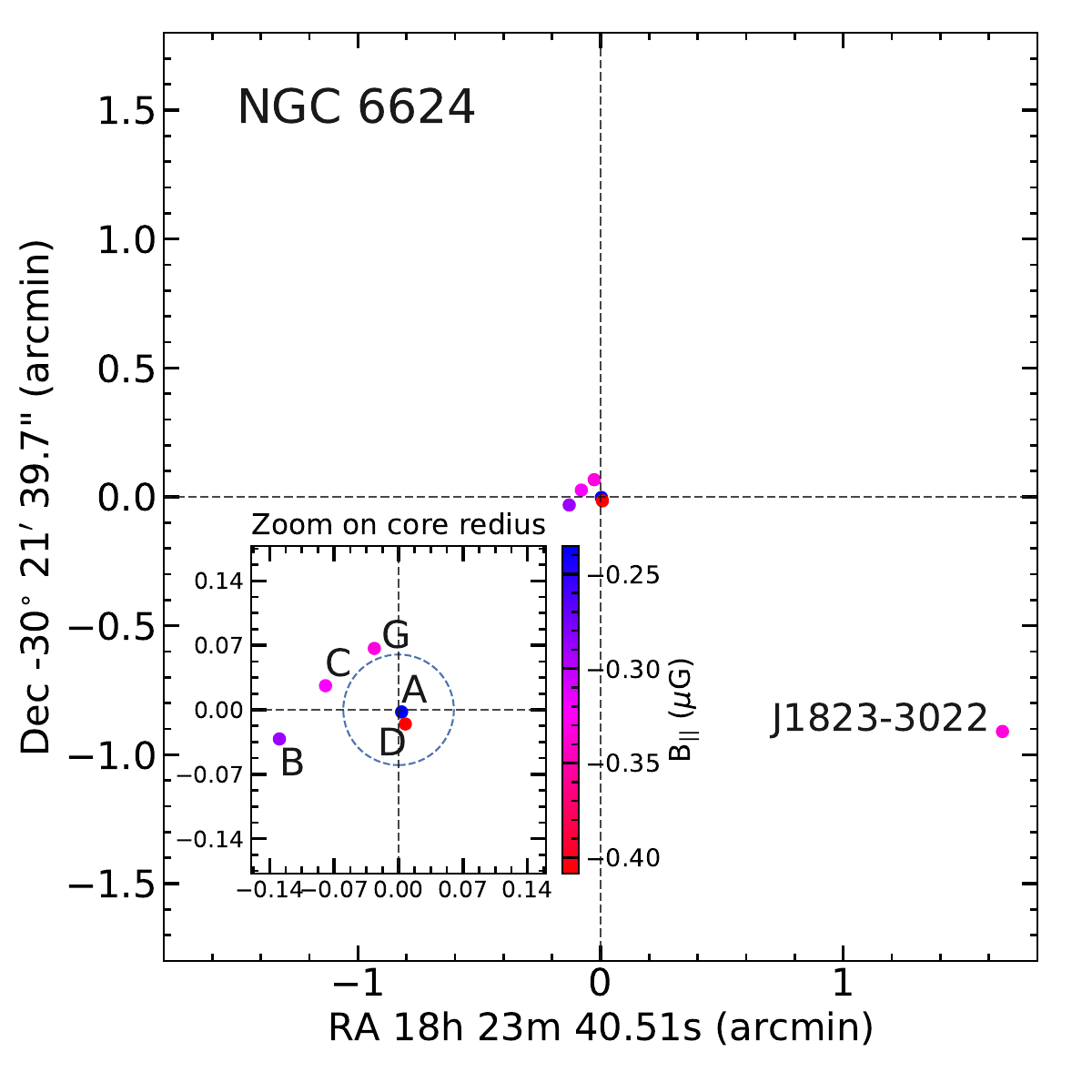}  
     \includegraphics[height=6.4cm,width=6.4cm]{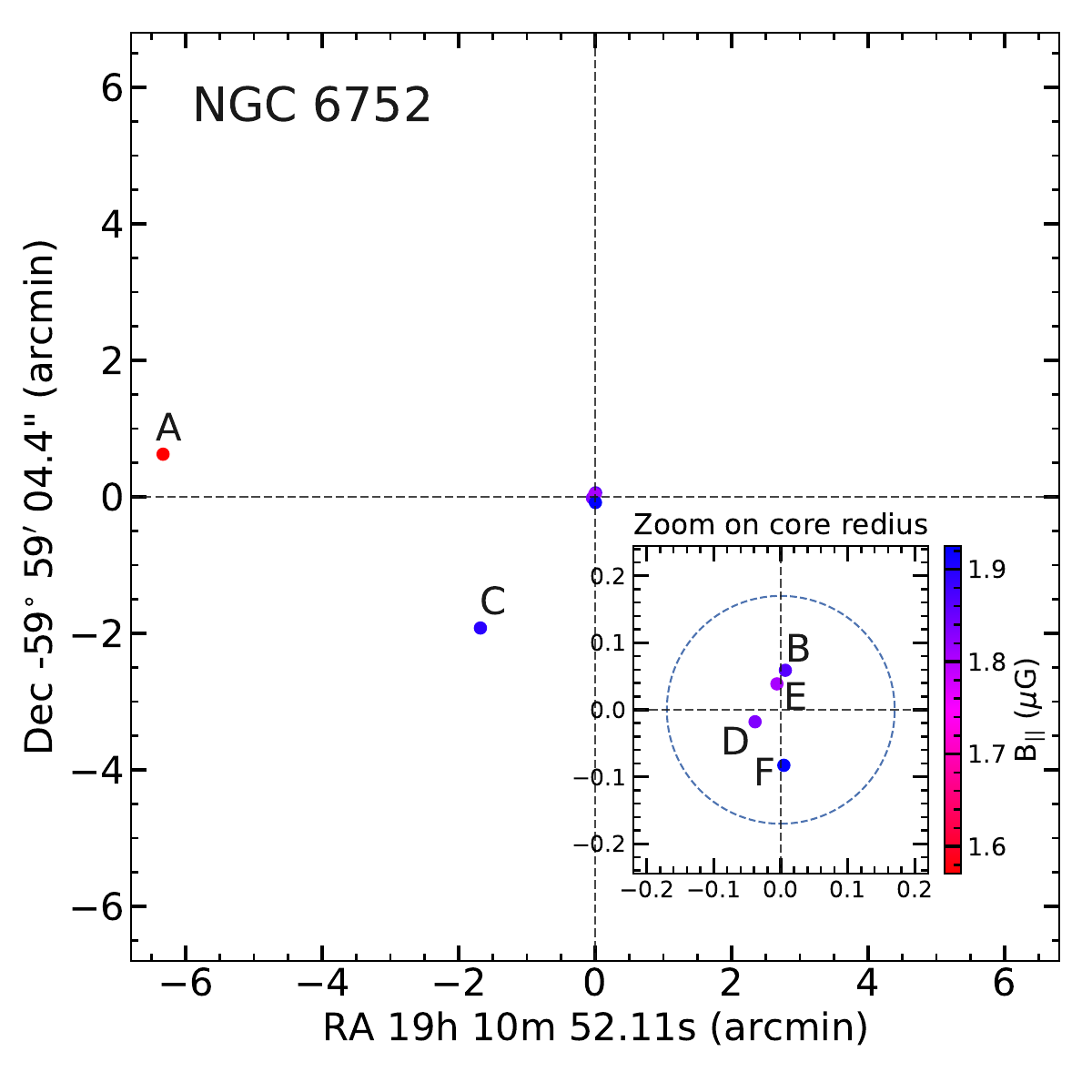}
    \caption{Map of eight GCs showing the position of the pulsars with measured B$_{\parallel}$ listed in the Table~\ref{tab:RMs_GCs}. The core radius of each cluster is indicated by a blue dashed circle. The colour of the pulsars represents the value of B$_{\parallel}$ according to the colour bar at the right of the plot. North is at the top, and east is to the left.}
    \label{fig.Bs_GCs}
\end{figure*}

We show in Figure~\ref{fig.Bs_GCs} the positions of the 46 pulsars in eight different GCs in our sample, coloured by their derived $\left<B_{||}\right>$. As can be seen from Figure~\ref{fig.Bs_GCs}, we measure significant variations in $\left<B_{||}\right>$ across the clusters. We try to look for the presence of a linear gradient of $\left<B_{||}\right>$ with a Monte Carlo Markov Chain (MCMC) code. We model the gradient as a linear dependence of $\left<B_{||}\right>$ as a function of the position of the pulsars in the cluster. The direction of the gradient in the plane of the sky is explored by projecting the positions of the pulsars along a direction that forms an angle $\theta$ with the positive Right Ascension axis and changing $\theta$ from 0 to $\pi$ radians. To account for isotropic variations caused by the turbulent ISM, we add an extra parameter, $\sigma$ that describes the standard deviation of the the variations, assumed to be gaussian. We only performed the fits for the clusters that contain more than 6 pulsars. The globular cluster NGC 6624 is also excluded even if it contains 6 pulsars since the large uncertainty of the RM of pulsar J1823$-$3021D makes it unusable for this kind of test. The best-fitting value of the intensity of the gradients are shown in Table \ref{tab:magnetic-field}. In addition to the already known case of Terzan 5 \citep{Martsen22}, also M62 shows a gradient that is statistically significant at the 3-$\sigma$ level.

The observed variations in $\left<B_{||}\right>$ can be explained by the turbulence in the Galactic magnetic field extending down to the scales sampled by the pulsars. This turbulence is thought to be coupled to the turbulence in the velocity field described in \cite{Armstrong95} and further extended in \cite{Chepurnov10}. In this theory, the energy is transferred down from the energy injection scale down to smaller scales. The physical size of this energy injection scale is thought to be between $1-100$ pc \citep{Minter96,Haverkorn08}, larger than the typical distances between GC pulsars. Therefore, we expect to see regions of coherent magnetic field of both larger and smaller size. The larger regions will cause a linear gradient in $\left<B_{||}\right>$ common to all the pulsars while the smaller regions will be responsible for variations among the single pulsars.

\subsection{Internal Ionized Gas Upper Limits}\label{sec:gas}
The study of GC pulsars also allows us to study the presence and the properties of gas within the GCs. Despite the presence of evolved stars that produce large quantities of gas and dust, there is very little observational evidence of the gas. Concrete evidence of dust and neutral has been seen only in M15 \citep{Evans03,vanLoon06} while ionized gas has only been detected in 47 Tucanae using pulsars \citep{Freire01,Abbate18}. This has puzzled astronomers for decades \citep{Smith90,vanLoon06,Barmby09}. This controversy has recently been solved reconciling theoretical predictions with observations \citep{Pancino24} by assuming that the gas follows a different spatial distribution than the stars. 

Using the measurements of DM of the pulsars we can test the presence of ionized gas within the GCs. We follow the technique developed by  \cite{Freire01} and further improved in \cite{Abbate18}. The technique consists in looking for a relation between the value of DM and the line-of-sight position of the pulsars with respect to the center of the GC. If ionized gas is present within the GC, the pulsars on the far side should have a larger DM than the pulsars in the near side. Assuming a constant gas density, the difference in DM will reveal the density of the ionized gas. This technique has proven successful for the case of 47 Tucanae detecting the presence of ionized gas with a density of $0.23 \pm 0.05$ pc cm$^{-3}$ \citep{Abbate18}. To apply the same technique to our sample of clusters, we need to first determine the line-of-sight position of the pulsars. The first rotational period derivative of the pulsar, obtained from publicly available timing solutions, is affected by the acceleration caused by the gravitational potential of the GC through the equation:
\begin{equation}
    \left( \frac{\dot P}{P}\right)_{\rm meas} = \left( \frac{\dot P}{P}\right)_{\rm int} + \frac{a_c}{c} + \frac{a_g}{c} + \frac{\mu^2 d}{c},
\end{equation}
where $\left( \frac{\dot P}{P}\right)_{\rm int}$ is the intrinsic spin down, $a_c$ is the acceleration along the line of sight caused by the cluster potential, $a_g$ is the acceleration caused by the Milky Way potential, $\mu$ is the proper motion, $d$ is the distance of the cluster and $c$ is the speed of light. In this equation the biggest contributors to the observed period derivatives are the intrinsic spin-down and the cluster acceleration \citep{Prager17}. Using appropriate estimates of the intrinsic spin down of the pulsar, following the approach of \citep{Abbate18}, the cluster acceleration can be estimated using a Monte Carlo algorithm. The acceleration is then used to estimate the line-of-sight position using mass models of the GCs derived from optical observations \footnote{the structural parameters of the GCs can be found here: \url{https://people.smp.uq.edu.au/HolgerBaumgardt/globular/parameter.html}}.

From the DM values and the estimates of the line-of-sight positions, we can look for the presence of ionized gas in the clusters. In a first approximation, we assume that the internal gas, if present, has a homogeneous distribution within the region of the cluster populated by the pulsars. We thus perform a fit assuming an internal gas of constant density and isotropic variation of DM due to the ISM, that are assumed to be gaussian and parametrized through the value of their standard deviation. The only cluster that shows hints of an internal gas is 47 Tucanae, as already shown in \cite{Abbate18}. For the rest of the clusters the value of the gas density is compatible with zero and can thus only derive upper limits on the density. We determine the maximum value of internal gas density that would be masked by the observed variations of DM caused by the ISM. To determine this we divide the standard deviations of the variations caused by the ISM by the scale length over which pulsars are distributed. We take this value as the 1-sigma upper limit of the gas density. These values are reported in Table~\ref{tab:gas-limits}.

These upper limits can be compared to the simulations run by \cite{Chantereau20} to determine the dominant gas clearing processes. In this work, the authors run simulations of the gas in GCs including several processes that can potentially clear the gas like ram-pressure stripping and UV-ionizing sources. The amount of retained gas varies from 0.1$-$8.5 M$_{\odot}$ within the central 2.5 parsecs which is equivalent to constant gas densities of 0.05$-$4.2 cm$^{-3}$. The upper limits from the clusters M5, M13 and Terzan 5 are not compatible with the simulations where the UV-ionizing flux is turned off. This suggests that the UV flux from newly born white dwarfs is necessary to remove the gas from the within the clusters.

\section{Extended data tables and figures }
\newpage
\begin{table*}
\caption{List of GCs observed in the context of this work, and with their basic parameters. Dist.: the distance from the GCs to the Sun~\citep{Baumgardt2021}; D$_{z}$: the distance from the globular clusters to the Galactic plane~\citep{Harris1996}; Core: core-collapsed clusters or not; \textit{r$_{\text{c}}$}: core radius; \textit{r$_{\text{h}}$}: half-light radius;  $\left<\text{DM}\right>$: median DM of the known pulsars; RM$_{\text{G}}$: the Galactic contribution to RM estimated from the background sources in the area surrounding the cluster~\citep{Hutschenreuter2023}.}
\footnotesize
\label{tab:GCsinf}
\setlength{\tabcolsep}{0.5mm}{
\begin{tabular}{lccccccccc}
\hline
\textbf{Cluster} & \textbf{Centre coordinate}  & \textbf{Centre coordinate}  & \textbf{Dist.} & \textbf{D$_{z}$} & \textbf{Core} & \textbf{\textit{r$_{\text{c}}$}} & \textbf{\textit{r$_{\text{h}}$}} &  \textbf{$\left<\text{DM}\right>$} & \textbf{RM$_{\text{G}}$} \\
name             & (RA, Dec.)                   & (gl, gb)      & (kpc)      & (kpc)      & (collapsed?)   & (arcmin)    & (arcmin) &  (pc cm$^{-3}$)  & (rad m$^{-2}$) \\ \hline
\multicolumn{8}{c}{\textbf{Observed with MeerKAT}}                                                                                                    \\ \hline
NGC 6266 (M62)   & $17^{\rm h}\,01^{\rm m}\,12^{\rm s}.90, -30^{\circ}\,06^{'}\,48^{''}.2$ & $353.57^{\circ},  7.32^{\circ}$   & 6.03(9)  &  0.9  & Yes   & 0.22  & 0.92 & 114.0 &  $-40\pm47$\\
NGC 6624         & $18^{\rm h}\,23^{\rm m}\,40^{\rm s}.51, -30^{\circ}\,21^{'}\,39^{''}.7$ & $2.79^{\circ},   -7.91^{\circ}$   & 8.0(1)   &  -1.1 & Yes   & 0.06  & 0.82 & 86.6  &  $-14\pm45$\\
NGC 6626 (M28)   & $18^{\rm h}\,24^{\rm m}\,32^{\rm s}.81, -24^{\circ}\,52^{'}\,11^{''}.2$ & $7.80^{\circ},   -5.58^{\circ}$   & 5.4(1)   &  -0.5 & No    & 0.24  & 1.97 & 119.7 &  $60\pm62$\\
NGC 6656 (M22)   & $18^{\rm h}\,36^{\rm m}\,23^{\rm s}.94, -23^{\circ}\,54^{'}\,17^{''}.1$ & $9.89^{\circ},   -7.55^{\circ}$   & 3.30(4)  &  -0.4 & No    & 1.33  & 3.36 & 89.9  &  $51\pm33$\\
NGC 6752         & $19^{\rm h}\,10^{\rm m}\,52^{\rm s}.11, -59^{\circ}\,59^{'}\,04^{''}.4$ & $336.49^{\circ}, -25.63^{\circ}$  & 4.13(4)  &  -1.7 & Yes   & 0.17  & 1.91 & 33.3  &  $50\pm12$\\ \hline
\multicolumn{8}{c}{\textbf{Observed with FAST}}                                                                                                       \\ \hline
NGC 5904 (M5)    & $15^{\rm h}\,18^{\rm m}\,33^{\rm s}.22, +02^{\circ}\,04^{'}\,51^{''}.7$ & $3.86^{\circ},  46.80^{\circ}$    & 7.48(6)   &  5.5 & No   & 0.44  & 1.77  & 29.5 &  $7\pm7$\\
NGC 6205 (M13)   & $16^{\rm h}\,41^{\rm m}\,41^{\rm s}.24, +36^{\circ}\,27^{'}\,35^{''}c.5$ & $59.01^{\circ}, 40.91^{\circ}$    & 7.42(8)   &  4.7 & No   & 0.62  & 1.69  & 30.2  &  $16\pm3$\\
NGC 7078 (M15)   & $21^{\rm h}\,29^{\rm m}\,58^{\rm s}.33, +12^{\circ}\,10^{'}\,01^{''}.2$ & $65.01^{\circ}, -27.31^{\circ}$   & 10.7(1)   & -4.9 & Yes  & 0.14  & 1.00  & 67.2  &  $-47\pm14$\\ \hline
\end{tabular}}
\end{table*}
\clearpage
\clearpage
\begin{table*}
\caption{Summary of the MeerKAT observations used in this work.}
\scriptsize
\label{tab:MeerKATobs}
\begin{tabular}{llccccccccc}
\hline
Observation                  & Target      & Beam & Mode   & Number of  & Number of & Duration & DM             & Sampling time & Central Frequency & Bandwidth \\
Date (UTC)                   &      &             &        & antennas   & channels  & (hours)  & (pc cm$^{-3}$) & ($\mu$s)      & (MHz)             & (MHz)\\\hline
\multicolumn{8}{l}{\textbf{NGC 6266 (M62)}}\\ 
2021 Sep 21                  & J1701-3006D & 2    & Search & 38   & 1024     & 2.0      & 114.0  & 9.57  & 1283.896      & 856 \\
2021 Dec 30                  & J1701-3006D & 2    & Search & 44   & 1024     & 2.0      & 114.0  & 9.57  & 1283.896      & 856 \\
2022 Apr 21                  & J1701-3006D & 4    & Search & 45   & 1024     & 2.0      & 114.0  & 9.57  & 1283.896      & 856 \\
2023 Jun 14                  & J1701-3006B & 2    & Search & 31   & 1024     & 2.0      & 113.4  & 19.14 & 1283.896      & 856 \\
2023 Jun 14                  & J1701-3006B & 2    & Search & 31   & 1024     & 2.0      & 113.4  & 19.14 & 1283.896      & 856 \\
2023 Aug 12                  & J1701-3006G & 1    & Search & 63   & 512      & 2.0      & 113.6  & 7.52  & 815.934       & 544 \\
                             & J1701-3006H & 2    & Search & 63   & 512      & 2.0      & 114.7  & 7.52  & 815.934       & 544 \\
                             & J1701-3006I & 3    & Search & 63   & 512      & 2.0      & 113.3  & 7.52  & 815.934       & 544 \\
                             & J1701-3006J & 4    & Search & 63   & 512      & 2.0      & 111.9  & 7.52  & 815.934       & 544 \\
2023 Nov 22                  & J1701-3006G & 1    & Search & 60   & 512      & 2.0      & 113.6  & 7.52  & 815.934       & 544 \\
                             & J1701-3006H & 2    & Search & 60   & 512      & 2.0      & 114.7  & 7.52  & 815.934       & 544 \\
                             & J1701-3006I & 3    & Search & 60   & 512      & 2.0      & 113.3  & 7.52  & 815.934       & 544 \\
                             & J1701-3006J & 4    & Search & 60   & 512      & 2.0      & 111.9  & 7.52  & 815.934       & 544 \\
2024 Mar 18                  & J1701-3006G & 1    & Search & 58   & 512      & 2.0      & 113.6  & 7.52  & 1283.896      & 856 \\
                             & J1701-3006H & 2    & Search & 58   & 512      & 2.0      & 114.7  & 7.52  & 1283.896      & 856 \\ 
                             & J1701-3006I & 3    & Search & 58   & 512      & 2.0      & 113.3  & 7.52  & 1283.896      & 856 \\
                             & J1701-3006J & 4    & Search & 58   & 512      & 2.0      & 111.9  & 7.52  & 1283.896      & 856 \\
\\
\multicolumn{8}{l}{\textbf{NGC 6624}}\\   
2021 Jun 18                  & J1823-3021G & 1    & Timing & 58   & 4096     & 3.5      & 86.2   & 9.57  & 1283.896      & 856 \\
2021 Jun 18                  & J1823-3021G & 1    & Timing & 58   & 4096     & 3.5      & 86.2   & 9.57  & 1283.896      & 856 \\
2021 Jun 19                  & J1823-3021G & 1    & Timing & 60   & 4096     & 2.0      & 86.2   & 9.57  & 1283.896      & 856 \\
2022 Aug 28                  & Centre      & 2    & Search & 60   & 256      & 3.5      & 86.2   & 9.57  & 1283.896      & 856 \\
                             & J1823-3021G & 1    & Timing & 60   & 4096     & 3.5      & 86.2   & 9.57  & 1283.896      & 856 \\
                             & J1823-3022  & 4    & Timing & 60   & 4096     & 3.5      & 96.8   & 9.57  & 1283.896      & 856 \\
2022 Aug 28                  & Centre      & 2    & Search & 60   & 256      & 3.5      & 86.2   & 9.57  & 1283.896      & 856 \\
                             & J1823-3021G & 1    & Timing & 60   & 4096     & 3.5      & 86.2   & 9.57  & 1283.896      & 856 \\
                             & J1823-3022  & 4    & Timing & 60   & 4096     & 3.5      & 96.8   & 9.57  & 1283.896      & 856 \\
2022 Aug 29                  & Centre      & 2    & Search & 59   & 256      & 2.0      & 86.2   & 9.57  & 1283.896      & 856 \\
                             & J1823-3021G & 1    & Timing & 50   & 4096     & 2.0      & 86.2   & 9.57  & 1283.896      & 856 \\
                             & J1823-3022  & 4    & Timing & 59   & 4096     & 2.0      & 96.8   & 9.57  & 1283.896      & 856 \\
2023 Jun 08                  & Centre      & 2    & Search & 59   & 256      & 2.0      & 86.2   & 9.57  & 1283.896      & 856 \\
                             & J1823-3021A & 3    & Search & 43   & 256      & 2.0      & 87.5   & 9.57  & 1283.896      & 856 \\
                             & J1823-3021G & 1    & Timing & 59   & 4096     & 2.0      & 86.2   & 9.57  & 1283.896      & 856 \\
                             & J1823-3022  & 4    & Timing & 59   & 4096     & 2.0      & 96.8   & 9.57  & 1283.896      & 856 \\
2023 Spe 07                  & Centre      & 2    & Search & 62   & 256      & 2.0      & 86.2   & 9.57  & 1283.896      & 856 \\
                             & J1823-3021A & 3    & Search & 43   & 256      & 2.0      & 87.5   & 9.57  & 1283.896      & 856 \\
                             & J1823-3021G & 1    & Timing & 62   & 4096     & 2.0      & 86.2   & 9.57  & 1283.896      & 856 \\
                             & J1823-3022  & 4    & Timing & 62   & 4096     & 2.0      & 96.8   & 9.57  & 1283.896      & 856 \\
2023 Oct 11                  & J1823-3021G & 1    & Timing & 58   & 4096     & 0.8      & 86.2   & 9.57  & 1283.896      & 856 \\
                             & J1823-3022  & 4    & Timing & 62   & 4096     & 0.8      & 96.8   & 9.57  & 1283.896      & 856 \\          
2023 Oct 12                  & Centre      & 1    & Search & 57   & 256      & 2.0      & 86.2   & 9.57  & 1283.896      & 856 \\
                             & J1823-3021A & 2    & Search & 40   & 256      & 2.0      & 87.5   & 9.57  & 1283.896      & 856 \\
                             & J1823-3021G & 4    & Timing & 57   & 4096     & 2.0      & 86.2   & 9.57  & 1283.896      & 856 \\
\\               
\multicolumn{8}{l}{\textbf{NGC 6626 (M28)}}\\                            
2019 Jul 19                  & Centre & 2    & Search & 60        & 768      & 2.5      & 119.9  &  9.57 & 1283.582   & 642\\
2020 Feb 04                  & Centre & 2    & Search & 60        & 768      & 2.5      & 119.9  &  9.57 & 1283.582   & 642\\
\\ 
\multicolumn{8}{l}{\textbf{NGC 6656 (M22)}}\\                            
2022 Feb 22                  & Centre & 4    & Search & 57        & 1024     & 4.0      & 89.1   &  9.57 & 1283.896   & 856\\
\\
\multicolumn{8}{l}{\textbf{NGC 6752}}\\
2023 Oct 04                  & Centre      & 1   & Search & 59   & 256       & 1.0      & 33.3   &  15.05 & 815.934           & 544\\
                             & J1910-5959A & 2   & Timing & 59   & 4096      & 1.0      & 33.7   &  15.05 & 815.934           & 544\\
                             & J1910-5959C & 4   & Timing & 59   & 4096      & 1.0      & 33.3   &  15.05 & 815.934           & 544\\
\hline
\end{tabular}
\end{table*}
\begin{table*}
\caption{Summary of the FAST observations used in this work.}  
\centering
\label{tab:FASTobs}
\begin{tabular}{lccccc}
\hline
 Observation       & Target         & Sampling time  & Number of & Observation & Project \\
 Date (UTC)        &                & ($\mu s$)      & channels  & Length (hr) & ID \\\hline
\multicolumn{4}{l}{\textbf{NGC 5904 (M5)}}\\ 
2020 Nov 16        & J1518+0204B    & 49.15          & 4096      & 0.5         &  PT2020\_0074\\
2020 Nov 18        & J1518+0204B    & 49.15             & 4096      & 0.5         &  PT2020\_0074\\
2020 Nov 19        & J1518+0204B    & 49.15             & 4096      & 0.5         &  PT2020\_0074\\
2020 Nov 27        & J1518+0204B    & 49.15             & 4096      & 0.5         &  PT2020\_0074\\
2020 Nov 28        & J1518+0204B    & 49.15             & 4096      & 0.5         &  PT2020\_0074\\
2020 Nov 29        & J1518+0204B    & 49.15             & 4096      & 0.5         &  PT2020\_0074\\
2020 Dec 01        & J1518+0204B    & 49.15             & 4096      & 0.5         &  PT2020\_0074\\
2021 Mar 06        & J1518+0204B    & 49.15             & 4096      & 2.0         &  PT2020\_0074\\
2022 Fer 04        & Centre         & 49.15             & 4096      & 4.0         &  PT2021\_0061\\
2022 Fer 06        & Centre         & 49.15             & 4096      & 4.0         &  PT2021\_0061\\
2022 Fer 07        & Centre         & 49.15             & 4096      & 1.0         &  PT2021\_0061\\
2022 Fer 08        & Centre         & 49.15             & 4096      & 1.0         &  PT2021\_0061\\
2022 Fer 09        & Centre         & 49.15             & 4096      & 1.0         &  PT2021\_0061\\
2022 Fer 10        & Centre         & 49.15             & 4096      & 2.0         &  PT2021\_0061\\
2022 Fer 15        & Centre         & 49.15             & 4096      & 2.0         &  PT2021\_0061\\
2022 Aug 21        & Centre         & 49.15             & 4096      & 1.0         &  PT2022\_0062\\
2022 Aug 29        & Centre         & 49.15             & 4096      & 1.0         &  PT2022\_0062\\
2022 Aug 30        & Centre         & 49.15             & 4096      & 1.0         &  PT2022\_0062\\
2022 Aug 31        & Centre         & 49.15             & 4096      & 1.0         &  PT2022\_0062\\
2022 Sep 03        & Centre         & 49.15             & 4096      & 1.0         &  PT2022\_0062\\
2022 Sep 06        & Centre         & 49.15             & 4096      & 1.0         &  PT2022\_0062\\
2022 Sep 09        & Centre         & 49.15             & 4096      & 1.0         &  PT2022\_0062\\
2022 Dec 04        & Centre         & 49.15             & 4096      & 1.0         &  PT2022\_0062\\
2022 Dec 05        & Centre         & 49.15             & 4096      & 1.0         &  PT2022\_0062\\
2022 Dec 06        & Centre         & 49.15             & 4096      & 1.0         &  PT2022\_0062\\
2022 Dec 09        & Centre         & 49.15             & 4096      & 1.0         &  PT2022\_0062\\
2022 Dec 14        & Centre         & 49.15             & 4096      & 1.0         &  PT2022\_0062\\
\\   
\multicolumn{4}{l}{\textbf{NGC 6205 (M13)}}\\ 
2021 Aug 21        & Centre         & 49.15             & 4096      & 1.0         &  PT2021\_0131\\
2021 Spe 07        & Centre         & 49.15             & 4096      & 1.0         &  PT2021\_0131\\
2021 Oct 12        & Centre         & 49.15             & 4096      & 1.0         &  PT2021\_0131\\
2022 Jan 02        & Centre         & 49.15             & 4096      & 1.0         &  PT2021\_0131\\
2022 Feb 01        & Centre         & 49.15             & 4096      & 1.0         &  PT2021\_0131\\
2022 Feb 18        & Centre         & 49.15             & 4096      & 1.0         &  PT2021\_0131\\
2022 Apr 04        & Centre         & 49.15             & 4096      & 1.0         &  PT2021\_0131\\
2022 May 23        & Centre         & 49.15             & 4096      & 1.0         &  PT2021\_0131\\
2022 Jun 25        & Centre         & 49.15             & 4096      & 1.0         &  PT2021\_0131\\
2022 Jul 20        & Centre         & 49.15             & 4096      & 1.0         &  PT2021\_0131\\
2022 Aug 23        & Centre         & 49.15             & 4096      & 1.5         &  PT2022\_0050\\
2022 Sep 23        & Centre         & 49.15             & 4096      & 1.5         &  PT2022\_0050\\
2022 Oct 23        & Centre         & 49.15             & 4096      & 1.5         &  PT2022\_0050\\
\\
\multicolumn{4}{l}{\textbf{NGC 7078 (M15)}}\\ 
2020 Sep 02        & Centre         & 49.15             & 4096      & 3.0         &  PT2020\_0161\\
2021 Mar 09        & Centre         & 49.15             & 4096      & 2.3         &  PT2020\_0028\\
2022 Apr 28        & J2129+1210A    & 98.30             & 8192      & 2.0         &  PT2021\_0004\\
2022 Oct 15        & J2129+1210A    & 98.30             & 8192      & 0.8         &  PT2022\_0164\\
2023 Feb 28        & J2129+1210A    & 98.30             & 8192      & 0.8         &  PT2022\_0164\\
2023 May 15        & J2129+1210A    & 98.30             & 8192      & 0.8         &  PT2022\_0164\\
2023 Jul 15        & J2129+1210A    & 98.30             & 8192      & 0.8         &  PT2022\_0164\\
2023 Oct 24        & Centre         & 49.15             & 4096      & 3.3         &  PT2023\_0122\\
\hline
\end{tabular}
\end{table*}
\begin{table*}
\centering
\caption{The basic parameters, DMs, RMs, $\left<B_{||}\right>$, the percentage of linear polarization (L/I), circular polarisation (V/I) and absolute circular polarisation ($|V|$/I) of 43 pulsars in eight GCs determined from this work.}
\label{tab:RMs_GCs}
\setlength{\tabcolsep}{3.2mm}{
\begin{tabular}{cccccccc}\hline
Pulsar &  Period  & DM             & RM    &  $\left<B_{||}\right>$ & L/I  & V/I  & $|V|$/I\\
       &  (ms)    & (pc cm$^{-3}$) & (rad m$^{-2}$) &  ($\mu$G)     & (\%) & (\%) & (\%)\\\hline
\multicolumn{8}{c}{\textbf{NGC 5904 (M5) (7 Pulsars, FAST Lband)}} \\
A  &  5.55 &  30.0546(5) & 1.85(9)$^{\dagger}$ &  0.076(4) &  18.45(5) &  2.68(5) & 4.62(5)\\
B  &  7.94 &  29.469(4)  & 3.7(8)  &  0.15(3)  &  11.1(3)  &  -5.2(3) & 6.5(3)\\
C  &  2.48 &  29.3109(3) & 2.4(5)  &  0.10(2)  &  11.5(4)  &  0.8(4)  & 2.5(4)\\
D  &  2.98 &  29.371(2)  & 2(2)    &  0.08(8)  &  14(1)    &  6(1)    & 14(1)\\
E  &  3.18 &  29.310(1)  & 1.6(2)  &  0.067(8) &  47.1(3)  &  5.4(3)  & 7.2(3)\\
F  &  3.18 &  29.409(1)  & 2(1)    &  0.08(4)  &  26.9(6)  &  -2.3(8) & 11.0(8)\\
G  &  2.75 &  29.3945(7) & 2.8(7)  &  0.12(3)  &  27(1)    &  -6(1)   & 10(1)\\
\multicolumn{8}{c}{\textbf{NGC 6205 (M13) (6 Pulsars, FAST Lband)}}\\
A  &  10.37  & 30.4387(8) & 12.0(2)$^{\dagger}$ & 0.485(8) & 21.0(1) & -25.2(1) & 25.4(1)\\
B  &  3.52   & 29.4508(9) & 15.0(6) & 0.63(3)   & 26.9(8)  & 19.9(7) &  20.5(7)\\
C  &  3.72   & 30.1318(6) & 13.3(2)$^{\dagger}$  & 0.543(8)  & 37.2(3)  & -8.8(3) &  11.6(3)\\
D  &  3.11  & 30.4527(4)  & 13.1(8)$^{\dagger}$ & 0.53(3)   & 17.7(5)  & 0.0(5)  &  3.3(5)\\
E  &  2.48  & 30.5383(4)  & 15.2(9) & 0.61(4)   & 33.6(5)  & -4.2(5) &  5.1(5)\\
F  &  3.00  & 30.3732(7)  & 18(3) & 0.7(1)    & 20(1)    & 1(1)    &  12(1)\\
\multicolumn{8}{c}{\textbf{NGC 6266 (M62) (6) Pulsars, MeerKAT Lband)}}\\
A  &  5.24 &  114.992(1)  & -45.4(9) & -0.5(2)   &  19.0(9) &  -1.8(9) &  11.9(9)\\
B  &  3.59 &  113.3249(5) & -43.2(2) & -0.456(3) &  29.1(5) &  -2.9(5) &  1 6.2(5)\\
C  &  7.61 &  114.555(1)  & -43.2(3) & -0.472(4) &  51(1)   &  2.4(9)  &  1 14.8(9)\\
D  &  3.41 &  114.214(1)  & -46.1(3) & -0.502(4) &  51(1)   &  9.2(9)  &  1 13.8(9)\\
E  &  3.23 &  113.7543(5) & -47(1)   & -0.51(1)  &  11(1)   &  -1(1)   &  1 7(1)\\
F  &  2.29 &  113.316(1)  & -43(3)   & -0.47(3)  &  8(1)    &  10(1)   &  1 13(1)\\
\multicolumn{8}{c}{\textbf{NGC 6624 (6 Pulsars, MeerKAT Lband)}}\\
A   & 5.44   & 86.8963(3) & -17(4)$^{\dagger}$ & -0.24(6)   &  1.3(2)  &  3.6(2)   & 3.9(2)\\
B   & 378.59 & 86.892(6)  & -20.3(2) &  -0.287(3) &  17.8(1) &  -10.4(2) & 12.9(2)\\
C   & 405.93 & 86.567(8)  & -22.5(4) &  -0.320(6) &  16.3(4) &  9.6(4)   & 11.0(4)\\
D   & 3.02   & 86.924(1)  & -29(5)   &  -0.41(7)  &  32(2)   &  -1(2)    & 19(2)\\
G   & 6.09   & 86.189(1)  & -23.3(6) &  -0.333(9) &  41(1)   &  1(1)     & 6(1)\\
J1823-3022 & 2497.78 & 96.7(1) & -26(2)  &  -0.33(3)  &  10.4(7) & -0.9(7)  & 3.4(7)\\
\multicolumn{8}{c}{\textbf{NGC 6626 (M28) (7 Pulsars, MeerKAT Lband)}}\\
A &  3.05   &  119.89918(9) &  82.90(3)$^{\dagger}$ &  0.8504(3) & 75.8(2) &  -0.3(1) & 3.3(1)\\
B &  6.54)  &  119.313(4)  &  81(1   &  0.84(1)   & 77(3)   &  -2(2)   & 14(2)\\
C &  4.15   &  120.5558(7) &  76(1)  &  0.78(1)   & 27(1)   &  4(1)    & 11(1)\\
D &  79.83  &  120.01(7)   &  77(1)  &  0.79(1)   & 64(3)   &  -2(2)   & 12(2)\\
G &  5.90   &  119.668(6)  &  75(1)  &  0.77(1)   & 23(3)   &  -3(3)   & 24(3)\\
J &  4.03   &  119.219(2)  &  77(12) &  0.7(1)    & 17(2)   &  1(2)    & 12(2)\\
K &  4.46   &  119.519(4)  &  82(3)  &  0.84(3)   & 50(3)   &  -2.2(3) & 13(3)\\
\multicolumn{8}{c}{\textbf{NGC 6656 (M22) (1 Pulsar, MeerKAT Lband)}}\\
A  & 3.35 &  89.1252(3) & 38(2) & 0.52(3)  & 11(1) & 2(1)  & 8(1)\\
\multicolumn{8}{c}{\textbf{NGC 6752 (6 Pulsars, MeerKAT UHF band)}}\\
A  & 3.26  & 33.6768(3) &  43.0(2)$^{\dagger}$ &  1.571(7) & 30.1(4) & 6.8(4) & 10.9(4)\\
B  & 8.35  & 33.295(1)  &  50.5(2) &  1.885(7) & 48(3)   & -14(3) & 29(3)\\
C  & 5.27  & 33.2817(4) &  51.3(4) &  1.90(1)  & 19.5(4) & 0.3(4) & 6.2(4)\\
D  & 9.03) & 33.297(1)  &  49.7(5) &  1.84(2)  & 30(2)   & -9(2)  & 11(2)\\
E  & 4.57  & 33.3250(5) &  49(1) &  1.81(4)  & 12(2)   & -3(2)  & 11(2)\\
F  & 8.48  & 33.2224(5) &  52(2) &  1.93(7)  & 13(3)   & -4(3)  & 20(3)\\
\multicolumn{8}{c}{\textbf{NGC 7078 (M15) (4 Pulsars, FAST Lband)}}\\
A & 110.66 &  67.244(4) & -70.6(3)$^{\dagger}$ & -1.291(5) & 14.7(1) & -0.37(1) &  4.3(1)\\
B & 56.13  &  67.752(5) & -65.9(3)$^{\dagger}$ & -1.196(5) & 28.5(3) & 2.2(3)   &  10.5(3)\\
D & 4.80   &  67.303(2) & -71(1)$^{\dagger}$   & -1.30(2)  & 23.6(8) & 8.1(6)   &  13.3(6)\\
E & 4.65   &  66.593(2) & -70(1)$^{\dagger}$   & -1.29(2)  & 24.7(9) & -1.3(8)  &  8.9(8)\\
\hline 
\multicolumn{8}{l}{{\bf Note:} The value in the bracket shows the 1$\sigma$ error on the last digit.}\\
\multicolumn{8}{l}{$^{\dagger}$ The pulsar with previously published RM value~\citep{Dai2015,Abbate20,Corongiu23,Wang23}.}\\
\end{tabular}}
\end{table*}
\begin{table*}
\centering
\caption{Summary of 49 pulsars and measurements derived from previous work~\protect\citep{Martsen22,Abbate23}. The $\left<B_{||}\right>$, percentage of linear polarization (L/I), circular polarisation (V/I) and absolute circular polarisation ($|V|$/I) towards each pulsar used the RM, DM and polarization pulse profile presented in~\protect\cite{Martsen22,Abbate23}.}
\label{tab:RMs_47Tuc-Ter5}
\setlength{\tabcolsep}{3.2mm}{
\begin{tabular}{cccccccc}\hline
Pulsar &  Period  & DM    & RM         &  $\left<B_{||}\right>$ & L/I   &   V/I    &  $|V|$/I  \\
       &  (ms)    & (pc cm$^{-3}$) & (rad m$^{-2}$) &  ($\mu$G)   & (\%)  &   (\%)   &  (\%)     \\\hline
\multicolumn{8}{c}{\textbf{Terzan 5 (27 Pulsars, GBT Lband)}} \\
A  &  11.56 &  242.34 &  191.2(3) &  0.97(2)  &  14.3(2)  &  1.3(2)   &  14.1(2)\\
C  &  8.43  &  237.06  &  176.97(6) &  0.9182(3)&  84.0(2)  &  11.8(1)  &  13.2(1)\\
D  &  4.71  &  243.62  &  234(2) &  1.18(1)  &  13.3(7)  &  -5.7(7)  &  13.5(7)\\
E  &  2.19  &  236.63  &  171.0(7) &  0.89(4)  &  18.8(6)  &  -0.4(6)  &  6.9(6)\\
F  &  5.54  &  239.07  &  187.2(6) &  0.96(3)  &  61(1)    &  -4.2(9)  &  9.5(9)\\

G  &  21.67 &  237.34  &  196.4(4) &  1.018(2) &  78(1)    &  3(1)     &  6(1)\\
H  &  4.92  &  237.97  &  194(6) &  1.00(3)  &  14(2)    &  -2.4(18) &  12.1(18)\\
I  &  9.57  &  238.55  &  181.7(7) &  0.937(4) &  31.5(5)  &  1.8(5)   &  4.8(5)\\
J  &  80.33 &  234.21  &  172.4(5) &  0.905(3) &  27.3(7)  &  -4.1(7)  &  8.9(7)\\
K  &  2.96  &  234.46  &  175(1) &  0.918(5) &  52.8(13) &  -0.9(12) &  12.5(12)\\

L  &  2.24  &  237.50  &  167(7) &  0.86(4)  &  10.5(6)  &  -11.4(6) &  13.3(6)\\
M  &  3.56  &  238.49  &  169(2) &  0.87(1)  &  13.6(5)  &  -6.6(5)  &  9.0(5)\\
N  &  8.66  &  238.29  &  183(1) &  0.945(5) &  10.5(3)  &  -38.6(3) &  39.5(3)\\
O  &  1.67  &  236.20  &  176.2(3) &  0.918(2) &  12.8(80) &  -0.7(80) &  6.9(80)\\
Q  &  2.81  &  234.24  &  175.4(8) &  0.921(4) &  14.3(68) &  -6.0(67) &  12.6(68)\\

R  &  5.02  &  237.38  &  179(4) &  0.93(2)  &  18.4(20) &  -2.2(20) &  11.3(20)\\
V  &  2.07  &  238.71  &  186.1(8) &  0.959(4) &  34.6(9)  &  -5.0(8)  &  10.6(8)\\
W  &  4.20  &  238.92  &  174.3(9) &  0.897(5) &  23.2(9)  &  7.1(9)   &  11.1(9)\\
X  &  2.99  &  239.81  &  202(2) &  1.04(1)  &  23.9(16) &  3.0(15)  &  9.9(15)\\
Y  &  2.04  &  238.79  &  194(3) &  0.10(2)  &  28.8(16) &  -13.1(15)&  21.4(16)\\

ab &  5.11  &  238.40  &  171(1) &  0.882(5) &  51.1(20) &  0.1(18)  &  12.3(18)\\
ac &  5.08  &  238.69  &  203(3) &  1.05(2)  &  28.6(28) &  1.3(27)  &  12.0(27)\\
ae &  3.65  &  238.61  &  184(1) &  0.948(5) &  43.2(15) &  -1.1(14) &  15.4(14)\\
af &  3.30  &  237.35  &  184(3) &  0.95(2)  &  22.4(27) &  -3.1(26) &  13.9(26)\\
ag &  4.44  &  237.28  &  186(4) &  0.96(2)  &  24.8(29) &  -9.8(28) &  13.9(28)\\

ah &  4.96  &  237.70  &  184(4) &  0.95(2)  &  40.1(54) &  -12.3(51)&  29.9(52)\\
ai &  21.2  &  234.02  &  173.1(5) &  0.910(3) &  96.4(17) &  10.4(12) &  14.3(12)\\
\multicolumn{8}{c}{\textbf{NGC 104 (47Tuc) (22 Pulsars, MeerKAT UHF)}} \\
C  & 5.75  &  24.5909(3) &  27.3(2)  &  1.37(1)  &  12.5(3) &  26.7(3) &  27.0(3)\\
D  & 5.35  &  24.7412(3) &  26.0(2)  &  1.29(1)  &  30.3(5) &  0.5(5)  &  3.9(5)\\
E  & 3.53  &  24.2396(2) &  26.0(1)  &  1.319(5) &  34.1(2) &  0.9(2)  &  6.3(2)\\
F  & 2.62  &  24.3841(3) &  26.6(2)  &  1.34(1)  &  23.5(4) &  14.3(4) &  15.0(4)\\
G  & 4.04  &  24.4343(2) &  25.8(1)  &  1.299(5) &  37.7(8) &  0.5(8)  &  7.3(8)\\

H  & 3.21 &  24.3750(9) &  27.5(8)  &  1.39(4)  &  12.8(5) &  -0.5(5) &  8.0(5)\\
I  & 3.48 &  24.4303(3) &  26.2(1)  &  1.319(5) &  57.9(9) &  -2.4(8) &  6.6(8)\\
J  & 2.10 &  24.5937(5) &  24.0(3)  &  1.20(2)  &  5.53(9) &  -0.28(9)&  6.19(9)\\
L  & 4.34 &  24.3986(6) &  26.2(1)    &  1.321(5) &  47.2(4) &  0.5(4)  &  6.8(4)\\
M  & 3.67 &  24.426(3)  &  25.3(4)   &  1.27(2)  &  19(3)   &  -6(3)   &  26(3)\\

N  & 3.05 &  24.5568(2)  &  25.9(1)  &  1.297(5) &  70(1)   &  -1.0(9) &  17.2(9)\\
O  & 2.64 &  24.3580(1)  &  25.76(9) &  1.301(5) &  38.1(3) &  3.3(2)  &  9.9(2)\\
Q  & 4.03 &  24.2794(9)  &  25.2(1)  &  1.277(5) &  55.1(8) &  -0.7(7) &  9.1(7)\\
R  & 3.48 &  24.36100(8) & 25.7(2)   &  1.30(1)  &  17.9(5) &  6.1(5)  &  8.2(5)\\
S  & 2.83 &  24.38179(1) &  25.80(8) &  1.302(4) &  51.3(5) &  1.7(5)  &  2.5(5)\\

T  & 7.58 & 24.421(4)  &  26.0(7)    &  1.31(4)  &  12(2)   &  0(2)    &  11(2)\\
U  & 4.34 & 24.340(1)  &  26.26(4)   &  1.327(2) &  48.9(7) &  2.6(7)  &  13.7(7)\\
W  & 2.35 & 24.370(4)  &  26.0(2)    &  1.31(1)  &  60(3)   &  4(2)    &  24(2)\\
X  & 4.77 & 24.538(2)  &  26.4(2)    &  1.32(1)  &  69(1)   &  2(1)    &  15(1)\\
Y  & 2.19 & 24.475(3)  &  26.0(3)    &  1.31(2)  &  33.8(7) &  2.5(7)  &  9.1(7)\\

aa & 3.69 & 24.921(2)  &  28.9(6)    &  1.43(3)  &  22(4)   &  -6(4)   &  21(4)\\
ab & 3.70 & 24.3256(5) &  26.4(6)    &  1.33(3)  &  35(2)   &  2(2)    &  28(2)\\
\hline 
\multicolumn{8}{l}{{\bf Note:} The value in the bracket shows the 1$\sigma$ error on the last digit.}\\
\end{tabular}}
\end{table*}

\begin{table*}
\setlength{\tabcolsep}{4mm}{
\centering
\caption{The linear gradient and maximum variation of the magnetic in front of the cluster.}
\label{tab:magnetic-field}
\begin{tabular}{lcc}\hline
Name            &  Linear gradient    & Maximum variation\\
                & (nG arsec$^{-1}$)   &  (nG arsec$^{-1}$)\\\hline
\multicolumn{3}{c}{\bf{Halo GCs}}\\\hline
NGC 104 (47Tuc) & $0.0_{-0.1}^{+0.1}$     &    $174 \pm 117$ \\\\
NGC 5904 (M5)   & $8.8_{-5.4}^{+4.6}$   &    $114 \pm 2$ \\\\
NGC 6205 (M13)  & $3.3_{-3.8}^{+2.6}$     &    $10 \pm 8$ \\\\
NGC 7078 (M15)  & -  &    $18 \pm 5$ \\\\
NGC 6752        & $0.4_{-2.9}^{+6.4}$    &    $39 \pm 35$ \\\hline
\multicolumn{3}{c}{\bf{Bulge GCs}}\\\hline
Terzan 5        & $4.2_{-0.6}^{+0.3}$    &    $194 \pm 5$ \\\\
NGC 6266 (M62)  & $2.1_{-0.7}^{+0.7}$     &    $5 \pm 5$ \\\\
NGC 6626 (M28)  & $3.3_{-2.6}^{+3.8}$     &    $17 \pm 3$ \\\\
NGC 6624        & -    &    $18 \pm 10$ \\\hline
\end{tabular}}
\end{table*}

\begin{table*}
\setlength{\tabcolsep}{13.8mm}{
\centering
\caption{The upper limits of ionized gas in the clusters.}
\label{tab:gas-limits}
\begin{tabular}{cc}\hline
Name            &  $n$\\
                & (cm$^{-3}$)       \\\hline
\multicolumn{2}{c}{\bf{Halo GCs}}      \\\hline
NGC 5904 (M5)   & 0.11            \\
NGC 6205 (M13)  & 0.19            \\\hline
\multicolumn{2}{c}{\bf{Bulge GCs}}     \\\hline
Terzan 5        & 1.01            \\
NGC6266 (M62)   & 19.73            \\
NGC 6626 (M28)  & 2.86           \\ \hline
\end{tabular}}
\end{table*}

\begin{figure*}
\centering 
    \centering
    \includegraphics[height=4.4cm,width=5.7cm]{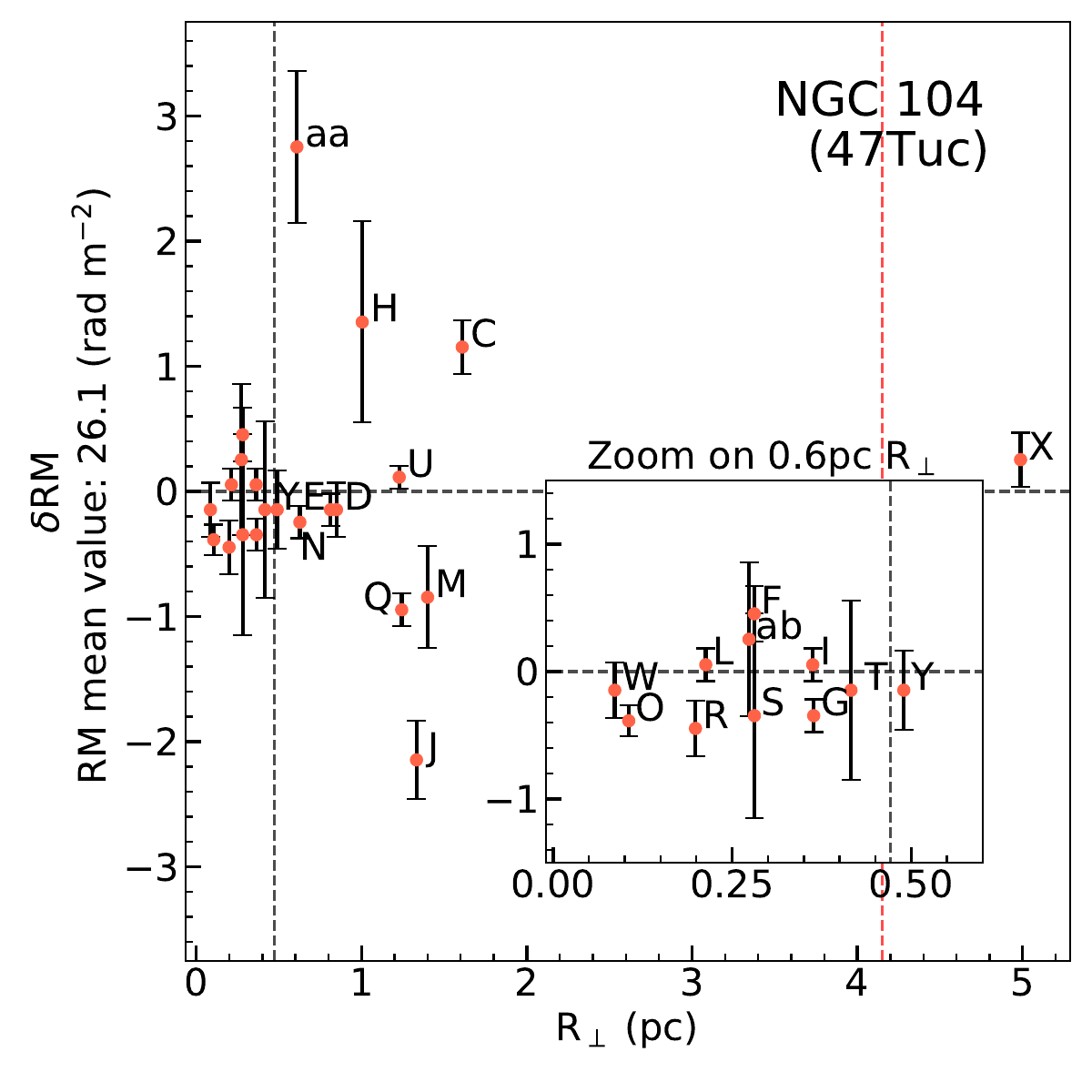}
    \includegraphics[height=4.4cm,width=5.7cm]{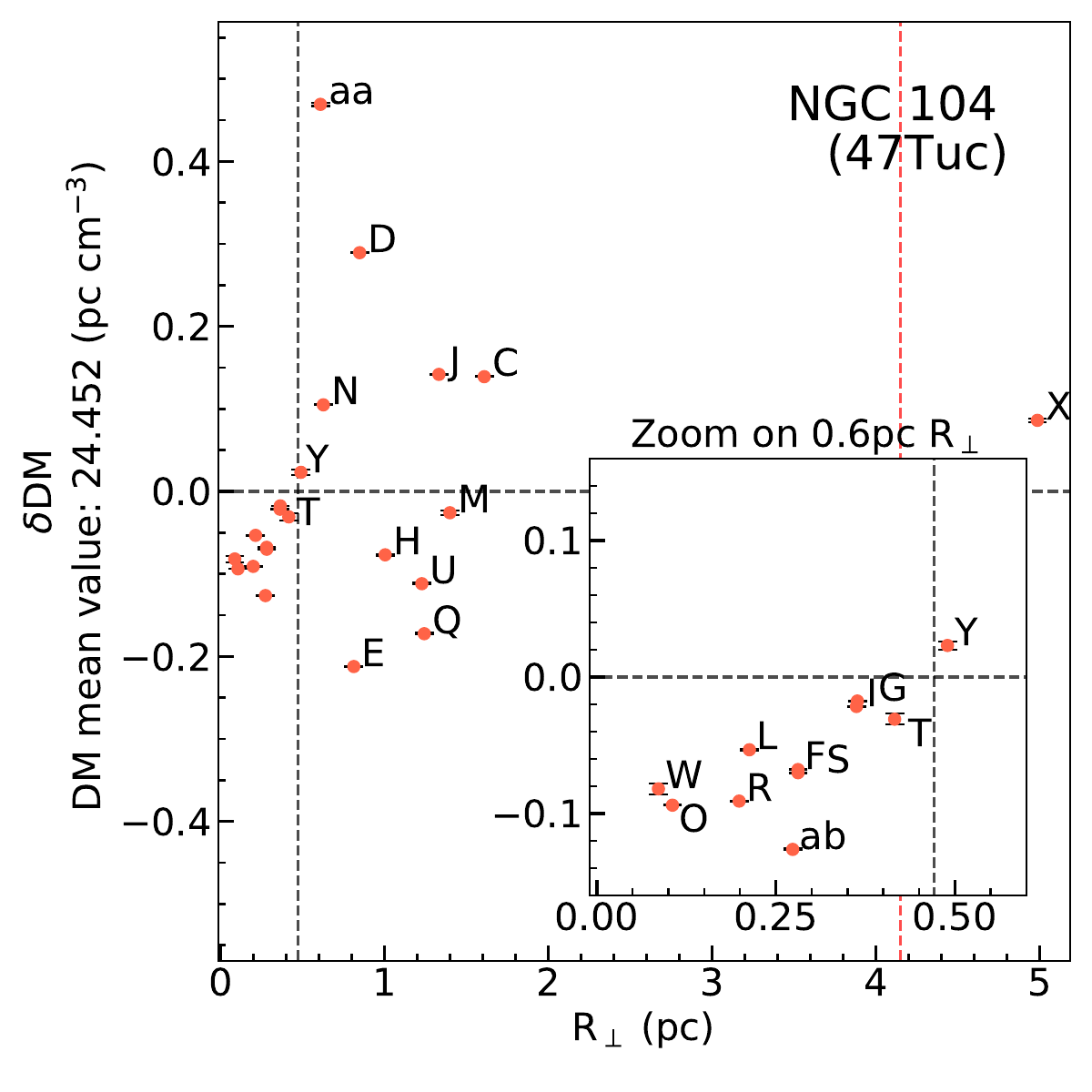}
    \includegraphics[height=4.4cm,width=5.7cm]{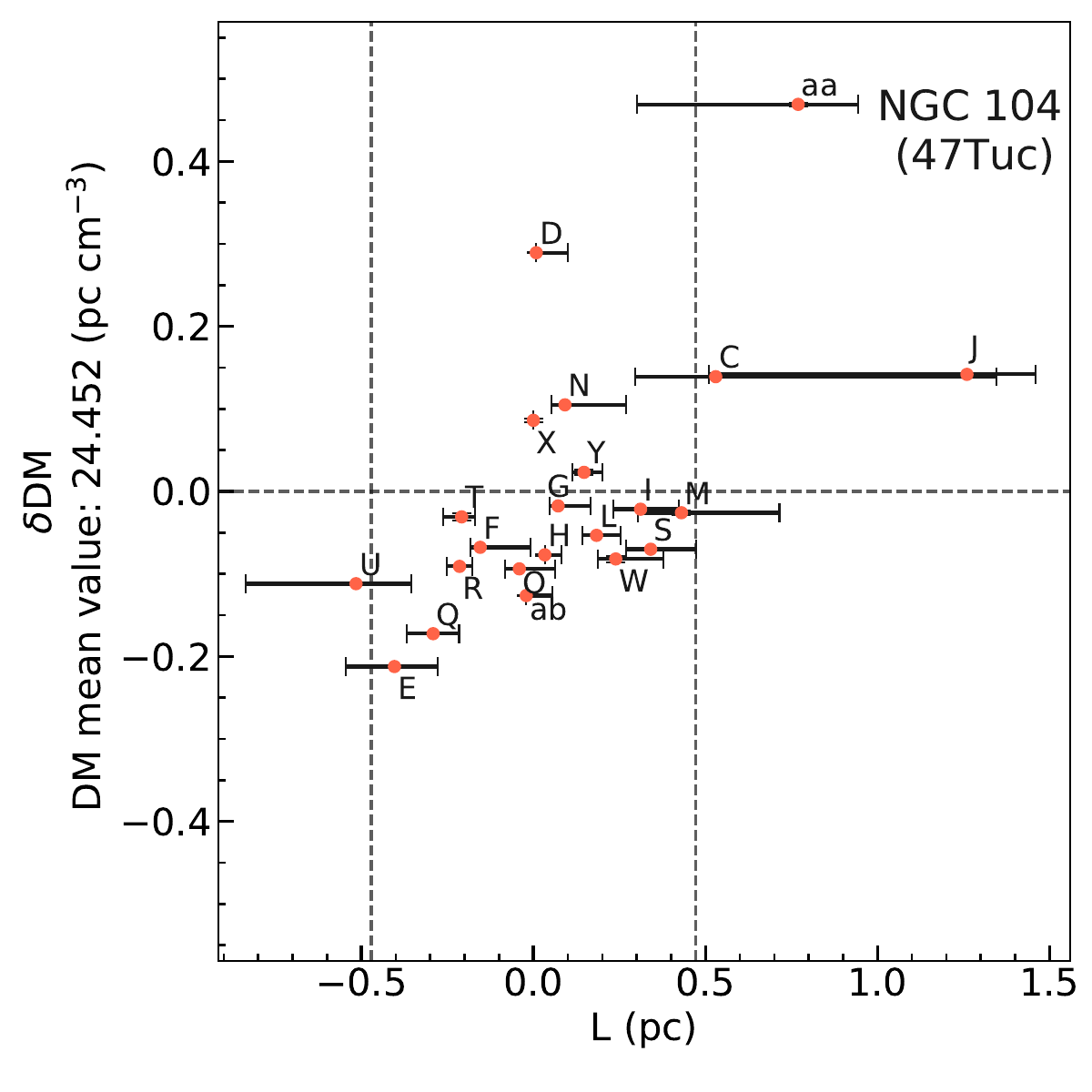}\\
    \includegraphics[height=4.4cm,width=5.7cm]{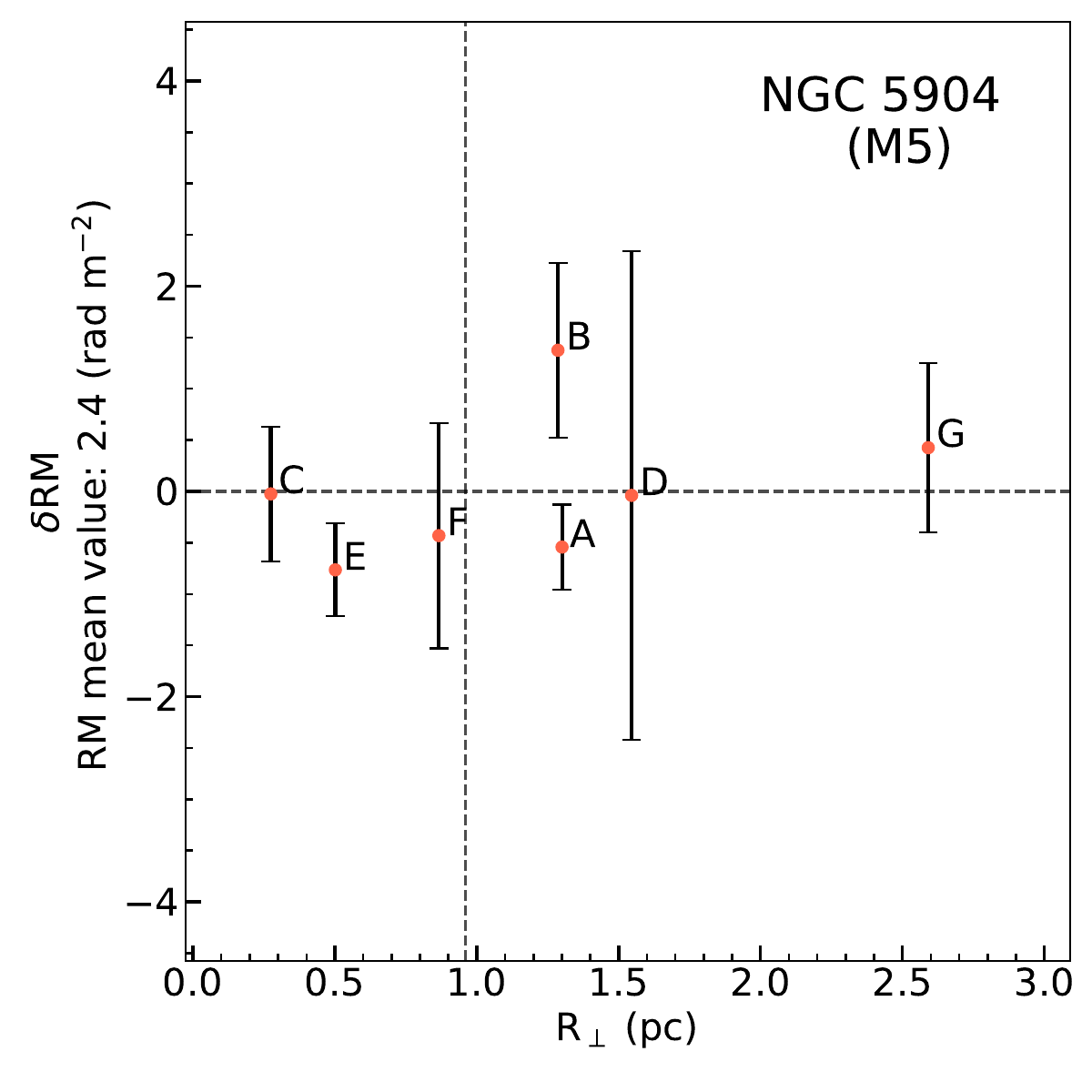}
    \includegraphics[height=4.4cm,width=5.7cm]{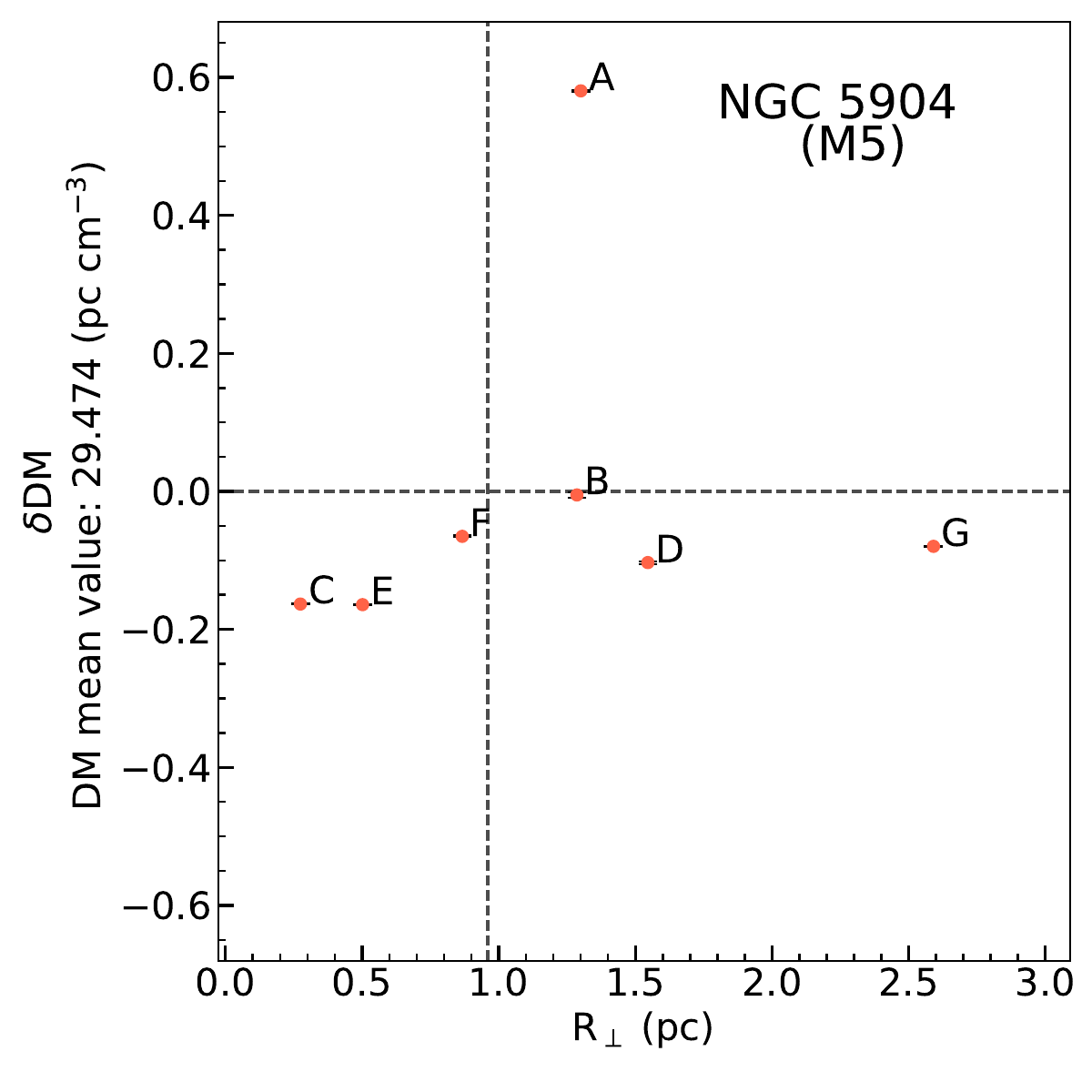}
    \includegraphics[height=4.4cm,width=5.7cm]{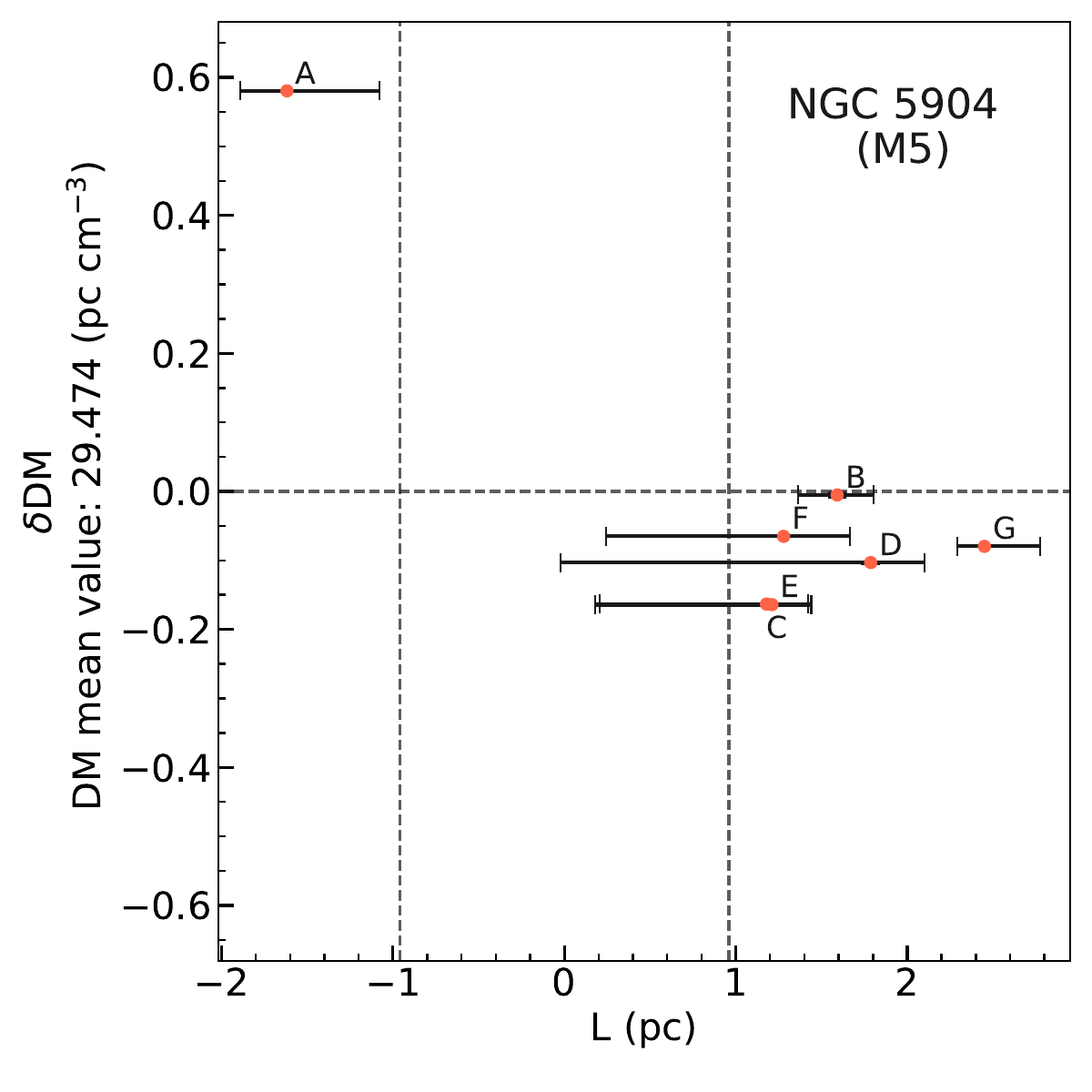}\\
    \includegraphics[height=4.4cm,width=5.7cm]{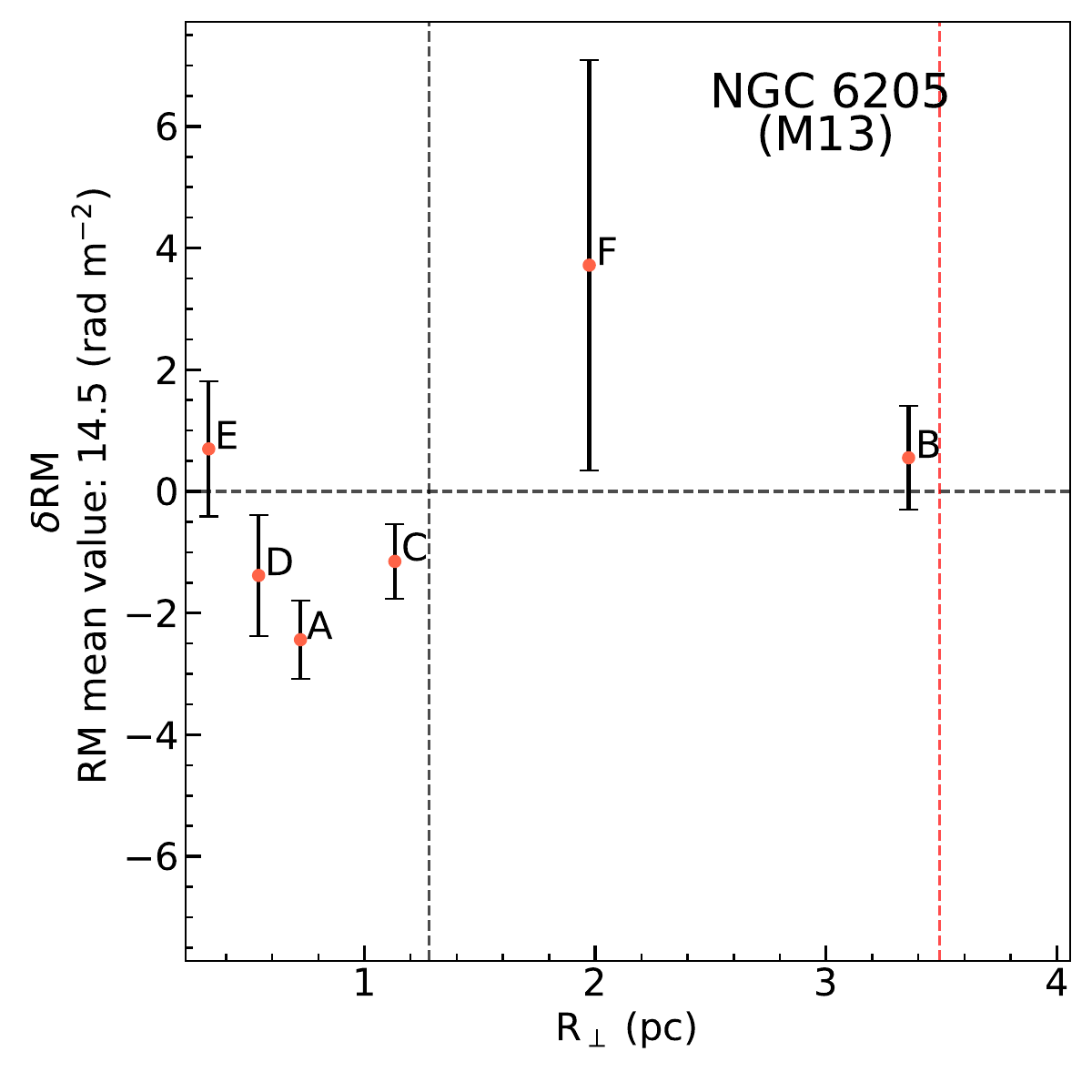}
    \includegraphics[height=4.4cm,width=5.7cm]{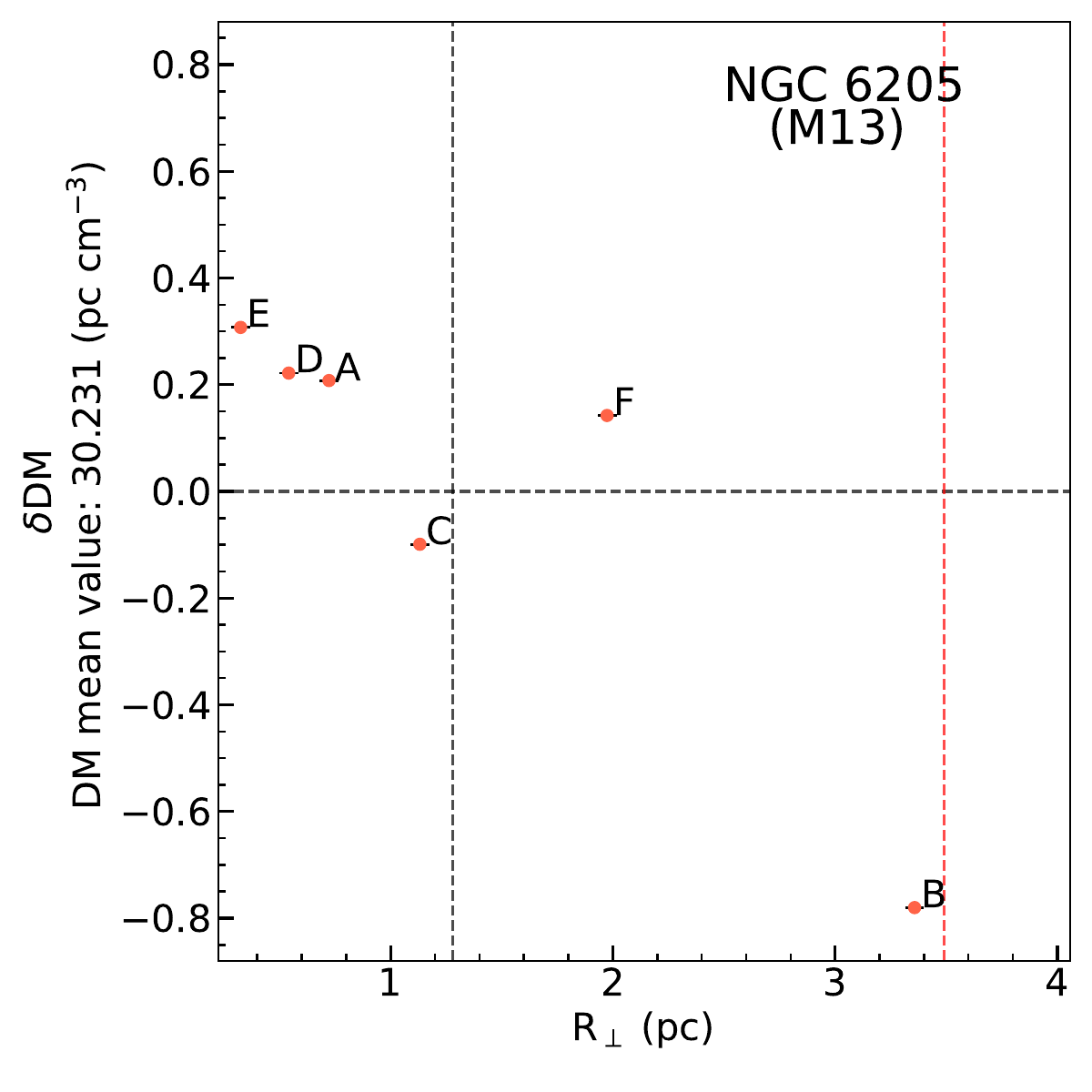}
    \includegraphics[height=4.4cm,width=5.7cm]{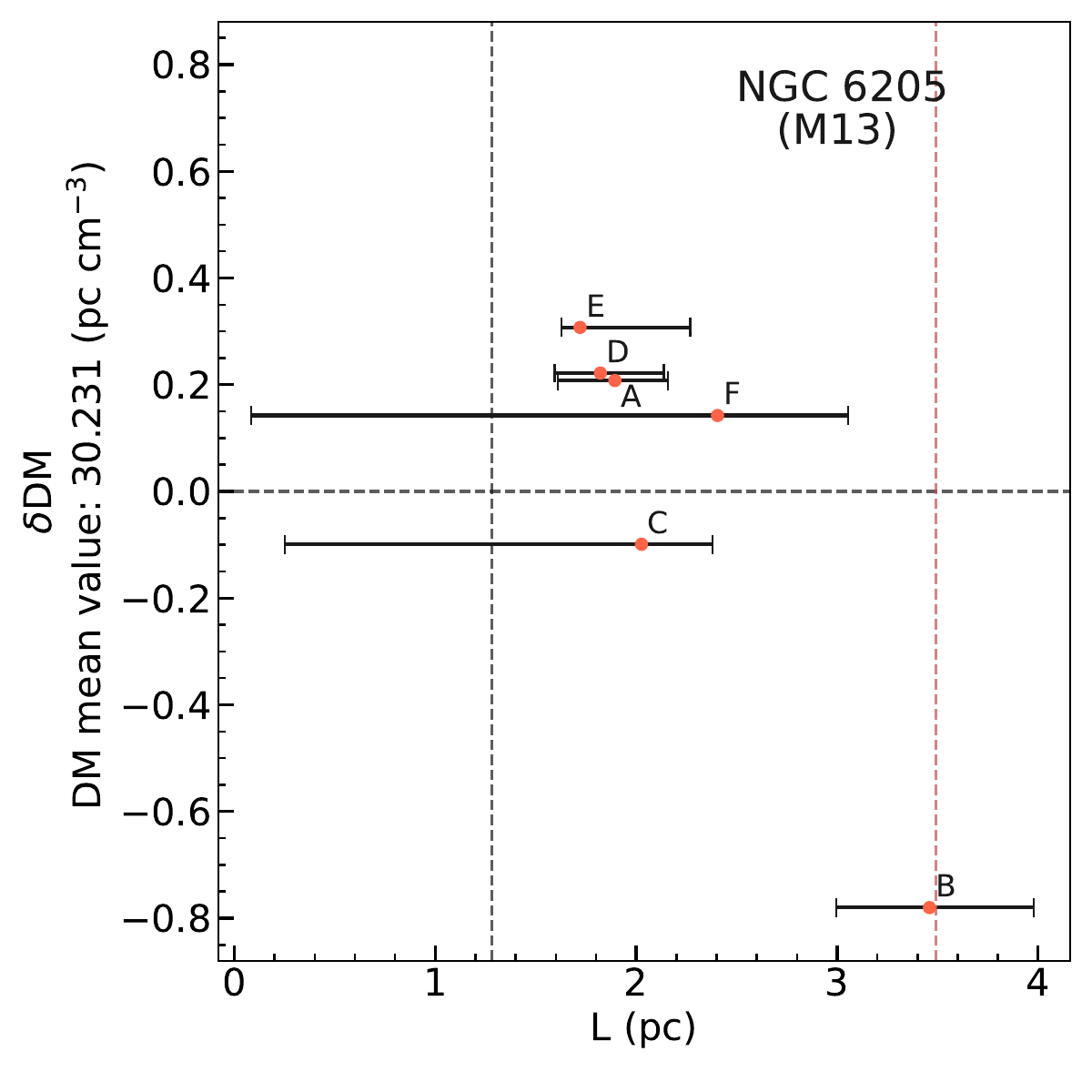}\\
    \includegraphics[height=4.4cm,width=5.7cm]{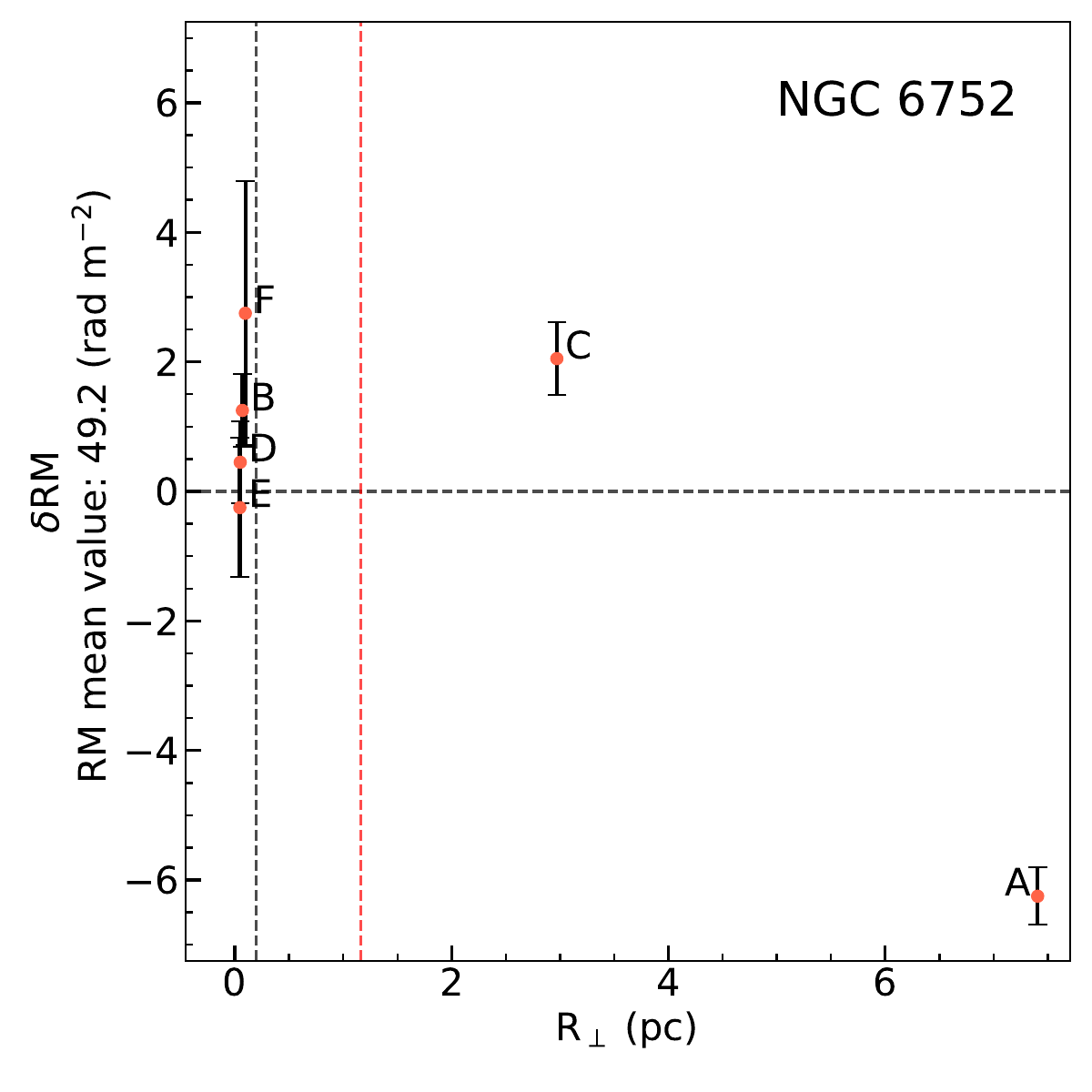}
    \includegraphics[height=4.4cm,width=5.7cm]{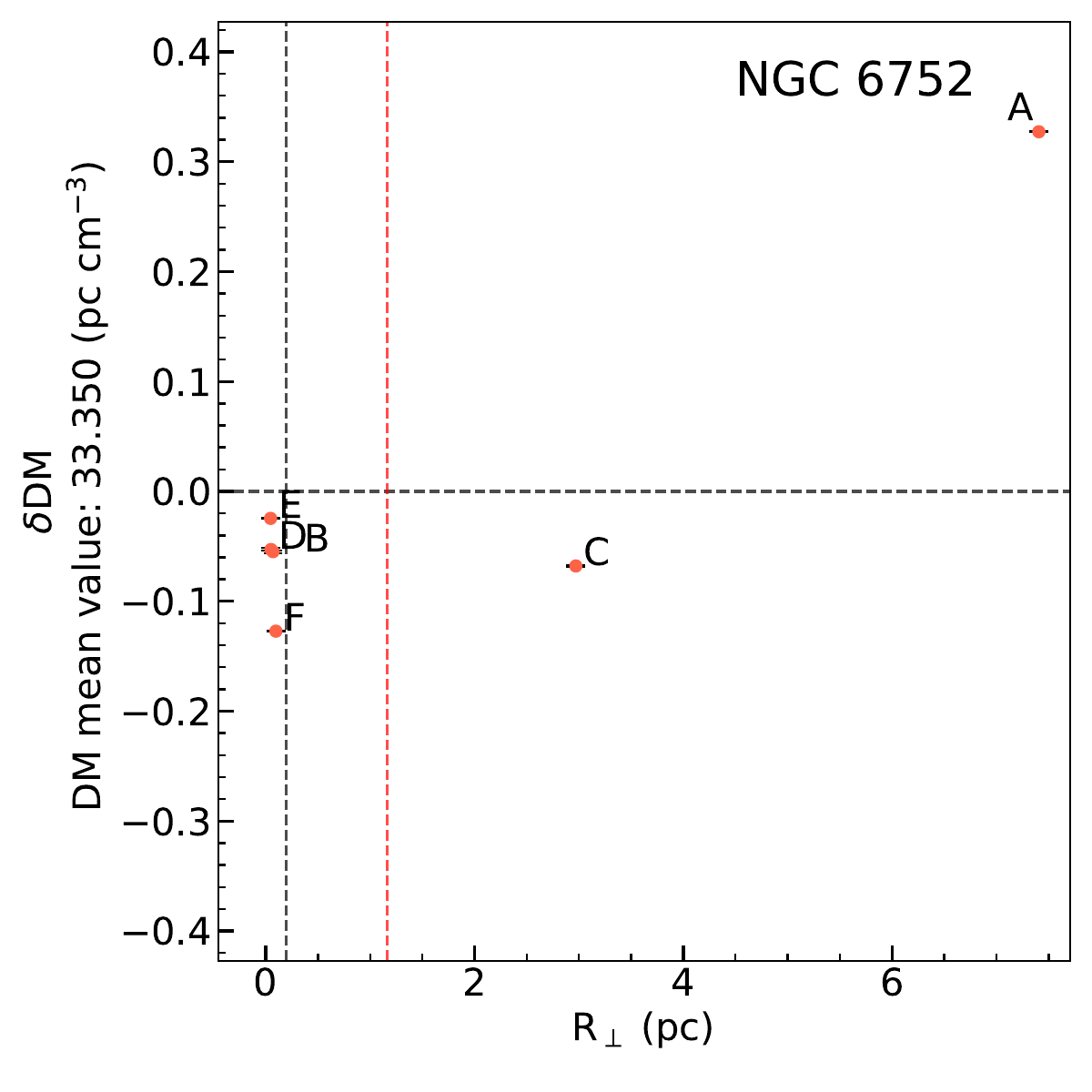}
    \includegraphics[height=4.4cm,width=5.7cm]{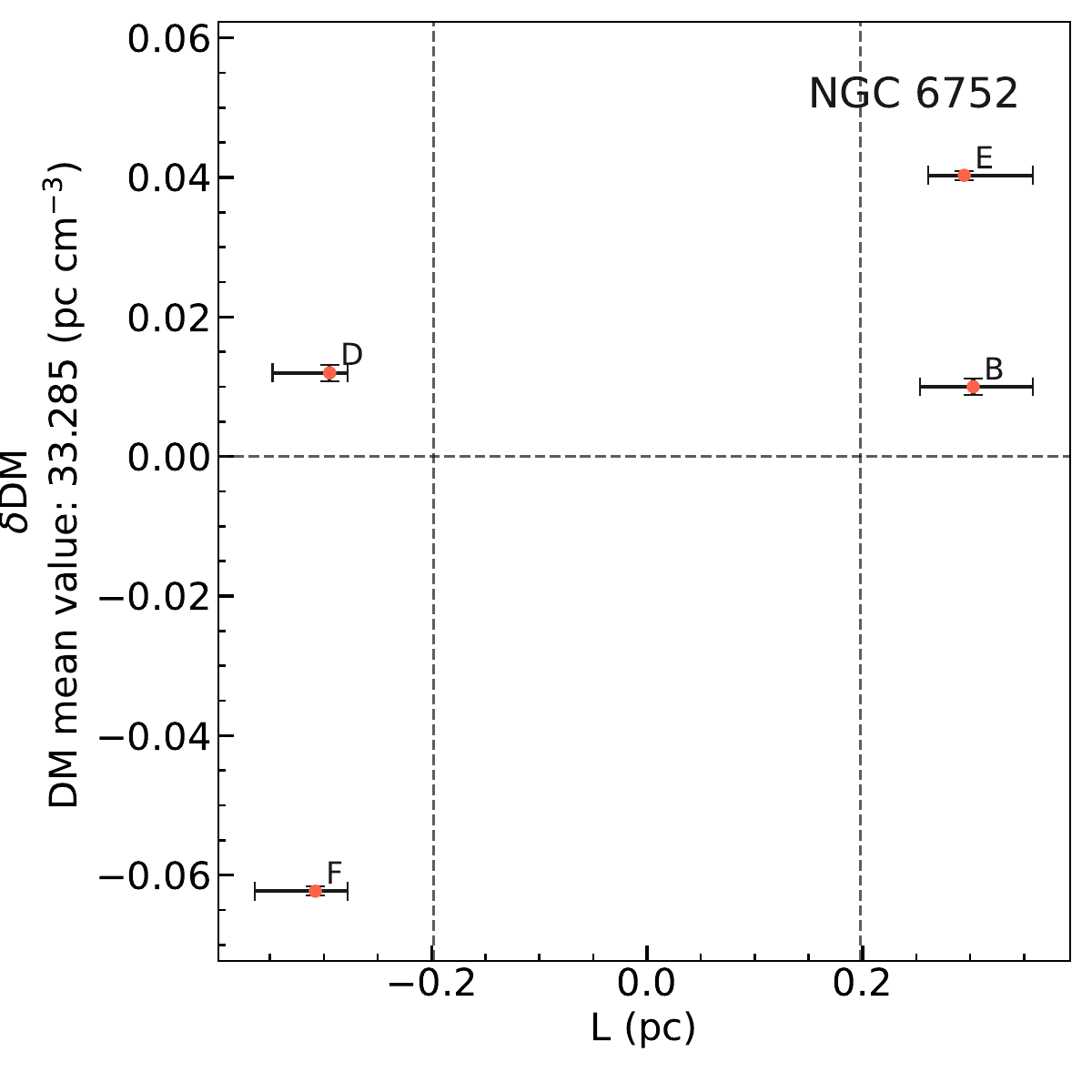}\\
    \includegraphics[height=4.4cm,width=5.7cm]{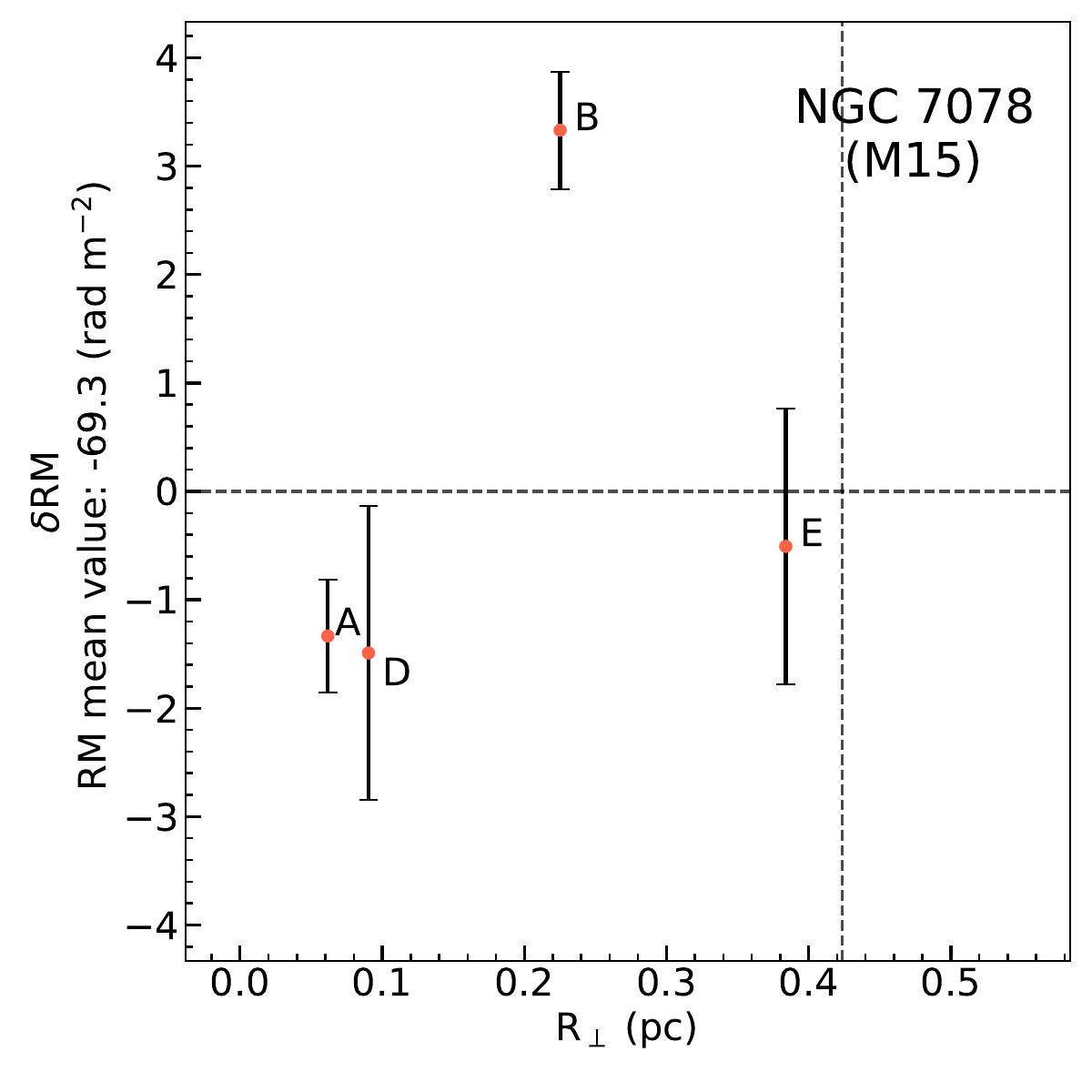}
    \includegraphics[height=4.4cm,width=5.7cm]{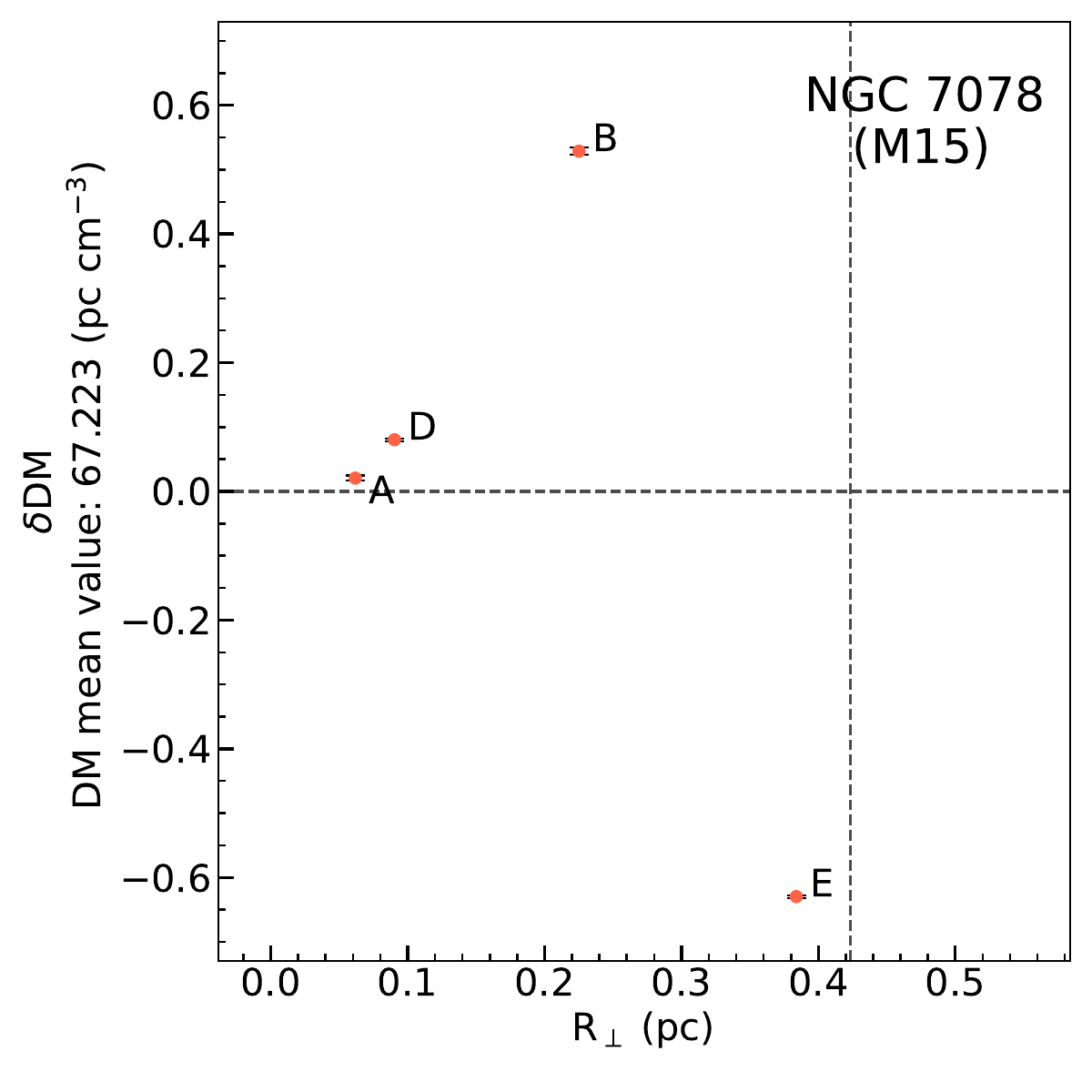}
    \includegraphics[height=4.4cm,width=5.7cm]{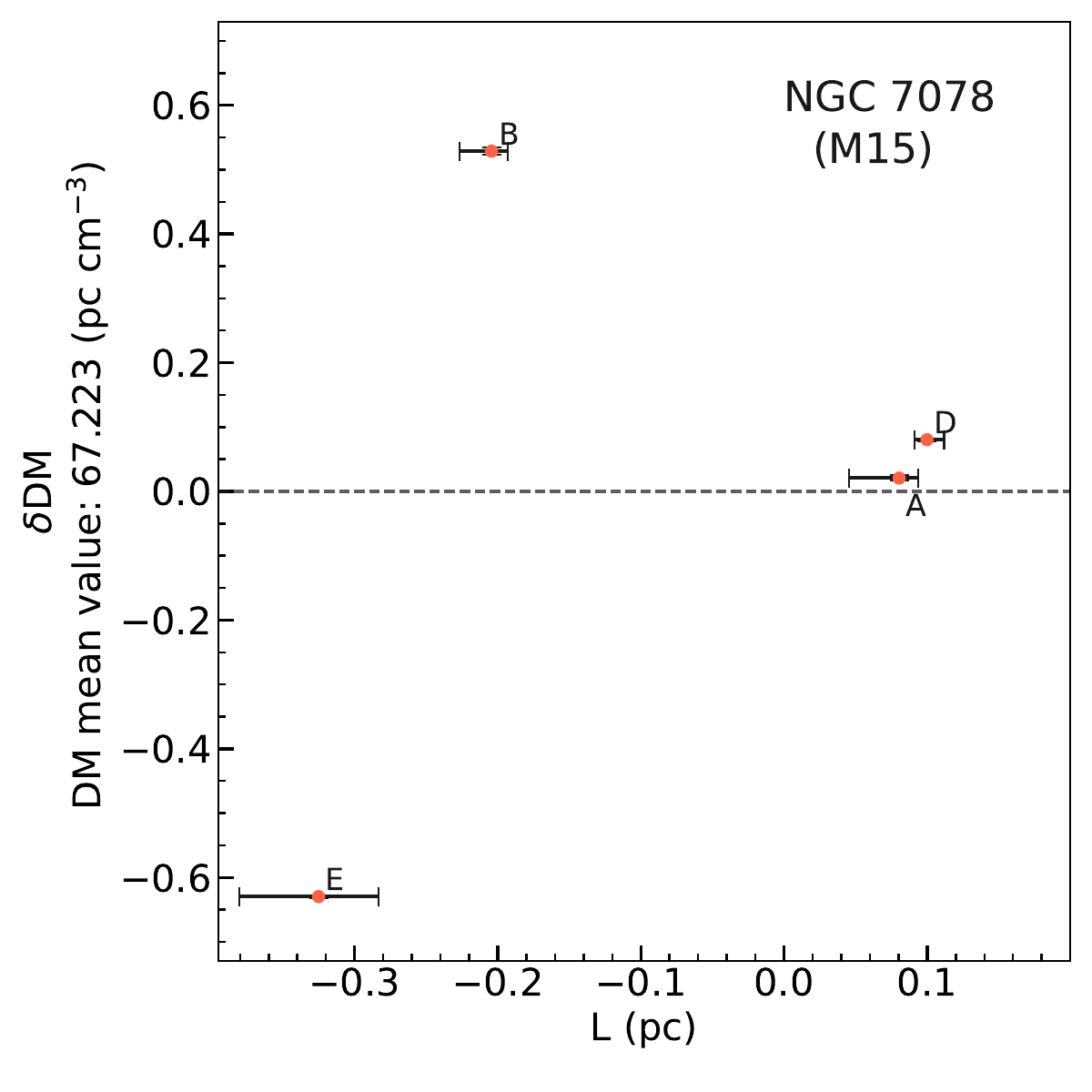}\\
\caption{Pulsars in five halo clusters. \textit{Left panel:} The difference between measured RM and the mean against position perpendicular offset from the center of GC. \textit{Middle panel:} The difference between the measured DM and the mean against position perpendicular offset from the center of GC. \textit{Right panel:} Measured DM and the mean plotted against the inferred line-of-sight distance (L) of pulsars within the plane of the sky containing the cluster center. Pulsars located far from the cluster center were excluded from the fits to the inferred line-of-sight distance. The vertical lines show the distance from the core radius (black dished line) and half-light radius (red dished line) to the cluster center.}
\label{fig.LpolRM_halo}
\end{figure*}

 \begin{figure*}
\centering 
    \centering
    \includegraphics[height=4.4cm,width=5.7cm]{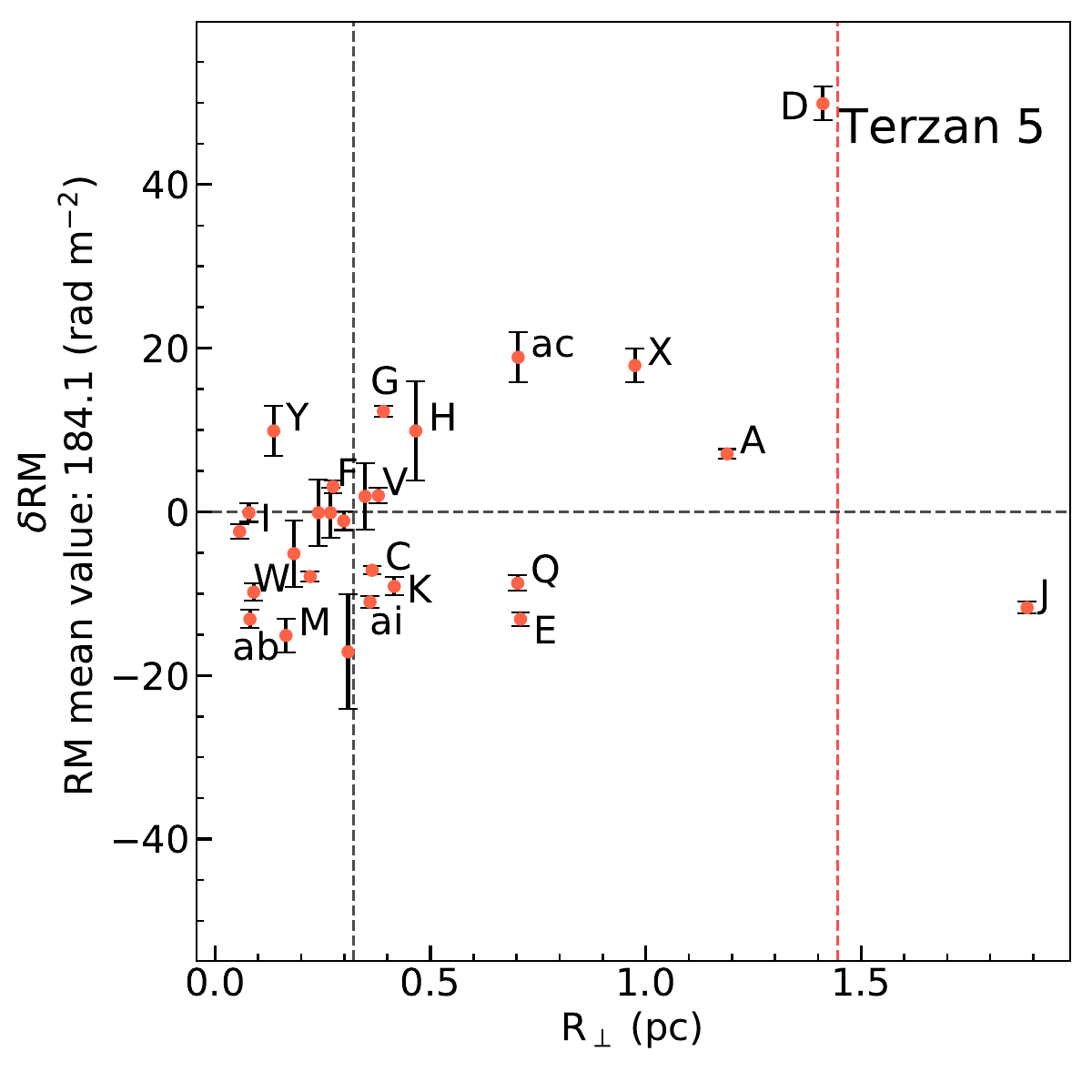}
    \includegraphics[height=4.4cm,width=5.7cm]{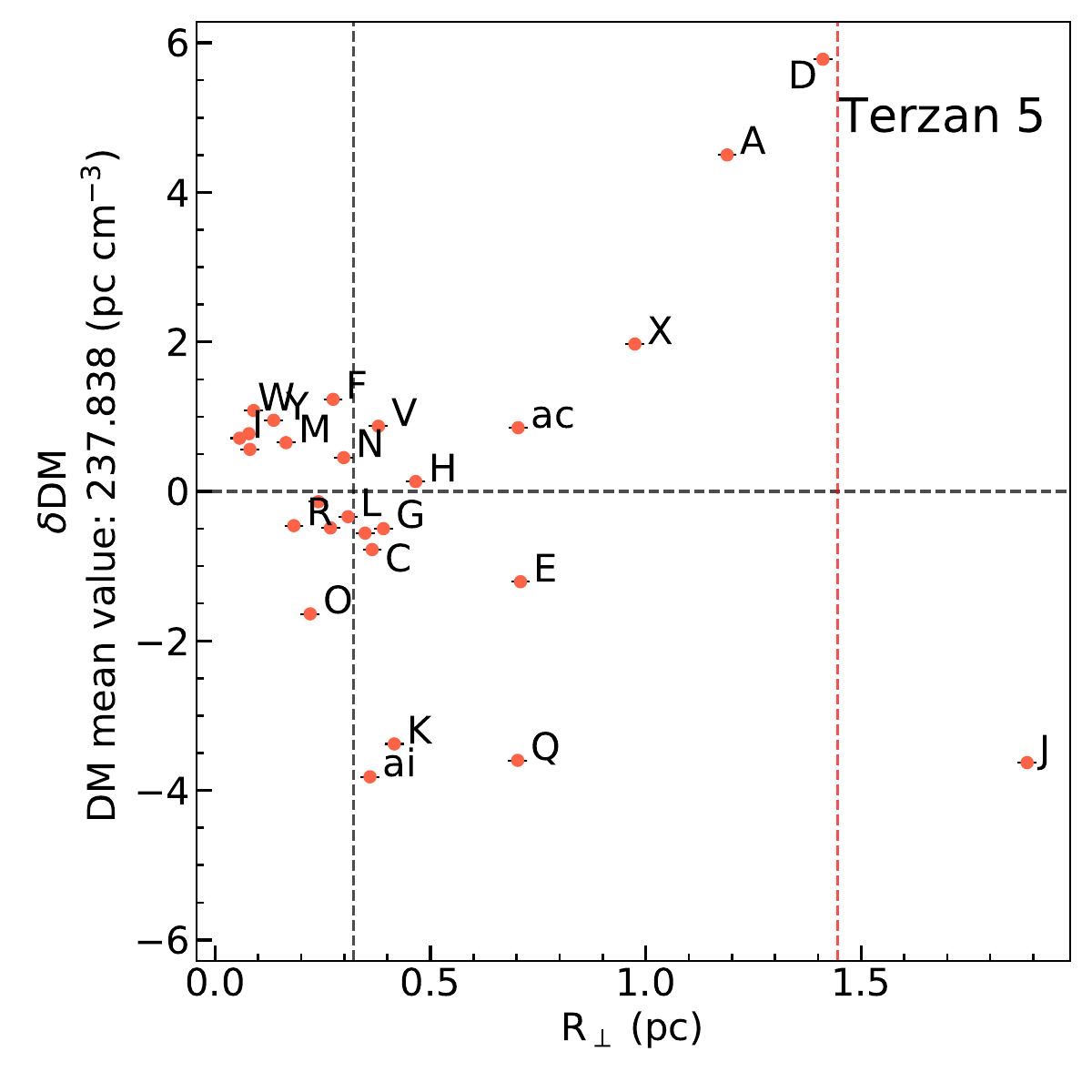}
    \includegraphics[height=4.4cm,width=5.7cm]{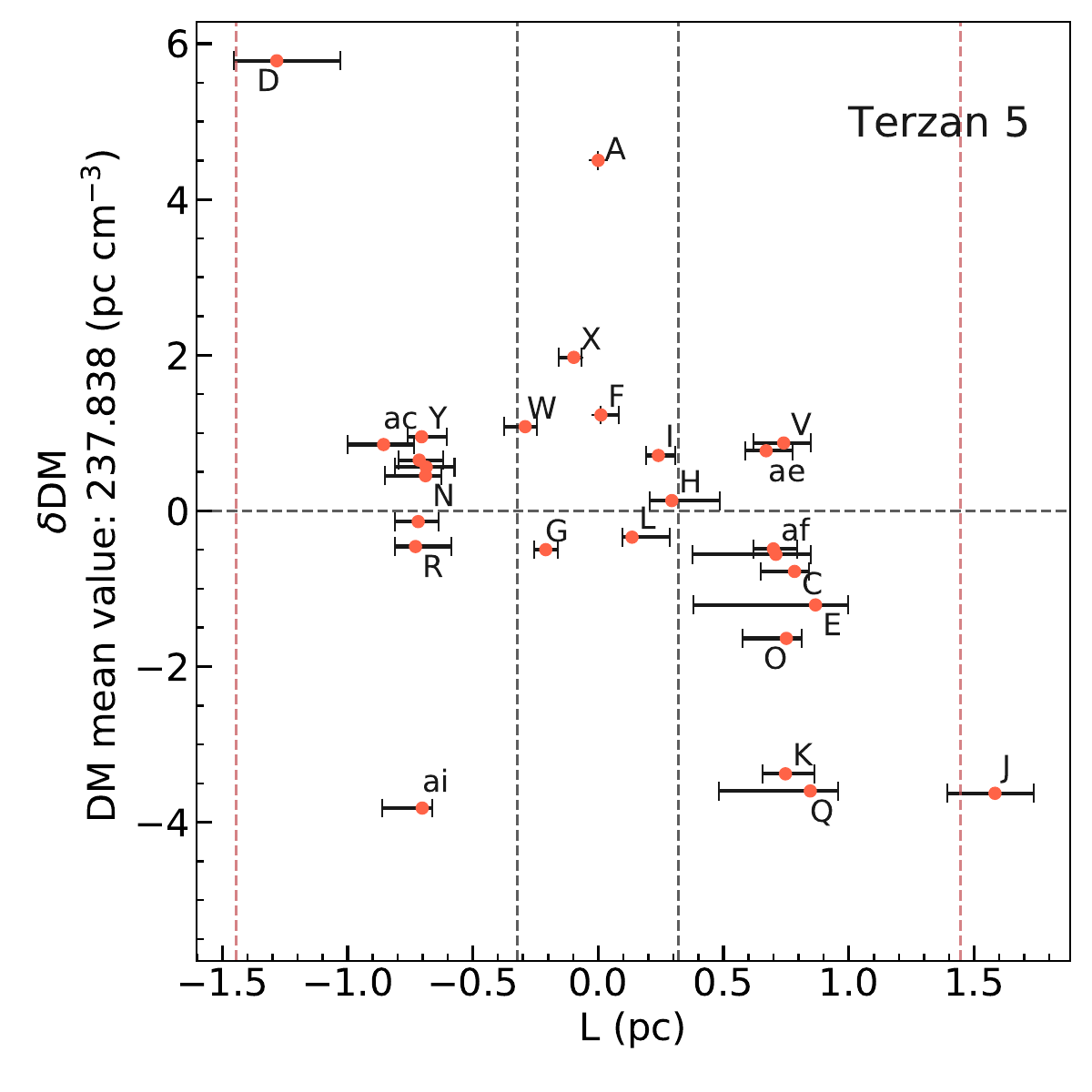}\\
    \includegraphics[height=4.4cm,width=5.7cm]{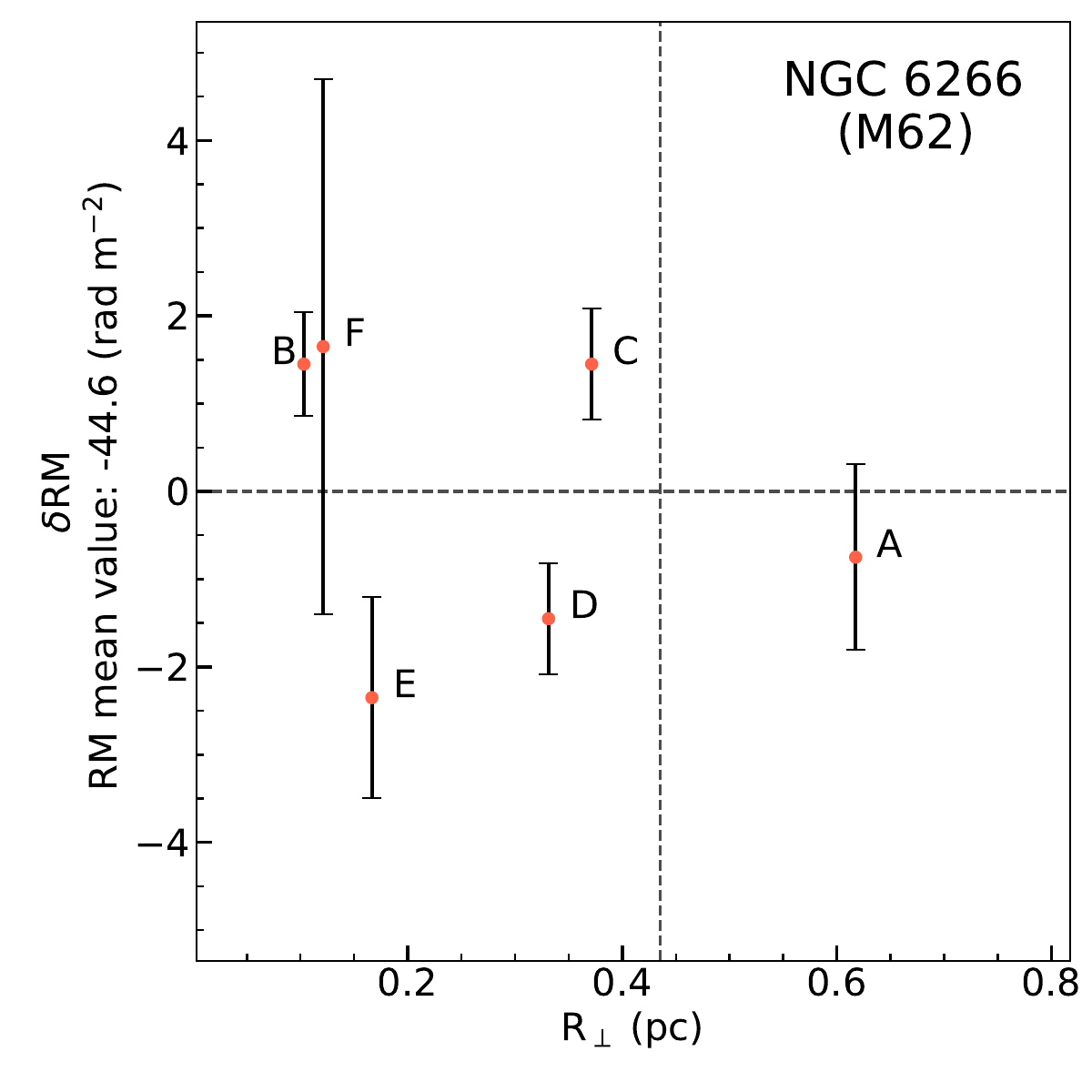}
    \includegraphics[height=4.4cm,width=5.7cm]{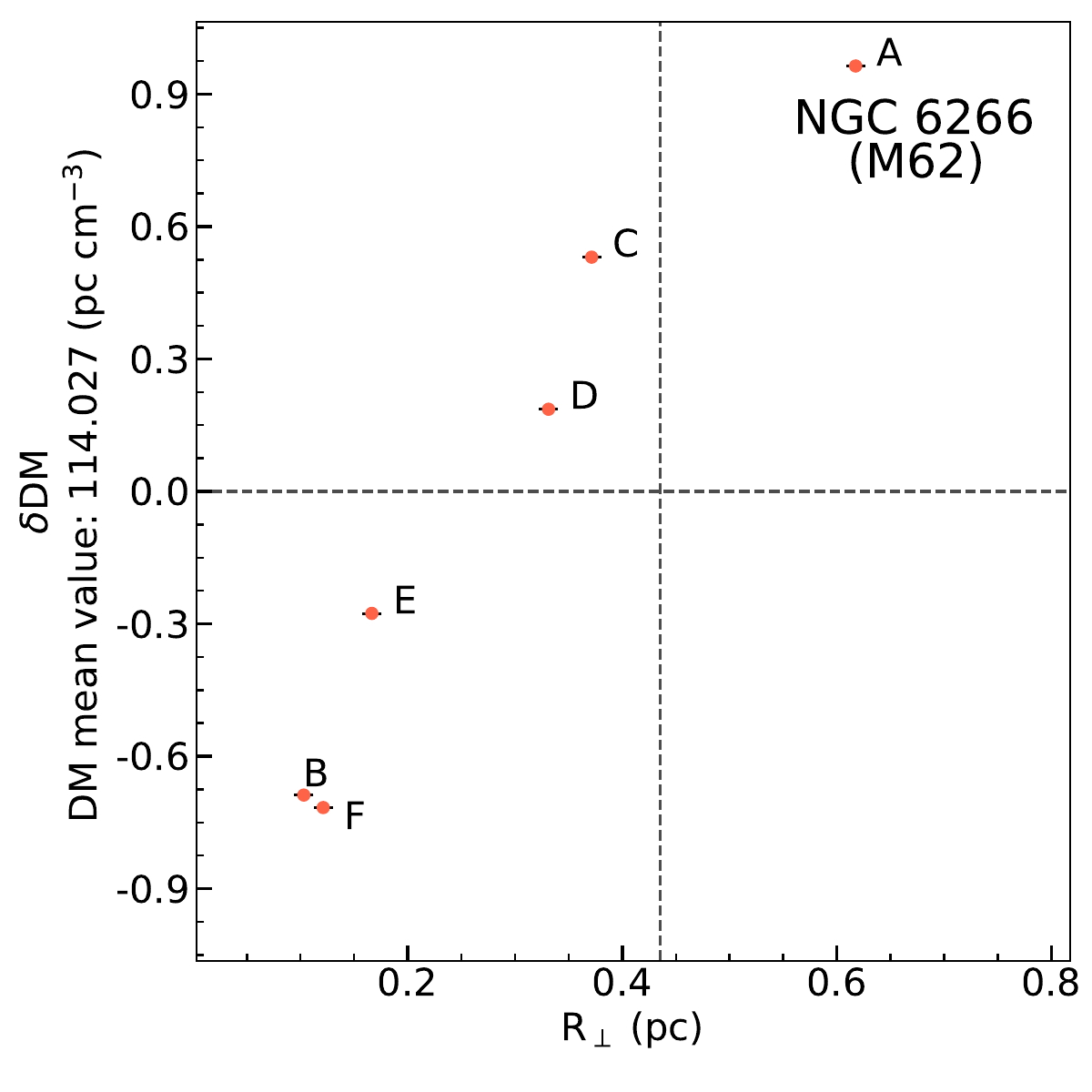}
    \includegraphics[height=4.4cm,width=5.7cm]{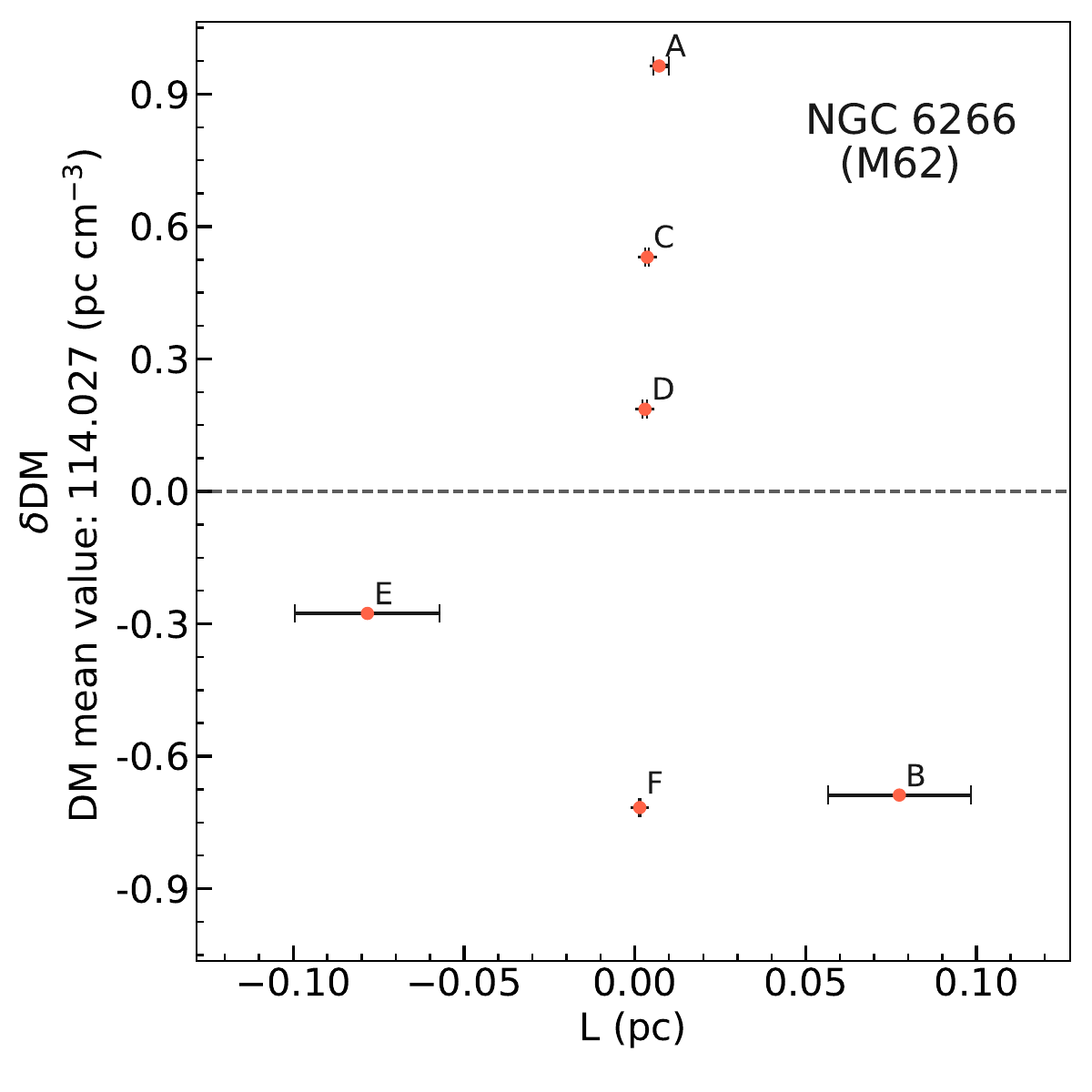}\\
    \includegraphics[height=4.4cm,width=5.7cm]{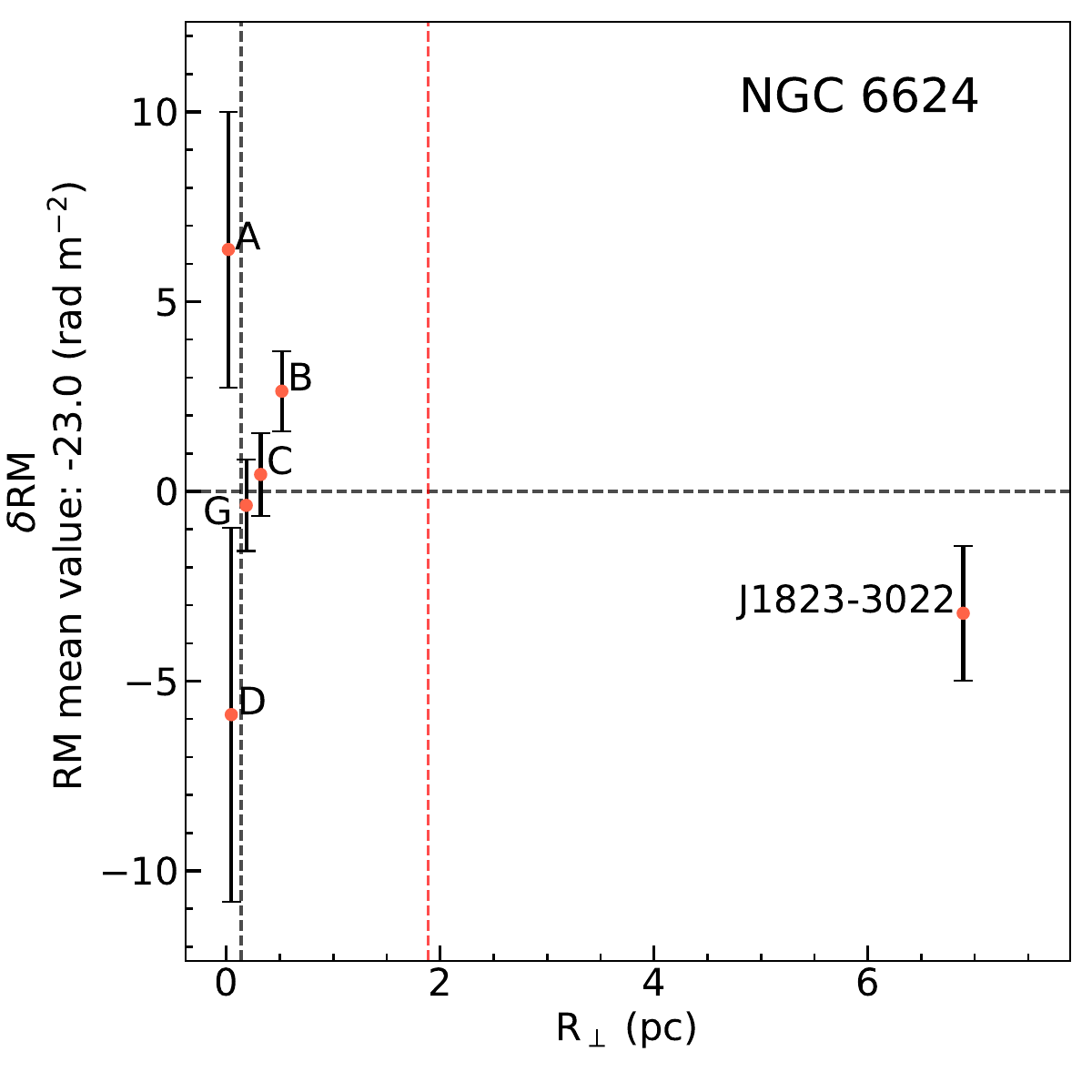}
    \includegraphics[height=4.4cm,width=5.7cm]{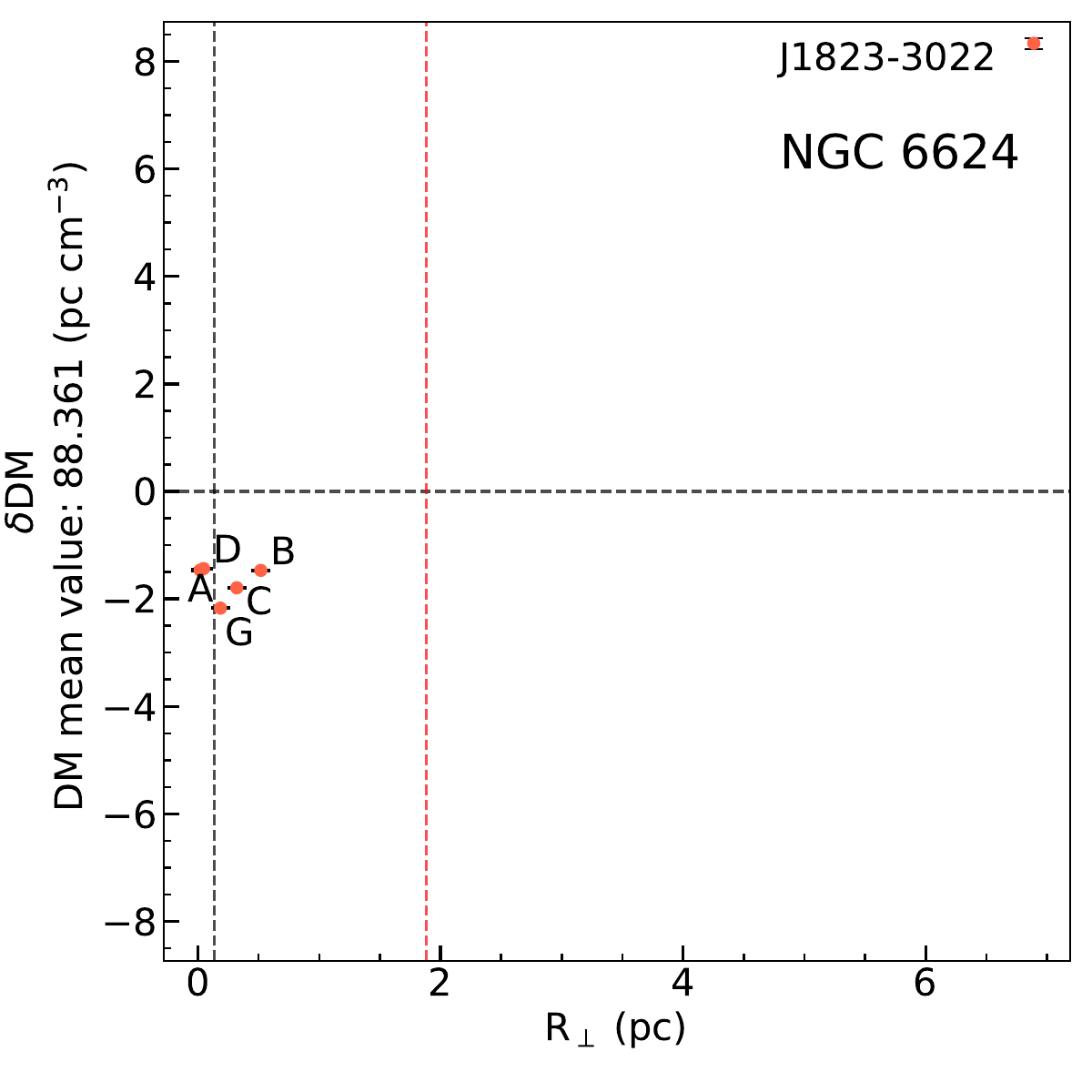}
    \includegraphics[height=4.4cm,width=5.7cm]{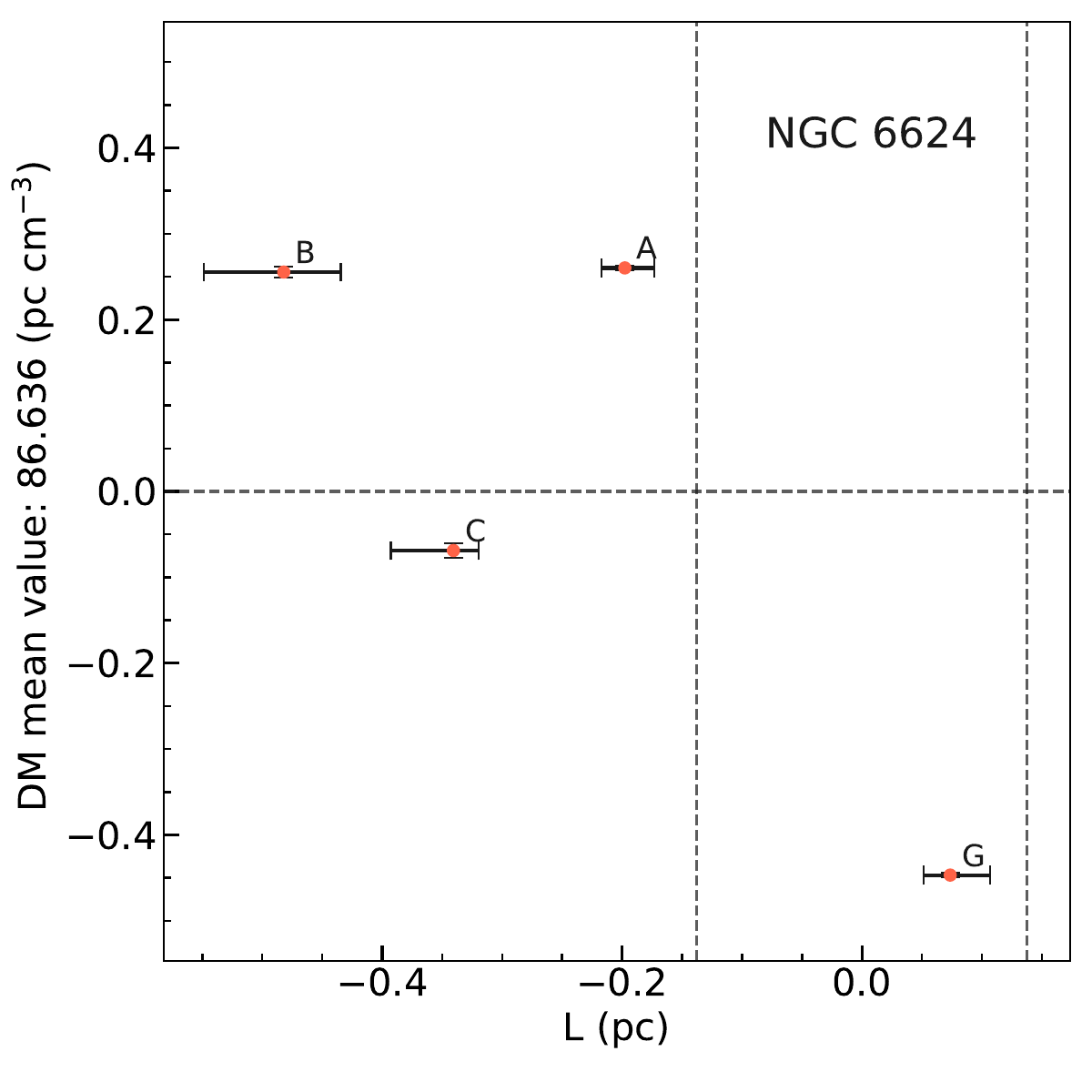}\\
    \includegraphics[height=4.4cm,width=5.7cm]{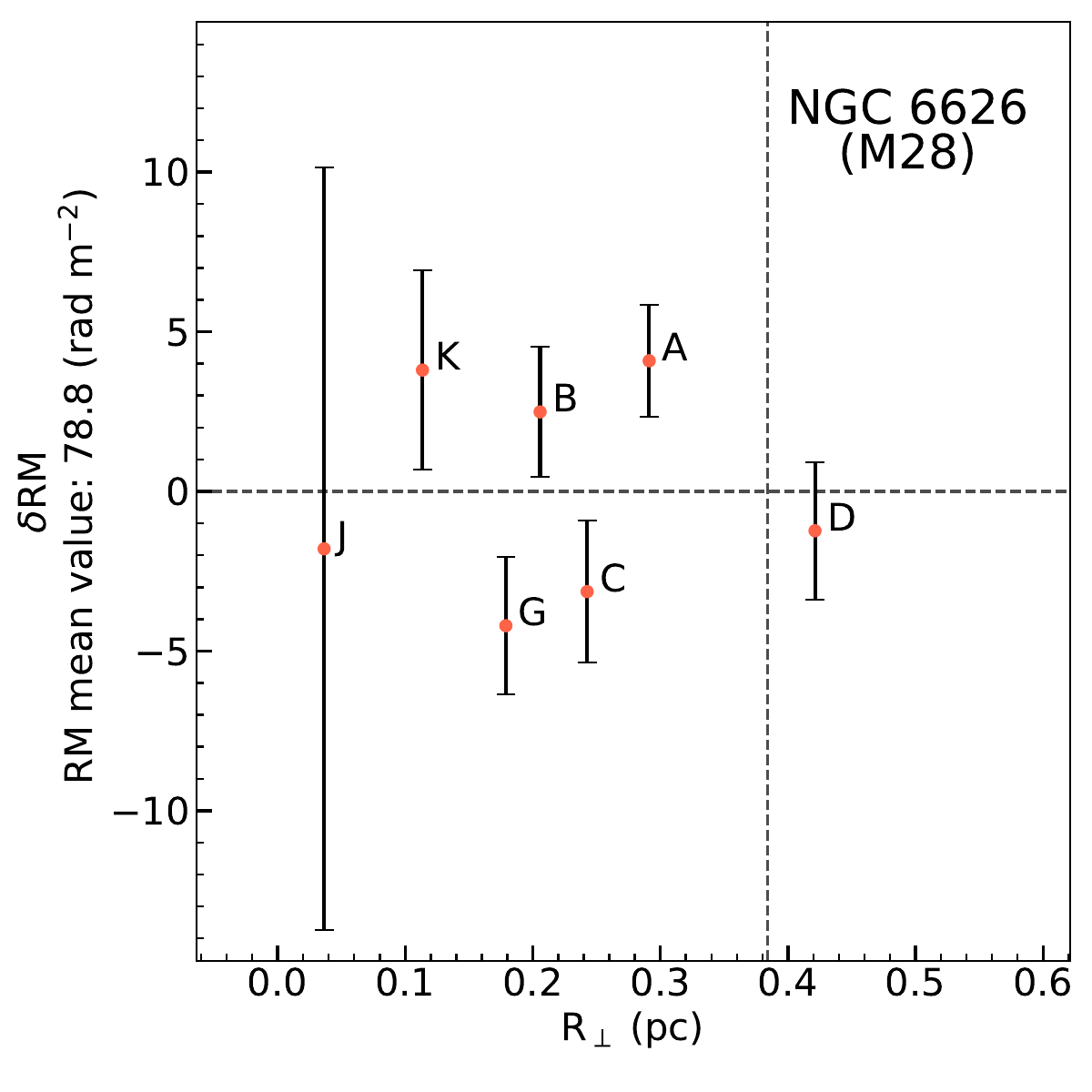}
    \includegraphics[height=4.4cm,width=5.7cm]{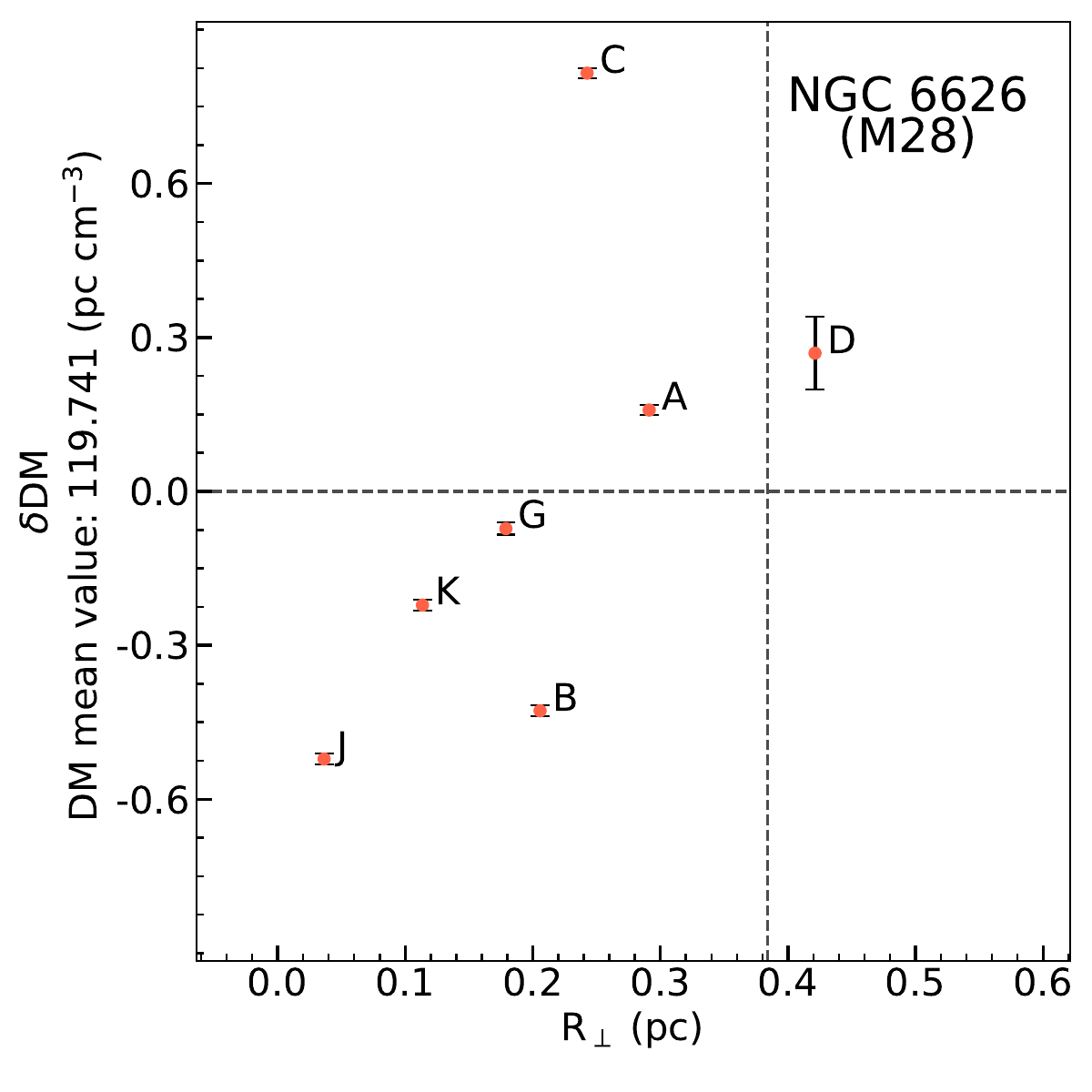}
    \includegraphics[height=4.4cm,width=5.7cm]{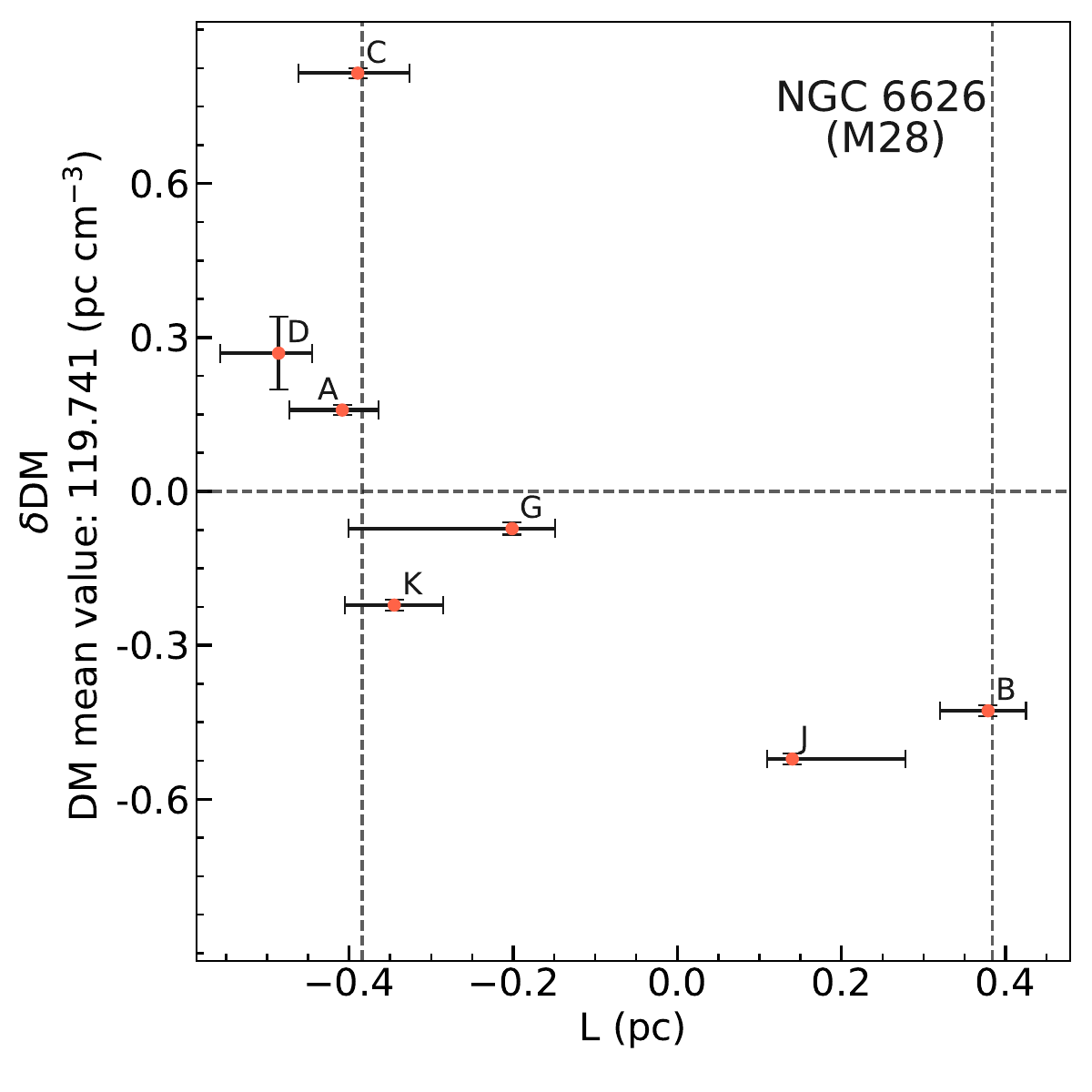}\\
    \includegraphics[height=4.4cm,width=6.1cm]{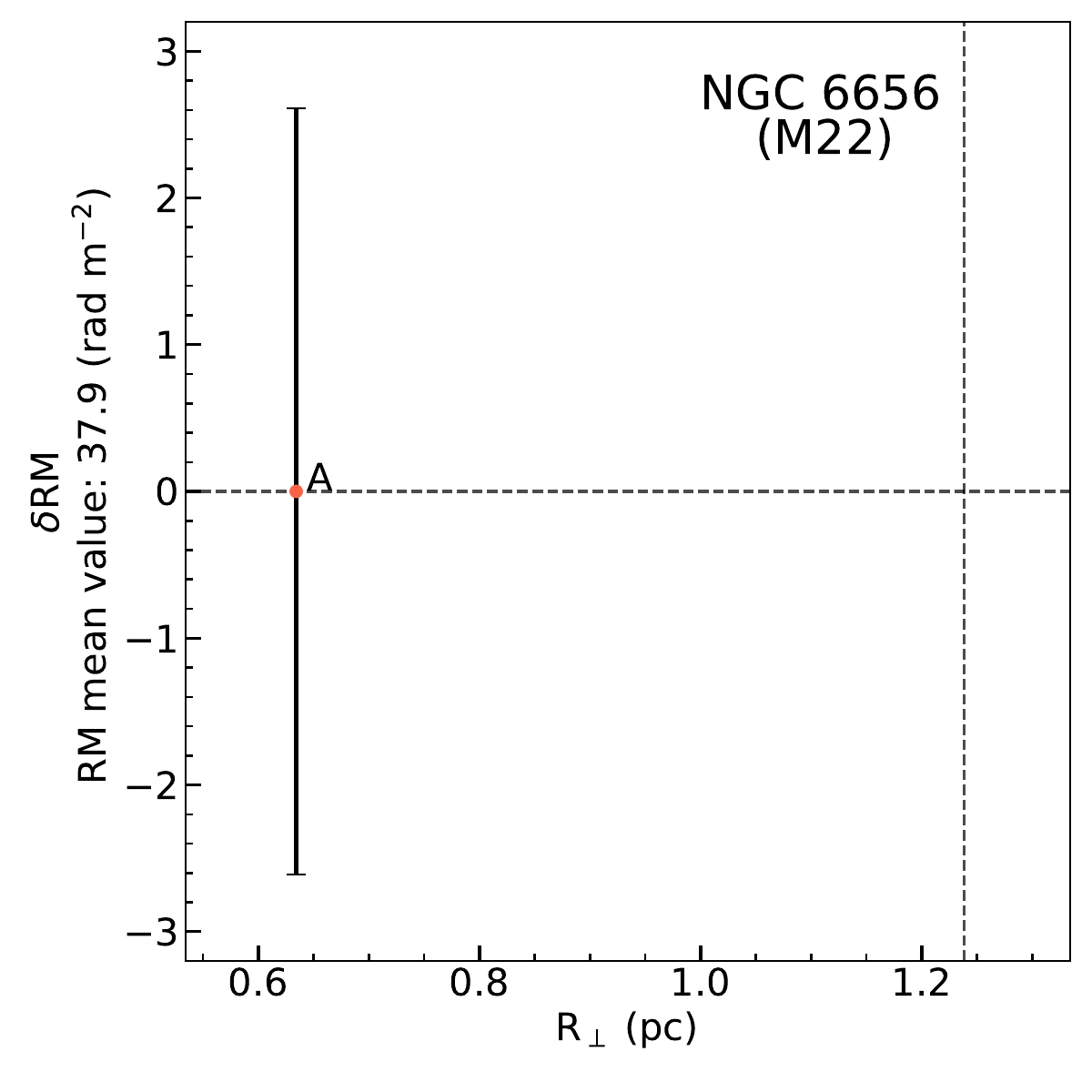}
    \includegraphics[height=4.4cm,width=6.1cm]{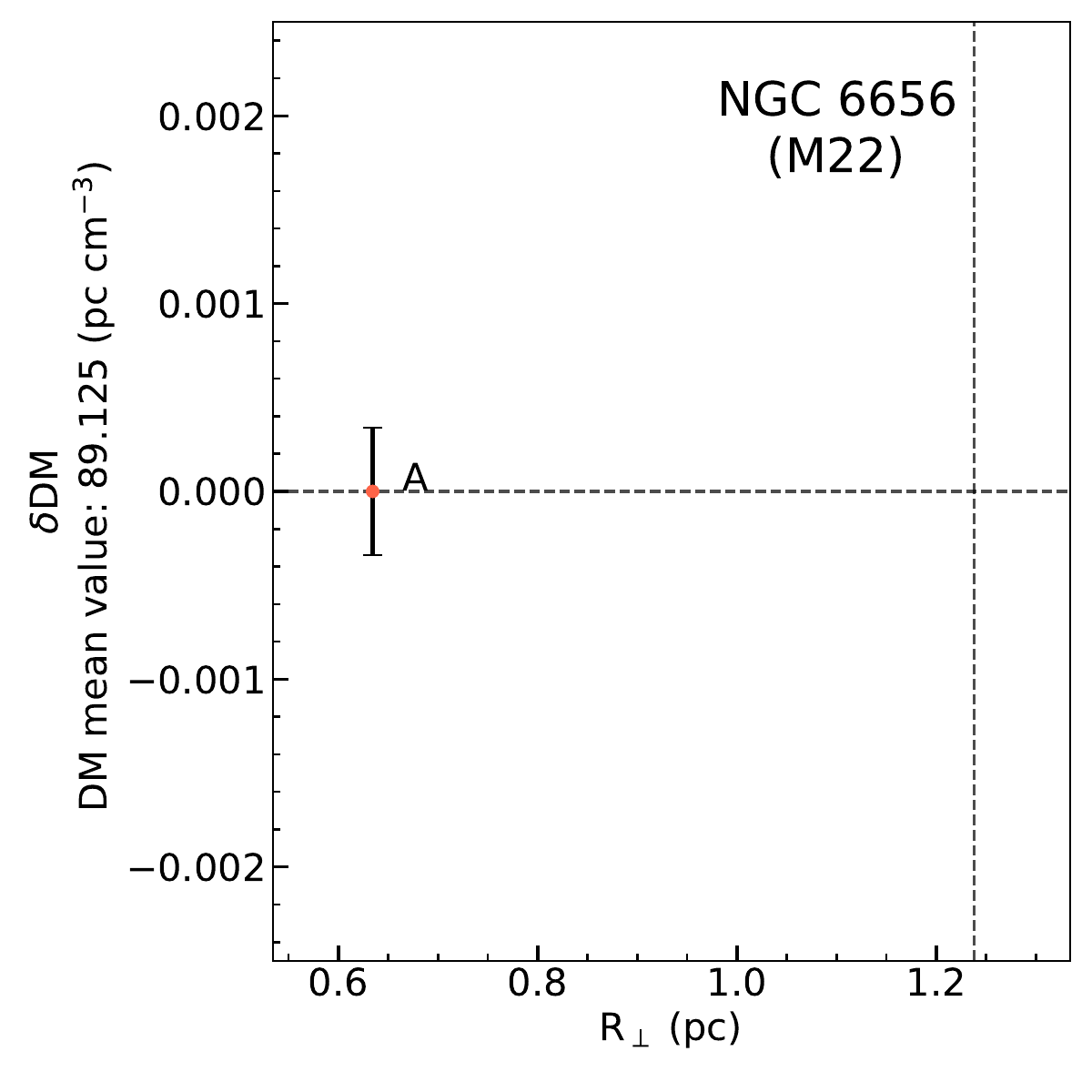}\\
\caption{Pulsars in five bulge clusters. \textit{Left panel:} The difference between measured RM and the mean against position perpendicular offset from the center of GC. \textit{Middle panel:} The difference between the measured DM and the mean against position perpendicular offset from the center of GC. \textit{Right panel:} Measured DM and the mean plotted against the inferred line-of-sight distance (L) of pulsars within the plane of the sky containing the cluster center. Pulsars located far from the cluster center were excluded from the fits to the inferred line-of-sight distance. The vertical lines show the distance from the core radius (black dished line) and half-light radius (red dished line) to the cluster center.}
\label{fig.LpolRM_bulge}
\end{figure*}


\end{document}